\documentclass[prd,showpacs,floatfix,amsmath,amssymb,floatfix]{revtex4}
\usepackage{graphicx,dcolumn,booktabs,bm}
\usepackage{longtable,lscape}
\usepackage{txfonts}
\usepackage{overpic}
\usepackage{multirow}
\usepackage{amssymb}
\usepackage{indentfirst}
\usepackage{feynmf}   %{feynmp}
\usepackage{slashed}  %for Feynman symbols
\usepackage{cases}
\usepackage{color,ulem}
\usepackage{graphicx}
\graphicspath{{Figures/}} %

\begin{document}
%\begin{CJK}{GBK}{}

%%%%%%%%%%%%%%%%%%%%%%%%%%%%%%%%%%%%%%%%%%%%%%%%%%%%%%%%%%%%%%%%%%%%%%%%%%%%%%%%%%%%%%%%%%%%%%%%%%%%%%%%%
%                                    The document begins here                                           %
%%%%%%%%%%%%%%%%%%%%%%%%%%%%%%%%%%%%%%%%%%%%%%%%%%%%%%%%%%%%%%%%%%%%%%%%%%%%%%%%%%%%%%%%%%%%%%%%%%%%%%%%%

\title{Strong decay patterns of the hidden-charm tetraquarks}
\author{Li Ma$^{1}$}\email{lima@pku.edu.cn}
\author{Wei-Zhen Deng$^{1}$}\email{dwz@pku.edu.cn}
\author{Xiao-Lin Chen$^{1}$}\email{chenxl@pku.edu.cn}
\author{Shi-Lin Zhu$^{1,2}$}\email{zhusl@pku.edu.cn}

\affiliation{
$^1$Department of Physics and State Key Laboratory of Nuclear Physics and Technology and Center of High Energy Physics, Peking University, Beijing 100871, China\\
$^2$Collaborative Innovation Center of Quantum Matter, Beijing
100871, China}

\date{\today}% It is always \today, today,

\begin{abstract}

With the spin rearrangement, we have performed a comprehensive
investigation of the decay patterns of the S-wave tetraquarks and
P-wave tetraquarks where the P-wave excitation exists either between
the diquark and anti-diquark pair or inside the diquark. Especially,
we compare the decay patterns of $Y(4260)$ with different inner
structures such as the conventional charmonium, the molecule, the
P-wave tetraquark and the hybrid charmonium. We notice the $J/\psi
\pi\pi$ mode is suppressed in the heavy quark symmetry limit if
$Y(4260)$ is a molecular state. Moreover the hybrid charmonium and
hidden-charm tetraquark have very similar decay patterns. Both of
them decay into the $J/\psi \pi\pi$ and open charm modes easily. We
also discuss the decay patterns of $X(3872)$, $Y(4360)$, and several
charged states such as $Z_c(4020)$. The $h_c\pi^{\pm}$ decay mode
disfavors the tetraquark assumption of $Z_c(4020)$.

\end{abstract}

 \maketitle

%%%%%%%%%%%%%%%%%%%%%%%%%%%%%%%%%%%%%%%%%%%%%%%%
\section{Introduction}\label{sec1}
%%%%%%%%%%%%%%%%%%%%%%%%%%%%%%%%%%%%%%%%%%%%%%%%

Since $X(3872)$ was observed by Belle Collaboration in the
$J/\psi\pi^+\pi^-$ invariant mass spectrum of $B\rightarrow
KJ/\psi\pi^+\pi^-$, many charmonium-like and bottomonium-like states
have been reported by the Belle, CMS, LHCb, CDF, D0, CLEO-c and
BESIII Collaborations
\cite{Beringer:1900zz,zhu-review,Liu:2013waa,belle1,belle2,lhcb,Aubert:2007zz,Wang:2007ea}.
Some states pose a great challenge to the conventional quark model.
Especially, the charged charmonium-like and bottomonium-like states
such as $Z_c(3900)$, $Z_c(4025)$ and $Z_1(4475)$ do not fit into the
conventional charmonium spectroscopy
\cite{Choi:2007wga,Ablikim:2013mio,Liu:2013dau,Xiao:2013iha,Ablikim:2013wzq,Yi:2010aa,Liu:2010hf,Abe:2004zs,Aaltonen:2009tz,Ablikim:2015vvn,Ablikim:2015gda}.
Up to now, these charged $Z_c$ and $Z_b$ states seem to be the best
candidates of the four-quark states. Very recently, two hidden-charm
pentaquarks $P_c(4380)$ and $P_c(4450)$ were reported by LHCb in the
$J/\psi p$ invariant mass spectrum of the $\Lambda\rightarrow J/\psi
pK$ process \cite{Aaij:2015tga}, which enriched our knowledge of the
hidden-charm multiquark systems
\cite{Chen:2015moa,Chen:2015loa,Wang:2015qlf}.

Many theoretical speculations have been proposed to interpret these
XYZ states, such as the kinematics artifacts, conventional
charmonium, hybrid charmonium, di-meson molecules and tetraquarks
\cite{Liu:2013waa,Swanson:2004pp,Maiani:2004vq,Bugg:2004rk,Rosner:2006vc,Li:2004sta,Hogaasen:2005jv,Ebert:2005nc,Barnea:2006sd,Cui:2006mp,Zhu:2005hp,Kou:2005gt,Close:2005iz,Chen:2011xk,Chen:2013wca,Li:2012cs,Chen:2013wva}.
Since some states are very close to the open charm or open bottom
threshold, the molecule assumption becomes quite popular among the
theoretical proposals. For instance, many authors suggest $X(3872)$
as a possible candidate of the $D\bar{D}^{\ast}$ molecules
\cite{Abe:2005ix,Close:2003sg,Voloshin:2003nt,Wong:2003xk,Swanson:2003tb,Tornqvist:2004qy,Suzuki:2005ha,Liu:2008fh,Thomas:2008ja,Lee:2009hy,Li:2012cs,Aubert:2008ae}.
$Y(3940)$, $Y(4140)$ and $Y(4274)$ were proposed as the
$D^{\ast}\bar{D}^{\ast}$, $D_s^{\ast}\bar{D}_s^{\ast}$ and
$D_s\bar{D}_{s0}(2317)$ molecular states in Ref.
\cite{Liu:2009ei,Liu:2008tn,Mahajan:2009pj,Branz:2009yt,Albuquerque:2009ak,Ding:2009vd,Zhang:2009st,Liu:2009iw,Liu:2009pu,Liu:2010hf,He:2011ed}.
The two charged bottomonium-like states $Z_b(10610)$ and
$Z_b(10650)$ were suggested as the loosely bound S-wave
$B\bar{B}^{\ast}$ and $B^{\ast}\bar{B}^{\ast}$ molecules
\cite{Sun:2011uh,Ohkoda:2012rj}. There were many discussions of the
possible molecular assignment of the XYZ states in literature
\cite{He:2013nwa,Zhao:2014gqa,Zhao:2015mga,Ma:2014ofa,Ma:2014zva,Ma:2014zua,Liu:2009ei,Liu:2008tn}.

The hidden-charm molecular states generally lie close to the
open-charm threshold. They were first observed in the hidden-charm
final states such as $J/\psi \pi$. However their open-charm decay
widths are much larger as in the case of $Z_c(3900/4020)$.

Besides the molecular scheme, the tightly bound tetraquark states
are also very interesting. They are expected to be very broad with a
width around several hundred MeV. Moreover, they tend to decay into
the hidden-charm modes easily. With a larger phase space, the
hidden-charm modes should be one of their main decay modes. For
example, the charged state $Z_c(4200)$ decays into $J/\psi\pi$. Its
decay width is more than three hundreds MeV \cite{Chilikin:2014bkk}.
The decay pattern of $Z_c(4200)$ as a tetraquark candidate was
investigated with the QCD sum rule formalism recently
\cite{Chen:2015fsa}. The spectrum of the hidden-charm tetraquarks
were calculated systematically in Refs. \cite{Du:2012wp,Chen:2012pe,Du:2012pn,Chen:2011qu,Chen:2010ze,Chen:2010jd,Jiao:2009ra}. There were other discussions of the
XYZ states as candidates of the tetraquarks
\cite{Maiani:2005pe,Braaten:2013boa}.

$Y(4260)$ was observed in the $J/\psi \pi\pi$ mode using the initial
state radiation technique. Many speculations of its inner structure
have been proposed in the past decade
\cite{Zhu:2005hp,LlanesEstrada:2005hz,Eichten:2005ga,Segovia:2008zz,Li:2009zu,Ebert:2008wm,Ding:2008gr,Wang:2013cya,Li:2013yka,Close:2010wq,Liu:2005ay,Yuan:2005dr,Qiao:2005av,MartinezTorres:2009xb,Chen:2010nv,Wang:2013kra,Cleven:2013mka}.
This vector charmonium-like state is quite broad. However, its main
decay modes have not been observed up to now. The non-observation of
$Y(4260)$ in many decay modes is very puzzling.

Especially, there is no evidence of the open charm decays for
$Y(4260)$, which poses a great challenge to the molecule assumption.
With the spin rearrangement scheme in heavy quark limit, Ma et al.
investigated the decay patterns of the hidden-charm molecular states
and their typical decay ratios \cite{Ma:2014ofa,Ma:2014zva}. The
authors found that the $1^{--}$ molecular states composed of
$D_0\bar{D}^{\ast}$, $D'_1\bar{D}$, $D_1\bar{D}$,
$D'_1\bar{D}^{\ast}$, $D_1\bar{D}^{\ast}$ or $D_2\bar{D}^{\ast}$ do
not decay into $J/\psi\pi^+\pi^-$ if the pion pair comes from
$\sigma$ or $f(980)$ in the heavy quark symmetry limit
\cite{Ma:2014zva}. In other words, the discovery mode
$J/\psi\pi^+\pi^-$ disfavors the molecule interpretation of
$Y(4260)$.

The observation of these exotic XYZ states provides us a platform to
investigate the possible multiquark states beyond the conventional
quark model. The experimental measurement of their decay behavior
may shed light on their underlying structures. Different decay
patterns may reveal their different inner structures.

In the heavy quark symmetry limit, the heavy quark spin within a
hadron can not be flipped in the decay process, which provides us a
useful handle to investigate the decay behavior of hadrons
containing heavy quarks. Within the molecule assumption, the decay
patterns of $Z_c(3900)$, $Z_c(4025)$ and $Z_b(10610)/Z_b(10650)$
were discussed via the spin rearrangement scheme under heavy quark
symmetry \cite{He:2013nwa}. The discussion on the selection rules in
the di-meson molecules was performed in Ref. \cite{Liu:2013rxa}.
Voloshin et al derived useful relations between the rate of the
radiative transitions from $\Upsilon(5S)$ to the hypothetical
isovector molecular bottomonium resonance with negative G-parity
using the spin rearrangement scheme \cite{Voloshin:2011qa}. An
extensive investigations of the decay patterns of the hidden-charm
molecular states with various quantum number can be found in Refs.
\cite{Ma:2014ofa,Ma:2014zva}.

In this work we extend the spin rearrangement scheme to investigate
the decay behavior of the XYZ states as the tetraquark candidates.
We present the results of the hidden-charm decay patterns of the
tetraquark states in the text since the expressions are slightly
simpler. The decay matrix elements of their open charm decays
contain many terms. In order to avoid the complicated and lengthy
expressions, we collect the spin configurations of various open
charm final states in Appendix D. One can check the spin
configurations of the tetraquark states in Appendix A and different
open charm final states in Appendix D. If the initial and final
states have one or more common spin configurations, such a strong
mode is allowed under the heavy quark symmetry. Otherwise, such a
decay is suppressed.

One can compare the experimental decay patterns of the XYZ states
with the theoretical predictions within the conventional quark
model, the molecular and tetraquark schemes to determine the
underlying structures of the XYZ states.

This paper is organized as follows. After the introduction, we give
the general expressions about the color and spin structures of the
tetraquarks in the heavy quark limit in Sec. \ref{sec2}. We present
the decay patterns of the S-wave tetraquarks in Sec. \ref{sec3}. And
we list the decay patterns of the P-wave tetraquarks in Sec.
\ref{sec4}. In Sec. \ref{sec5}, we focus on $Y(4260)$. We discuss
the decay patterns of $X(3872)$, $Z_c(3900)$ etc in Sec. \ref{sec6}.
The last section is the summary. The lengthy expressions and
definitions are collected in Appendix A, B and C.

%%%%%%%%%%%%%%%%%%%%%%%%%%%%%%%%%%%%%%%%%%%%%%%%
\section{The formalism}
\label{sec2}
%%%%%%%%%%%%%%%%%%%%%%%%%%%%%%%%%%%%%%%

In the heavy quark limit, the spin-dependent chromomagnetic
interaction proportional to $1/m_{Q}$ is suppressed, while the
 chromoelectric interaction is spin-independent and dominant.
For a hadron that contains heavy quarks, the total spin of heavy
quarks $S_H$ which is named as the "heavy spin",  and the total
angular momentum $J$ are conserved. The sum of all the angular
momenta other than the heavy spin is referred to the "light spin",
which is also conserved with the definition $\vec{S}_l\equiv
\vec{J}-\vec{S}_H$. The "heavy spin" and "light spin" provide an
effective tool in the study of the structures of hadrons containing
heavy quarks.

Considering a hidden charm $cq\bar{c}\bar{q}$ tetraquark, the
interaction Hamiltonian derived from the one-gluon exchange in the
MIT bag model reads
\begin{equation}
H_{eff}=\sum_{i}m_i-\sum_{i>j}f_{ij}\vec{\lambda}_i\cdot\vec{\lambda}_j\vec{\sigma}_i\cdot\vec{\sigma}_j,\label{k1}
\end{equation}
where $m_i$ refers to the i-th constituent quark mass,
$\vec{\lambda}_i$ and $\vec{\sigma}_i$ are the color and spin
operator respectively, $f_{ij}$ are the coefficients depending on
the specific models and systems
\cite{Jaffe:1976ig,Jaffe:1976ih,Jaffe:1976yi,DeGrand:1975cf}. The
second term is the color-spin interaction which satisfies the
$SU(6)$ symmetry. For the anti-quarks, we have
$\vec{\lambda}_q=-\vec{\lambda}_q^{\star}$ and
$\vec{\sigma}_q=-\vec{\sigma}_q^{\star}$. In general, $f_{ij}$ have
different values for different systems \cite{Cui:2005az}. Here we
use the spin rearrangement scheme to probe the strong decay behavior
of the hidden charm tetraquarks. The specific values of $f_{ij}$ are
irrelevant in this work.

The eigenstates of the effective Hamiltonian in Eq. (\ref{k1}) can
be chosen as $|D_{6cs},D_{3c},J,N\rangle$, where $D_{6cs}$ and
$D_{3c}$ are the $SU(6)_{cs}$ color-spin and $SU(3)_c$ color
representations of the multiquark system respectively. $J$ is the
total angular momentum and $N$ is the total number of quarks and
antiquarks. We want to emphasize that the following discussions also
hold for the general interaction in Eq. (\ref{VVV}).

We employ the diquark models to construct the color spin wave
functions of the hidden charm tetraquarks, which are denoted as
$|cq,D_{6cs},D_{3c},J,2\rangle\otimes|\bar{c}\bar{q},D_{6cs},D_{3c},J,2\rangle$.
The $SU(6)_{cs}$ color-spin representations of the diquark are
$6_{cs}\otimes6_{cs} = 21_{cs}\oplus\bar{15}_{cs}$. For the
anti-diquark, we have $\bar{6}_{cs}\otimes\bar{6}_{cs} =
\bar{21}_{cs}\oplus15_{cs}$. The $SU(3)_c$ color representations of
the diquark are $3_c\otimes3_c = 6_c\oplus\bar{3}_c$, and the
anti-diquark parts are $\bar{3}_c\otimes\bar{3}_c =
\bar{6}_c\oplus3_c$. The diquark couples with the anti-diquark to
form a tetraquark,
\begin{eqnarray*}
21_{cs}\otimes\bar{21}_{cs} &=& 405_{cs}\oplus280_{cs}\oplus35_{cs}\oplus1_{cs} \\
21_{cs}\otimes\bar{15}_{cs} &=& 280_{cs}\oplus35_{cs} \\
15_{cs}\otimes\bar{15}_{cs} &=& 189_{cs}\oplus35_{cs}\oplus1_{cs}.
\end{eqnarray*}

The color-spin wave functions of the S-wave hidden charm tetraquarks
with the quantum numbers $J^P=0^+,1^+,2^+$ are
\begin{eqnarray*}
|1_{cs},1_c,0,4\rangle &=& \sqrt{\frac{6}{7}} |cq,21_{cs},6_c,1,2\rangle|\bar{c}\bar{q},\bar{21}_{cs},\bar{6}_c,1,2\rangle+\sqrt{\frac{1}{7}} |cq,21_{cs},\bar{3}_c,0,2\rangle|\bar{c}\bar{q},\bar{21}_{cs},3_c,0,2\rangle \\
|405_{cs},1_c,0,4\rangle &=& \sqrt{\frac{1}{7}} |cq,21_{cs},6_c,1,2\rangle|\bar{c}\bar{q},\bar{21}_{cs},\bar{6}_c,1,2\rangle-\sqrt{\frac{6}{7}} |cq,21_{cs},\bar{3}_c,0,2\rangle|\bar{c}\bar{q},\bar{21}_{cs},3_c,0,2\rangle\\
|1_{cs},1_c,0,4\rangle &=& \sqrt{\frac{3}{5}} |cq,15_{cs},\bar{3}_c,1,2\rangle|\bar{c}\bar{q},\bar{15}_{cs},3_c,1,2\rangle+\sqrt{\frac{2}{5}} |cq,15_{cs},6_c,0,2\rangle|\bar{c}\bar{q},\bar{15}_{cs},\bar{6}_c,0,2\rangle\\
|189_{cs},1_c,0,4\rangle &=& \sqrt{\frac{2}{5}}
|cq,15_{cs},\bar{3}_c,1,2\rangle|\bar{c}\bar{q},\bar{15}_{cs},3_c,1,2\rangle-\sqrt{\frac{3}{5}}
|cq,15_{cs},6_c,0,2\rangle|\bar{c}\bar{q},\bar{15}_{cs},\bar{6}_c,0,2\rangle,
\end{eqnarray*}
\begin{eqnarray*}
|35_{cs},1_c,1,4\rangle &=& |cq,21_{cs},6_c,1,2\rangle|\bar{c}\bar{q},\bar{21}_{cs},\bar{6}_c,1,2\rangle \\
|35_{cs},1_c,1,4\rangle &=&
|cq,15_{cs},\bar{3}_c,1,2\rangle|\bar{c}\bar{q},\bar{15}_{cs},3_c,1,2\rangle,
\end{eqnarray*}
\begin{eqnarray*}
|35_{cs},1_c,1,4\rangle &=& \sqrt{\frac{1}{3}} |cq,21_{cs},\bar{3}_c,0,2\rangle|\bar{c}\bar{q},\bar{15}_{cs},3_c,1,2\rangle-\sqrt{\frac{2}{3}} |cq,21_{cs},6_c,1,2\rangle|\bar{c}\bar{q},\bar{15}_{cs},\bar{6}_c,0,2\rangle \\
|280_{cs},1_c,1,4\rangle &=& \sqrt{\frac{2}{3}} |cq,21_{cs},\bar{3}_c,0,2\rangle|\bar{c}\bar{q},\bar{15}_{cs},3_c,1,2\rangle+\sqrt{\frac{1}{3}} |cq,21_{cs},6_c,1,2\rangle|\bar{c}\bar{q},\bar{15}_{cs},\bar{6}_c,0,2\rangle\\
|35_{cs},1_c,1,4\rangle &=& \sqrt{\frac{1}{3}} |cq,15_{cs},\bar{3}_c,1,2\rangle|\bar{c}\bar{q},\bar{21}_{cs},3_c,0,2\rangle-\sqrt{\frac{2}{3}} |cq,15_{cs},6_c,0,2\rangle|\bar{c}\bar{q},\bar{21}_{cs},\bar{6}_c,1,2\rangle\\
|280_{cs},1_c,1,4\rangle &=& \sqrt{\frac{2}{3}}
|cq,15_{cs},\bar{3}_c,1,2\rangle|\bar{c}\bar{q},\bar{21}_{cs},3_c,0,2\rangle+\sqrt{\frac{1}{3}}
|cq,15_{cs},6_c,0,2\rangle|\bar{c}\bar{q},\bar{21}_{cs},\bar{6}_c,1,2\rangle,
\end{eqnarray*}
\begin{eqnarray*}
|405_{cs},1_c,2,4\rangle &=& |cq,21_{cs},6_c,1,2\rangle|\bar{c}\bar{q},\bar{21}_{cs},\bar{6}_c,1,2\rangle\\
|189_{cs},1_c,2,4\rangle &=&
|cq,15_{cs},\bar{3}_c,1,2\rangle|\bar{c}\bar{q},\bar{15}_{cs},3_c,1,2\rangle.
\end{eqnarray*}

Besides the heavy and light spin of a hadron, the isospin is another
conserved quantity in the strong decays. The hidden charm
tetraquarks $cq\bar{c}\bar{q}$ can be categorized as the isovector
and isoscalar states with the corresponding color-spin wave
functions
\begin{eqnarray*}
|cq\bar{c}\bar{q}\rangle_1^+ &=& |cu\rangle_{(D_{6cs},D_{3c},J)}\otimes|\bar{c}\bar{d}\rangle_{(D'_{6cs},D'_{3c},J')}\\
|cq\bar{c}\bar{q}\rangle_1^0 &=& \frac{1}{\sqrt{2}}\Big[|cd\rangle_{(D_{6cs},D_{3c},J)}\otimes|\bar{c}\bar{d}\rangle_{(D'_{6cs},D'_{3c},J')}-|cu\rangle_{(D_{6cs},D_{3c},J)}\otimes|\bar{c}\bar{u}\rangle_{(D'_{6cs},D'_{3c},J')}\Big] \\
|cq\bar{c}\bar{q}\rangle_1^- &=& |cd\rangle_{(D_{6cs},D_{3c},J)}\otimes|-\bar{c}\bar{u}\rangle_{(D'_{6cs},D'_{3c},J')},\\
|cq\bar{c}\bar{q}\rangle_1^0 &=&
\frac{1}{\sqrt{2}}\Big[|cd\rangle_{(D_{6cs},D_{3c},J)}\otimes|\bar{c}\bar{d}\rangle_{(D'_{6cs},D'_{3c},J')}+|cu\rangle_{(D_{6cs},D_{3c},J)}\otimes|\bar{c}\bar{u}\rangle_{(D'_{6cs},D'_{3c},J')}\Big].
\end{eqnarray*}
The heavy quark spin is not flipped in the strong decays. The heavy
spin, light spin, total angular momentum and isospin are conserved
quantities. For a neutral hidden charm system, its G-parity and
C-parity are also conserved. We define the C-parity eigenstates for
the neutral partners of the hidden charm tetraquarks following the
standard convention
\begin{eqnarray*}
\frac{1}{\sqrt{2}}\Bigg\{\Big[\big([cq]_{(d_6,d_3,S)}\otimes[\bar{c}\bar{q}]_{(d'_6,d'_3,S')}\big)_{(D_{6},D_{3},K)}\otimes
L\Big]_{(D_{6},D_{3},J)}\pm\Big[\big([cq]_{(d'_6,d'_3,S')}\otimes[\bar{c}\bar{q}]_{(d_6,d_3,S)}\big)_{(D_{6},D_{3},K)}\otimes
L\Big]_{(D_{6},D_{3},J)}\Bigg\},
\end{eqnarray*}
where $S$, $d_6$ and $d_3$ denote the total spin, $SU(6)_{cs}$
color-spin representations and $SU(3)_c$ color representations of
the diquark $[cq]$, and $S'$, $d'_6$ and $d'_3$ for the
anti-diquarks $[\bar{c}\bar{q}]$. $K$ represents the total spin of
the four quarks. $L$ is the orbital angular momenta within the
tetraquark. $J$, $D_6$ and $D_3$ denote the total spin, $SU(6){cs}$
color-spin and $SU(3)_c$ color representations of tetraquarks
$[cq\bar{c}\bar{q}]$. The $\pm$ corresponds to C-parity
$C\doteq\pm$.

The possible $J^{PC}$ quantum numbers of the neutral partners of a
hidden charm tetraquark without any orbital excitation are $0^{++}$,
$1^{+-}$, $1^{++}$ and $2^{++}$. Within the framework of the spin
rearrangement scheme, the total spin can be recoupled from the total
spin of the heavy quark pair $c\bar{c}$ and the total spin of the
light quark pair $q\bar{q}$. If the total spin of either the heavy
quark pair or light quark pair is $1$ and the total spin of the
other quark pair is zero, the total spin of the tetraquark is $1$
while it has negative C-parity. If both the heavy quark pair and
light quark pair have the total spin $1$, the total spin of the
tetraquark can also be $1$ but it has positive C-parity. The S-wave
tetraquark can carry either positive or negative C-parity when its
total spin is $1$. This argument can be illustrated by the expanded
color-spin structures in the following.

If there is a P-wave excitation within the hidden charm tetraquark,
its parity is negative. Now the C-parity of the neutral partner of a
P-wave hidden charm tetraquark is rather complicated. The charm
quark $c$ and the light quark $q$ constitute the diquark $[cq]$,
$\bar{c}$ and $\bar{q}$ constitute the anti-diquark
$[\bar{c}\bar{q}]$. If the P-wave excitation exists between the
charm quark and light quark in the diquark (or anti-diquark), the
P-wave excitation does not correlate with the charge conjugation
transformation and does not contribute to its $C$ parity. However,
if the P-wave excitation exists between the diquark and anti-quark
pair, the $C$ transformation and P-wave excitation are correlated
with each other, which is the same as in the neutral meson case
where $C=(-)^{L+S}$. If one treats the antidiquark and diquark as an
effective quark and antiquark, one arrives at the same C-parity
formula when there exists an orbital excitation between the diquark
and antidiquark pair. In other words, the different location of the
P-wave excitation affects the C-parity of the tetraquark. We need to
distinguish these two kinds of the P-wave excitations and give an
elaborate discussion in the next section.

The decomposition of the total spin of the above tetraquarks into
the heavy spin and light spin will cast light on their decay
behavior. We employ the color-spin re-coupling formula in analyzing
the general color-spin structure as in Ref. \cite{Cui:2005az}. For
instance, a physical hidden charm tetraquark must be singlet in
color $SU(3)_c$ presentation. Its color-spin structures can be
written as
\begin{eqnarray*}
&&\quad\Bigg\{\Big[[c_1q_2D_{6cs}^{12}D_{3c}^{12}S^{12}]\otimes[\bar{c}_3\bar{q}_4D_{6cs}^{34}D_{3c}^{34}S^{34}]\Big]_{(D_{6cs},K)}\otimes L\Bigg\}_{(D_{6cs},J)} \\
&&=\sum_{D_{3c}^{13},D_{3c}^{24}}R_c[D_{3c}^{12},D_{3c}^{34};D_{3c}^{13},D_{3c}^{24};D_{6cs}]\sum_{S^{24},S_H,S_l}R_s[S^{12},S^{34},K,L;S^{24},S_H,S_l;J] \\
&&\quad\times\Bigg\{\Big[c_1\bar{c}_3D_{6cs}^{13}D_{3c}^{13}S_H\Big]\otimes\bigg[\Big([q_2\bar{q}_4S^{24}]\otimes
L\Big)D_{6cs}^{24}D_{3c}^{24}S_l\bigg]\Bigg\}_{(D_{6cs},J)},
\end{eqnarray*}
where the color-recoupling coefficients $R_c$ are
\begin{eqnarray*}
&&\quad R_c[(\lambda^{12}\mu^{12})(\lambda^{34}\mu^{34});(\lambda^{13}\mu^{13})(\lambda^{24}\mu^{24})] \\
&&=(-1)^{\lambda^{12}+\mu^{12}+\lambda^{13}+\mu^{13}}U[(10)(10)(10)(01);(\lambda^{12}\mu^{12})(\lambda^{13}\mu^{13})],
\end{eqnarray*}
with $SU(3)_c$ Racah coefficients $U$ which were discussed in Refs.
\cite{Strottman:1979et,So:1979nw}. The angular momentum recoupling
coefficients $R_s$ are
\begin{eqnarray*}
&&\quad R_s[S^{12},S^{34},K,L;S^{24},S_H,S_l;J] \\
&&=(-1)^{S_H+S^{24}+L+J}\sqrt{(2S^{12}+1)(2S^{34}+1)(2S^{13}+1)(2S_H+1)(2S_l+1)(2K+1)} \\
&&\quad\times \left\{
             \begin{array}{ccc}
               \frac{1}{2} & \frac{1}{2} & S^{12} \\
               \frac{1}{2} & \frac{1}{2} & S^{34} \\
               S_H & S^{24} & K \\
             \end{array}
           \right\}
  \left\{
    \begin{array}{ccc}
      S^{24} & S_H & K \\
      J & L & S_l \\
    \end{array}
  \right\},
\end{eqnarray*}
where $q_2$ and $q_4$ are the light quarks $u$ and $d$. We use the
subscripts $H$ and $l$ to distinguish the heavy and light spins of a
tetraquark. $S_H$ and $S_l$ denote the heavy and light spins after
the color-spin rearrangement. The notation
$\{[c_1\bar{c}_3D_{6cs}^{13}D_{3c}^{13}S_H]\otimes[([q_2\bar{q}_4S^{24}]\otimes
L)D_{6cs}^{24}D_{3c}^{24}S_l]\}_{(D_{6cs},J)}$ means that the spins
of the $c$ and $\bar{c}$ quarks are coupled into the heavy quark
spin $S_H$ while the spins of the $q$ and $\bar{q}$ quarks are
coupled into the light quark spin $S^{24}$. And then $S^{24}$
couples with the orbital angular momentum $L$ to form the light spin
$S_l$. The notations $D_{6cs}^{ij}$ and $D_{3c}^{ij}$ denote the
color-spin and color representations of the corresponding parts.

Applying the conventional definition for the C-parity eigenstates,
we can express the neutral parts of the tetraquarks and its isospin
partners as
\begin{eqnarray*}
|cq\bar{c}\bar{q}\rangle_1^+ &=& \frac{1}{\sqrt{2}}\Bigg\{\bigg[\Big([cu]_{d_6d_3S}[\bar{c}\bar{d}]_{d'_6d'_3S'}\Big)_{D_6K}\otimes L\bigg]_{D_6J}+\tilde{c}\bigg[\Big([cu]_{d'_6d'_3S'}[\bar{c}\bar{d}]_{d_6d_3S}\Big)_{D_6K}\otimes L\bigg]_{D_6J}\Bigg\} \\
|cq\bar{c}\bar{q}\rangle_1^0 &=& \frac{1}{2}\Bigg\{\bigg[\Big([cd]_{d_6d_3S}[\bar{c}\bar{d}]_{d'_6d'_3S'}-[cu]_{d_6d_3S}[\bar{c}\bar{u}]_{d'_6d'_3S'}\Big)_{D_6K}\otimes L\bigg]_{D_6J}\\
&&+\tilde{c}\bigg[\Big([cd]_{d'_6d'_3S'}[\bar{c}\bar{d}]_{d_6d_3S}-[cu]_{d'_6d'_3S'}[\bar{c}\bar{u}]_{d_6d_3S}\Big)_{D_6K}\otimes L\bigg]_{D_6J}\Bigg\}\\
|cq\bar{c}\bar{q}\rangle_1^- &=&
\frac{1}{\sqrt{2}}\Bigg\{\bigg[\Big([cd]_{d_6d_3S}[-\bar{c}\bar{u}]_{d'_6d'_3S'}\Big)_{D_6K}\otimes
L\bigg]_{D_6J}+\tilde{c}\bigg[\Big([cd]_{d'_6d'_3S'}[-\bar{c}\bar{u}]_{d_6d_3S}\Big)_{D_6K}\otimes
L\bigg]_{D_6J}\Bigg\},
\end{eqnarray*}
and the isoscalar part of the hidden charm tetraquarks can be
rewritten as
\begin{eqnarray*}
|cq\bar{c}\bar{q}\rangle_1^0 &=& \frac{1}{2}\Bigg\{\bigg[\Big([cd]_{d_6d_3S}[\bar{c}\bar{d}]_{d'_6d'_3S'}+[cu]_{d_6d_3S}[\bar{c}\bar{u}]_{d'_6d'_3S'}\Big)_{D_6K}\otimes L\bigg]_{D_6J}\\
&&+\tilde{c}\bigg[\Big([cd]_{d'_6d'_3S'}[\bar{c}\bar{d}]_{d_6d_3S}+[cu]_{d'_6d'_3S'}[\bar{c}\bar{u}]_{d_6d_3S}\Big)_{D_6K}\otimes
L\bigg]_{D_6J}\Bigg\},
\end{eqnarray*}
where the coefficient $\tilde{c}$ is $+1$ or $-1$ for the positive
or negative C-parity respectively. The subscripts $d_6^{(\prime)}$,
$d_3^{(\prime)}$ and $S^{(\prime)}$ denote the color-spin
representations, color representations and total spins of
corresponding parts. $D_6$ represent the color-spin representations
of the total systems. $K$, $L$ and $J$ denote the total spin of the
four quarks, the orbital angular momenta inside the system and the
total angular momentum respectively.

With the color-spin rearrangement scheme, the total angular momentum
and $SU(3)_c$ color singlet of the system are decomposed into heavy
spin, light spin and their corresponding $SU(3)_c$ color
representions. The color-spin structures of the isovector states
after recoupling can be obtained as
\begin{eqnarray}
|cq\bar{c}\bar{q}\rangle_1^+ &=& \frac{1}{\sqrt{2}}\Bigg\{\sum R_cR_s \bigg[[c\bar{c}]_{D_6^{13}D_3^{13}S_H}\otimes\Big([u\bar{d}]_{S^{24}}\otimes L\Big)_{D_6^{24}D_3^{24}S_l}\bigg]_{D_6J}|(cu)_{d_6d_3S}(\bar{c}\bar{d})_{d'_6d'_3S'}\rangle \nonumber\\
&&\quad+\tilde{c}\sum R_cR_s \bigg[[c\bar{c}]_{\bar{D}_6^{13}\bar{D}_3^{13}\bar{S}_H}\otimes\Big([u\bar{d}]_{\bar{S}^{24}}\otimes L\Big)_{\bar{D}_6^{24}\bar{D}_3^{24}\bar{S}_l}\bigg]_{D_6J} |(cu)_{d'_6d'_3S'}(\bar{c}\bar{d})_{d_6d_3S}\rangle\Bigg\} \label{r1}\\
|cq\bar{c}\bar{q}\rangle_1^0 &=& \frac{1}{2}\Bigg\{\sum R_cR_s \bigg[[c\bar{c}]_{D_6^{13}D_3^{13}S_H}\otimes\Big([\frac{d\bar{d}-u\bar{u}}{\sqrt{2}}]_{S^{24}}\otimes L\Big)_{D_6^{24}D_3^{24}S_l}\bigg]_{D_6J} \nonumber\\
&&\quad\times \bigg(\frac{|(cd)_{d_6d_3S}(\bar{c}\bar{d})_{d'_6d'_3S'}\rangle-|(cu)_{d_6d_3S}(\bar{c}\bar{u})_{d'_6d'_3S'}\rangle}{\sqrt{2}}\bigg) \nonumber\\
&&\quad+\tilde{c}\sum R_cR_s \bigg[[c\bar{c}]_{\bar{D}_6^{13}\bar{D}_3^{13}\bar{S}_H}\otimes\Big([\frac{d\bar{d}-u\bar{u}}{\sqrt{2}}]_{\bar{S}^{24}}\otimes L\Big)_{\bar{D}_6^{24}\bar{D}_3^{24}\bar{S}_l}\bigg]_{D_6J}\nonumber\\
&&\quad\times \bigg(\frac{|(cd)_{d'_6d'_3S'}(\bar{c}\bar{d})_{d_6d_3S}\rangle-|(cu)_{d'_6d'_3S'}(\bar{c}\bar{u})_{d_6d_3S}\rangle}{\sqrt{2}}\bigg)\Bigg\} \\
|cq\bar{c}\bar{q}\rangle_1^- &=& \frac{1}{\sqrt{2}}\Bigg\{\sum R_cR_s \bigg[[c\bar{c}]_{D_6^{13}D_3^{13}S_H}\otimes\Big([-d\bar{u}]_{S^{24}}\otimes L\Big)_{D_6^{24}D_3^{24}S_l}\bigg]_{D_6J}\nonumber|-(cd)_{d_6d_3S}(\bar{c}\bar{u})_{d'_6d'_3S'}\rangle \nonumber\\
&&\quad+\tilde{c}\sum R_cR_s
\bigg[[c\bar{c}]_{\bar{D}_6^{13}\bar{D}_3^{13}\bar{S}_H}\otimes\Big([-d\bar{u}]_{\bar{S}^{24}}\otimes
L\Big)_{\bar{D}_6^{24}\bar{D}_3^{24}\bar{S}_l}\bigg]_{D_6J}|-(cd)_{d'_6d'_3S'}(\bar{c}\bar{u})_{d_6d_3S}\rangle\Bigg\}.
\end{eqnarray}
Similarly, the isoscalar partner of the states can be expanded as
\begin{eqnarray}
|cq\bar{c}\bar{q}\rangle_1^0 &=& \frac{1}{2}\Bigg\{\sum R_cR_s \bigg[[c\bar{c}]_{D_6^{13}D_3^{13}S_H}\otimes\Big([\frac{d\bar{d}+u\bar{u}}{\sqrt{2}}]_{S^{24}}\otimes L\Big)_{D_6^{24}D_3^{24}S_l}\bigg]_{D_6J} \nonumber\\
&&\quad\times \bigg(\frac{|(cd)_{d_6d_3S}(\bar{c}\bar{d})_{d'_6d'_3S'}\rangle+|(cu)_{d_6d_3S}(\bar{c}\bar{u})_{d'_6d'_3S'}\rangle}{\sqrt{2}}\bigg) \nonumber\\
&&\quad+\tilde{c}\sum R_cR_s \bigg[[c\bar{c}]_{\bar{D}_6^{13}\bar{D}_3^{13}\bar{S}_H}\otimes\Big([\frac{d\bar{d}+u\bar{u}}{\sqrt{2}}]_{\bar{S}^{24}}\otimes L\Big)_{\bar{D}_6^{24}\bar{D}_3^{24}\bar{S}_l}\bigg]_{D_6J}\nonumber\\
&&\quad\times
\bigg(\frac{|(cd)_{d'_6d'_3S'}(\bar{c}\bar{d})_{d_6d_3S}\rangle+|(cu)_{d'_6d'_3S'}(\bar{c}\bar{u})_{d_6d_3S}\rangle}{\sqrt{2}}\bigg)\Bigg\}.\label{r2}
\end{eqnarray}

In Eqs. (\ref{r1})-(\ref{r2}), we have explicitly included the
flavor wave function, for instance,
$|(cu)_{d_6d_3S}(\bar{c}\bar{d})_{d'_6d'_3S'}\rangle$. Here we use
the color-spin, color representations and the total spin of diquark
or anti-diquark to distinguish different flavor but with the same
constituents and total representations. If the tetraquarks have the
same constituents and total representations but their inner diquarks
or anti-diquarks have the different representations, these
tetraquarks should be different physical states. The
orthogonalization of these different states are guaranteed when we
label their flavor wave functions with the color-spin, color
representations and the total spin of their diquark or anti-diquark.

In order to probe the decay pattern of the hidden charm tetraquarks
under heavy quark symmetry, we also decompose the charmonium final
states into the heavy spin and light spin, which read
\begin{eqnarray}
    |\eta_c(1^1S_0) \rangle &=& |(0_H^-\otimes0_l^+)_0^{-+}\rangle|(c\bar{c})_{1_{cs}1_c0}\rangle,\label{f1}\\
  | J/\psi(1^3S_1)\rangle &=& |(1_H^-\otimes0_l^+)_1^{--}\rangle|(c\bar{c})_{35_{cs}1_c1}\rangle,\\
  | h_c(1^1P_1)\rangle &=&| (0_H^-\otimes1_l^-)_1^{+-}\rangle|(c\bar{c})_{1_{cs}1_c1}\rangle,\\
  | \chi_{c0}(1^3P_0) \rangle&=&| (1_H^-\otimes1_l^-)_0^{++}\rangle|(c\bar{c})_{35_{cs}1_c0}\rangle,\\
   |\chi_{c1}(1^3P_1) \rangle&=&| (1_H^-\otimes1_l^-)_1^{++}\rangle|(c\bar{c})_{35_{cs}1_c1}\rangle,\\
  | \chi_{c2}(1^3P_2)\rangle &=&| (1_H^-\otimes1_l^-)_2^{++}\rangle|(c\bar{c})_{35_{cs}1_c2}\rangle,\\
   |\eta_{c2}(1^1D_2)\rangle &=&| (0_H^-\otimes2_l^+)_2^{-+}\rangle|(c\bar{c})_{1_{cs}1_c2}\rangle,\\
   |\psi(1^3D_1)\rangle &=&| (1_H^-\otimes2_l^+)_1^{--}\rangle|(c\bar{c})_{35_{cs}1_c1}\rangle,\\
  | \psi(1^3D_2)\rangle &=&| (1_H^-\otimes2_l^+)_2^{--}\rangle|(c\bar{c})_{35_{cs}1_c2}\rangle,\\
  | \psi(1^3D_3)\rangle &=&| (1_H^-\otimes2_l^+)_3^{--}\rangle|(c\bar{c})_{35_{cs}1_c3}\rangle.\label{f2}
  \end{eqnarray}

Following our previous work, the flavor wave function is defined as
$|(c\bar{c})\rangle\equiv\frac{1}{\sqrt{2}}(|\bar{c}c\rangle+|c\bar{c}\rangle)$
\cite{Ma:2014ofa,Ma:2014zva}. The superscripts $+$ and $-$ inside
the parentheses represent the positive and negative parity of the
corresponding parts respectively, while the superscripts $-+$
outside the parentheses correspond to the quantum numbers $PC$. The
subscripts $0,1,2,3$ outside the parentheses denote the total
angular momentum $J$ of the charmonium. The subscripts outside in
the flavor wave function have the same meaning as in Eqs.
(\ref{f1})-(\ref{f2}). The $C$ parity of the charmonium is reflected
in the spin wave functions, i.e., $C=(-1)^{S_H+S_l}$. Obviously, the
color representation of those observed charmonium is singlet.

We also introduce the color-spin structures of the light mesons,
 \begin{eqnarray}
   |\pi^+ \rangle &=&| (0_H^+\otimes 0_l^-)_0^{-+}\rangle|(u\bar{d})_{1_{cs},1_c,0}  \rangle, \\
   |\pi^0 \rangle &=&| (0_H^+\otimes 0_l^-)_0^{-+}\rangle|\frac{1}{\sqrt{2}}(d\bar{d}-u\bar{u})_{1_{cs},1_c,0}  \rangle, \\
   |\pi^- \rangle &=&| (0_H^+\otimes 0_l^-)_0^{-+}\rangle|-(d\bar{u})_{1_{cs},1_c,0}  \rangle, \label{n1}\\
   |\rho^+\rangle &=& |(0_H^+\otimes 1_l^-)_1^{--}\rangle|(u\bar{d})_{35_{cs},1_c,1}  \rangle,\\
   |\rho^0\rangle &=& |(0_H^+\otimes 1_l^-)_1^{--}\rangle|\frac{1}{\sqrt{2}}(d\bar{d}-u\bar{u})_{35_{cs},1_c,1}  \rangle,\\
   |\rho^-\rangle &=& |(0_H^+\otimes 1_l^-)_1^{--}\rangle|-(d\bar{u})_{35_{cs},1_c,1}  \rangle,\label{n2}\\
   |\eta \rangle &=&| (0_H^+\otimes 0_l^-)_0^{-+}\rangle|\frac{1}{\sqrt{2}}(d\bar{d}+u\bar{u})_{1_{cs},1_c,0}  \rangle, \label{n3}\\
   |\omega\rangle &=& |(0_H^+\otimes 1_l^-)_1^{--}\rangle|\frac{1}{\sqrt{2}}(d\bar{d}+u\bar{u})_{35_{cs},1_c,1}  \rangle,\label{n4}\\
   |\sigma \rangle &=&| (0_H^+\otimes 0_l^+)_0^{++}\rangle|\frac{1}{\sqrt{2}}(d\bar{d}+u\bar{u})_{1_{cs},1_c,0}  \rangle. \label{n5}
 \end{eqnarray}
The orthogonalization of the color-spin wave functions leads to
 \begin{eqnarray*}
   &&\bigg\langle\bigg(a_{D_6^HD_3^H}^H\otimes b_{D_6^LD_3^L}^L\bigg)_{D_6D_3J}^{pc}\bigg|\bigg(c_{\bar{D}_6^H\bar{D}_3^H}^H\otimes d_{\bar{D}_6^L\bar{D}_3^L}^L\bigg)_{D'_6D'_3J'}^{p'c'}\bigg\rangle \\
   &&=\delta_{ac}\delta_{bd}\delta_{JJ'}\delta_{pp'}\delta_{cc'}\delta_{D_6D'_6}\delta_{D_3D'_3}\delta_{D_6^H\bar{D}_6^H}\delta_{D_3^H\bar{D}_3^H}\delta_{D_6^L\bar{D}_6^L}\delta_{D_3^L\bar{D}_3^L},
  \end{eqnarray*}
where the superscripts $p^{(\prime)}$ and $c^{(\prime)}$ represent
the parity and $C$ parity, respectively. The superscripts $H$ and
$L$ denote the heavy spin part and light spin part. $D^H_6$ and
$\bar{D}^H_6$ represent the color-spin $SU(6)_{cs}$ representations
of the heavy quarks. Similarly, $D^L_6$ and $\bar{D}^L_6$ represent
the color-spin $SU(6)_{cs}$ representations of the light quark.
$D^H_3$ and $\bar{D}^H_3$ denote the color $SU(3)_c$ representations
of the heavy quarks. Likewise, $D^L_3$ and $D^L_3$ denote the color
$SU(3)_c$ representations of the light quarks. The subscripts
$D^{(\prime)}_6$ and $D^{(\prime)}_3$ represent the color-spin
$SU(6)_{cs}$ representations and the color $SU(3)_3$ representations
of the whole systems respectively.

Moreover, the orthogonalization of the flavor wave functions are
defined as
\begin{eqnarray*}
 &&\bigg\langle \Big(cq_i\Big)_{\bar{d}_6\bar{d}_3\bar{S}}\Big(\bar{c}\bar{q}_m\Big)_{\bar{d}'_6\bar{d}'_3\bar{S}'}\bigg|\Big(cq_j\Big)_{d_6d_3S}\Big(\bar{c}\bar{q}_n\Big)_{d'_6d'_3S'}\bigg\rangle \\
 &&= \delta_{ij}\delta_{mn}\delta_{\bar{d}_6d_6}\delta_{\bar{d}_3d_3}\delta_{\bar{S}S}\delta_{\bar{d}'_6d'_6}\delta_{\bar{d}'_3d'_3}\delta_{\bar{S}'S'},
 \end{eqnarray*}
where $q_i$, $q_j$, $q_m$ and $q_n$ can be $u$ or $d$ quark. We need
to specify that the ordering of the $\bar{c}$ and $c$, $q_i$ and
$\bar{q_m}$ can not be interchanged \cite{Ma:2014ofa,Ma:2014zva}.
The above definition guarantees the orthogonalization of their total
wave functions.

In the heavy quark symmetry limit, the effective Hamiltonian
$H_{eff}$ for a decay process can be separated into the spatial and
flavor parts,
\begin{equation}
  H_{eff}=H_{eff}^{spatial}\otimes H_{eff}^{flavor},
\end{equation}
In the strong decays, the effective Hamiltonian $H_{eff}$ conserves
the heavy spin, light spin, isospin, parity, C parity and $G$-parity
separately, while it may change the color-spin $SU(6)_{cs}$ and
color $SU(3)_c$ representations. The general form of the decay
matrix can be expressed in terms of the reduced matrix,
\begin{eqnarray*}
   &&\bigg\langle\bigg(a_{D_6^HD_3^H}^H\otimes b_{D_6^LD_3^L}^L\bigg)_{D_6D_3J}^{pc}\bigg|H_{eff}^{spatial}\bigg|\bigg(c_{\bar{D}_6^H\bar{D}_3^H}^H\otimes d_{\bar{D}_6^L\bar{D}_3^L}^L\bigg)_{D'_6D'_3J'}^{p'c'}\bigg\rangle \\
   &&=\delta_{ac}\delta_{bd}\delta_{JJ'}\delta_{pp'}\delta_{cc'}\bigg\langle\phi_{D_6^HD_3^H;D_6^LD_3^L}^{D_6D_3}\bigg|\bigg|H_{eff}^{spatial}\bigg|\bigg|\phi_{\bar{D}_6^H\bar{D}_3^H;\bar{D}_6^L\bar{D}_3^L}^{D'_6D'_3}\bigg\rangle,
  \end{eqnarray*}
where $\phi$ in the reduced matrix denotes the radial wave
functions, which may depend on the different $SU(6)_{cs}$ and
$SU(3)_c$ representations. This formula reflects the conservation of
the parity, C parity, the total angular momentum, heavy spin, and
light spin. In addition, the flavor part of the decay matrix can be
written as
\begin{eqnarray*}
 \bigg\langle \Big(cq_i\Big)_{\bar{d}_6\bar{d}_3\bar{S}}\Big(\bar{c}\bar{q}_m\Big)_{\bar{d}'_6\bar{d}'_3\bar{S}'}\bigg|H_{eff}^{flavor}\bigg|\Big(cq_j\Big)_{d_6d_3S}\Big(\bar{c}\bar{q}_n\Big)_{d'_6d'_3S'}\bigg\rangle= \delta_{ij}\delta_{mn}.
 \end{eqnarray*}
For the decays
$|cq\bar{c}\bar{q}\rangle\rightarrow(c\bar{c})+light\,\,meson$, the
transition matrix elements related to the flavor wave functions can
be written as
\begin{eqnarray*}
 \bigg\langle \Big(c\bar{c}\Big)_{\bar{d}_6\bar{d}_3\bar{S}}\bigg|\bigg\langle\Big(q_i\bar{q}_m\Big)_{\bar{d}'_6\bar{d}'_3\bar{S}'}\bigg|H_{eff}^{flavor}\bigg|\Big(cq_j\Big)_{d_6d_3S}\Big(\bar{c}\bar{q}_n\Big)_{d'_6d'_3S'}\bigg\rangle = \delta_{ij}\delta_{mn}.
 \end{eqnarray*}
We also need the rearranged color-spin structures of the final
states. Its general expression reads
\begin{eqnarray}
 && |\textsf{charmonia}\rangle\otimes|\textsf{light}\,\, \textsf{meson}\rangle\nonumber\\
 & &= \left[[(c\bar{c})_{d_6d_3S_H}\otimes L ]_K\otimes Q_{d'_6d'_3}\right]_J |(c\bar{c})_{d_6d_3S}\rangle|(q_i\bar{q_j})_{d'_6d'_3S'}\rangle  \nonumber\nonumber\\
  &&=\sum_{S_l=|L-Q|}^{L+Q}  \mathcal{D}^{Q,L,K,J}_{S_H,S_l}\Bigg|\left[\left(c\bar{c}\right)_{d_6d_3S_H}\otimes\left[L\otimes Q\right]_{d'_6d'_3S_l}\right]_J \Bigg\rangle |(c\bar{c})_{d_6d_3S}\rangle|(q_i\bar{q_j})_{d'_6d'_3S'}\rangle,\label{11}%\nonumber\\
%  &&=\sum_{h=|L-Q|}^{L+Q}  \mathcal{D}^{g,L,K,J}_{g,h}(g_H\otimes h_l)_J  (\bar{b}b)\rangle|\gamma\rangle,
   \end{eqnarray}
where the indices $c$, $\bar{c}$ and $q_i$, $\bar{q_j}$ in the
square brackets denote the corresponding spin wave functions. And
$g$ and $L$ denote the heavy and light spins of the charmonium,
respectively. We collect the coefficients
$\mathcal{D}^{Q,L,K,J}_{S_H,S_l}$ in Table \ref{final state}.

\renewcommand{\arraystretch}{1.4}
\begin{table*}[htbp]
\caption{The coefficient $\mathcal{D}^{Q,L,K,J}_{S_H,S_l}$ in Eq.
(\ref{11}) corresponding to different combinations of $[S_H,S_l]$,
which are
$\left[\left(c\bar{c}\right)_{d_6d_3S_H}\otimes\left[L\otimes
Q\right]_{d'_6d'_3S_l}\right]$.}\label{final state}
\begin{center}
   \begin{tabular}{c|cc|c c c c|cccc} \toprule[1pt]
      %\multicolumn{2}{|c|}
     &\multicolumn{2}{c|}{$J=0$}  &&\multicolumn{2}{c}{$J=1$}&  &&\multicolumn{2}{c}{$J=2$}&\\\midrule[1pt]

    & $[0,0]$   &  $[1,1]$
      & $[0,1]$  &  $[1,0]$ &  $[1,1]$ &  $[1,2]$ & $[0,2]$   &  $[1,1]$ &  $[1,2]$  &  $[1,3]$\\\midrule[1pt]

      $|\eta_c(1^1S_0)\pi/\eta/\sigma\rangle$ & 1 & 0
      & -- & -- & -- & -- &--&--&--&--\\

      $|J/\psi(1^3S_1)\pi/\eta/\sigma\rangle$ & -- & --
       & 0 & 1 & 0 & 0
       & -- & -- & -- & --\\

      $|h_c(1^1P_1)\pi/\eta/\sigma\rangle$ & -- & --
       & 1 & 0 & 0 & 0
       & -- & -- & -- & --\\

      $|\chi_{c0}(1^3P_0)\pi/\eta/\sigma\rangle$ & 0 & 1
       & -- & -- & -- & -- &--&--&--&-- \\

      $|\chi_{c1}(1^3P_1)\pi/\eta/\sigma\rangle$ & -- & --
       & 0 & 0 & 1 & 0
       & -- & -- & -- & --\\

      $|\chi_{c2}(1^3P_2)\pi/\eta/\sigma\rangle$ &--&--
       & -- & -- & -- & --
       & 0 & 1 & 0 & 0\\

      $|\eta_{c2}(1^1D_2)\pi/\eta/\sigma\rangle$ &--&--
       & -- & -- & -- & --
       & 1 & 0 & 0 & 0\\

      $|\psi(1^3D_1)\pi/\eta/\sigma\rangle$ & -- &--
       & 0 & 0 & 0 & 1
       & -- & -- & -- & --\\

      $|\psi(1^3D_2)\pi/\eta/\sigma\rangle$ &--&--
       & -- & -- & -- & --
       & 0 & 0 & 1 & 0 \\\hline

      $|\eta_c(1^1S_0)\rho/\omega\rangle$ &--&--
      & 1 & 0 & 0 & 0 &--&--&--&--\\

      $|J/\psi(1^3S_1)\rho/\omega\rangle$& 0 & 1
       & 0 & 0 & 1 & 0
       & 0 & 1 & 0 & 0\\

      $|h_c(1^1P_1)\rho/\omega\rangle$& 1 & 0
       & 1 & 0 & 0 & 0
       & 1 & 0 & 0 & 0\\
      $|\chi_{c0}(1^3P_0)\rho/\omega\rangle$&--&--
       & 0 & $\frac{1}{3}$ & $-\frac{\sqrt{3}}{3}$ & $\frac{\sqrt{5}}{3}$&--&--&--&-- \\

      $|\chi_{c1}(1^3P_1)\rho/\omega\rangle$& 0 & 1
       & 0 & $-\frac{\sqrt{3}}{3}$ & $\frac{1}{2}$ & $\frac{\sqrt{15}}{6}$
       & 0 & $-\frac{1}{2}$ & $\frac{\sqrt{3}}{2}$ & 0\\

      $|\chi_{c2}(1^3P_2)\rho/\omega\rangle$&--&--
       & 0 & $\frac{\sqrt{5}}{3}$ & $\frac{\sqrt{15}}{6}$ & $\frac{1}{6}$
       & 0 & $\frac{\sqrt{3}}{2}$ & $\frac{1}{2}$ & 0\\

      $|\eta_{c2}(1^1D_2)\rho/\omega\rangle$&--&--
       & 1 & 0 & 0 & 0
       & 1 & 0 & 0 & 0\\

      $|\psi(1^3D_1)\rho/\omega\rangle$& 0 &1
       & 0 & 0 & $-\frac{1}{2}$ & $\frac{\sqrt{3}}{2}$
       & 0 & $\frac{1}{10}$ & $-\frac{\sqrt{15}}{10}$ & $\frac{\sqrt{21}}{5}$\\

      $|\psi(1^3D_2)\rho/\omega\rangle$&--&--
       & 0 & 0 & $\frac{\sqrt{3}}{2}$ & $\frac{1}{2}$
       & 0 & $-\frac{\sqrt{15}}{10}$ & $\frac{5}{6}$ & $\frac{\sqrt{35}}{15}$\\

      \bottomrule[1pt]
      \end{tabular}
  \end{center}
\end{table*}

%%%%%%%%%%%%%%%%%%%%%%%%%%%%%%%%%%%%%%%%%%%
\section{Decay patterns of the S-wave tetraquarks}\label{sec3}
%%%%%%%%%%%%%%%%%%%%%%%%%%%%%%%%%%%%%%%%%%%

%%%%%%%%%%%%%%%%%%%%%%%%
\subsection{Color-spin structures of the S-wave tetraquarks}\label{subsec1}
%%%%%%%%%%%%%%%%%%%%%%%%%

In this section we discuss the color-spin structures of the S-wave
hidden charm tetraquarks in Eqs. (\ref{r1})-(\ref{r2}), which are
collected in Appendix A. After the color-spin rearrangement, the
charm quark and anti-charm quark are coupled into the heavy spin,
color-spin and color representations. We use the combination
$(1,1,0)_H^{-+}\otimes(1,1,0)_l^{-+}$ to indicate that the $c\bar c$
quarks and light quarks couple into the color-spin $SU(6)_{cs}$
singlet, color $SU(3)_c$ singlet and the heavy/light spin singlet
respectively. Then both the heavy part and light part have the
negative parity and the positive C-parity. The subscripts
$(21,\bar{21})$ in $|1_{cs},1_c,0\rangle_{(21,\bar{21})}$ in Eq.
(\ref{eq27}) indicate that its diquark has the color-spin
representations $21$ and the anti-diquark has the color-spin
representations $\bar{21}$. We use the color-spin representations of
the diquark and anti-diquark to distinguish the tetraquark states
that have the same total representations and total spin.

There are four tetraquark states with $J^{PC}=0^{++}$. Their isospin
wave functions do not affect the results of the color-spin
re-coupling. Both the isoscalar and isovector tetraquarks have the
same color-spin wave functions, but have different flavor functions.
All the four states contain the color $SU(3)_c$ singlet and octet,
which differs from the color-spin wave functions of the molecular
states. In addition, all the four states include the heavy spin
singlet and triplet. The color octet terms in the color-spin wave
functions of the states $|1_{cs},1_c,0\rangle_{(21,\bar{21})}$ and
$|1_{cs},1_c,0\rangle_{(15,\bar{15})}$ contain the heavy spin
singlet and triplet. The color octet configurations of the states
$|405_{cs},1_c,0\rangle_{(21,\bar{21})}$ and
$|189_{cs},1_c,0\rangle_{(15,\bar{15})}$ contain the heavy spin
singlet only.

There are also four tetraquark states with $J^{PC}=1^{++}$. The
color singlet and color octet terms of the four states contain the
heavy and light spin triplet only. The states
$|35_{cs},1_c,1\rangle_{(21,\bar{21})}$ and
$|35_{cs},1_c,1\rangle_{(15,\bar{15})}$ have the same color-spin
wave functions. Their strong decay patterns are the same if we
ignore their phase space difference.

There are six S-wave $1^{+-}$ tetraquark states as shown in Eqs.
(\ref{eq35})-(\ref{eq36}). The states
$|35_{cs},1_c,1\rangle_{(21,\bar{21})}$ and
$|35_{cs},1_c,1\rangle_{(15,\bar{15})}$ contain the color singlet
configuration only and have the same re-coupling coefficients, while
the state $|280_{cs},1_c,1\rangle_{(21,\bar{15})}$ and
$|280_{cs},1_c,1\rangle_{(15,\bar{21})}$ have the color octet
configurations only. The two states also have the same re-coupling
coefficients.

%%%%%%%%%%%%%%%%%%%%%%%%%%
\subsection{Decay patterns under the heavy quark symmetry}\label{subsec2}
%%%%%%%%%%%%%%%%%%%%%%%%%%%

\renewcommand{\arraystretch}{1.5}
\begin{table*}[htbp]
\scriptsize
\begin{center}
\caption{The decay matrix elements of the S-wave tetraquarks under
the heavy quark symmetry. The reduced matrix elements $H_{\alpha}^{ij}\propto\langle
Q,i\|H_{eff}(\alpha)\|j\rangle$, where the indices $i$ and $j$
denote the light spin of the final and initial hadron respectively,
and $Q$ is the angular momentum of the final light meson. The
quantum numbers in the bracelets represent the total angular
momentum configurations of the final state particles.}\label{tab:1}
   \begin{tabular}{l ccccccccccc} \toprule[1pt]
        {Initial state} & \multicolumn{5}{c}{Final state} \\\midrule[1pt]
      $1^+(1^{+-})$ & $J/\psi(1^3S_1)\pi$ & $\psi(1^3D_1)\pi$ & $\eta_c\rho$ &  $\eta_{c2}\rho$ & $h_c\pi\,\,\{^3P_1\}$  \\
      %\midrule[1pt]
       $|35_{cs},1_c,1\rangle_{(21,\bar{21})}$     &$\frac{\sqrt{2}}{2}H_{\pi}^{00}$    &$0$      &$-\frac{\sqrt{2}}{2}H_{\rho}^{01}$     &$-\frac{\sqrt{2}}{2}H_{\rho}^{21}$ & $-\frac{\sqrt{2}}{2}H_{\pi}^{11}$  \\

       $|35_{cs},1_c,1\rangle_{(15,\bar{15})}$     &$\frac{\sqrt{2}}{2}H_{\pi}^{00}$    &$0$      &$-\frac{\sqrt{2}}{2}H_{\rho}^{01}$     &$-\frac{\sqrt{2}}{2}H_{\rho}^{21}$ & $-\frac{\sqrt{2}}{2}H_{\pi}^{11}$   \\

      $|35_{cs},1_c,1\rangle_{(21,\bar{15})}$     &$\frac{\sqrt{3}}{3}H_{\pi}^{00}+\frac{\sqrt{6}}{6}\tilde{H}_{\pi}^{00}$    &$0$      &$\frac{\sqrt{3}}{3}H_{\rho}^{01}+\frac{\sqrt{6}}{6}\tilde{H}_{\rho}^{01}$     &$\frac{\sqrt{3}}{3}H_{\rho}^{21}+\frac{\sqrt{6}}{6}\tilde{H}_{\rho}^{21}$ & $\frac{\sqrt{3}}{3}H_{\pi}^{11}+\frac{\sqrt{6}}{6}\tilde{H}_{\pi}^{11}$  \\

       $|35_{cs},1_c,1\rangle_{(15,\bar{21})}$     &$\frac{\sqrt{6}}{6}H_{\pi}^{00}-\frac{\sqrt{3}}{3}\tilde{H}_{\pi}^{00}$    &$0$      &$\frac{\sqrt{6}}{6}H_{\rho}^{01}-\frac{\sqrt{3}}{3}\tilde{H}_{\rho}^{01}$     &$\frac{\sqrt{6}}{6}H_{\rho}^{21}-\frac{\sqrt{3}}{3}\tilde{H}_{\rho}^{21}$ & $\frac{\sqrt{6}}{6}H_{\pi}^{11}-\frac{\sqrt{3}}{3}\tilde{H}_{\pi}^{11}$  \\

       $|280_{cs},1_c,1\rangle_{(21,\bar{15})}$     &$-\frac{\sqrt{2}}{2}\tilde{H}_{\pi}^{00}$    &$0$      &$\frac{\sqrt{2}}{2}\tilde{H}_{\rho}^{01}$     &$\frac{\sqrt{2}}{2}\tilde{H}_{\rho}^{21}$ & $\frac{\sqrt{2}}{2}\tilde{H}_{\pi}^{11}$   \\

       $|280_{cs},1_c,1\rangle_{(15,\bar{21})}$     &$-\frac{\sqrt{2}}{2}\tilde{H}_{\pi}^{00}$    &$0$      &$\frac{\sqrt{2}}{2}\tilde{H}_{\rho}^{01}$     &$\frac{\sqrt{2}}{2}\tilde{H}_{\rho}^{21}$ & $\frac{\sqrt{2}}{2}\tilde{H}_{\pi}^{11}$   \\

      \midrule[1pt]

      {Initial state} & \multicolumn{6}{c}{Final state} \\\midrule[1pt]

      $1^-(1^{++})$ & $J/\psi(1^3S_1)\rho$ & $\psi(1^3D_1)\rho$  & $\psi(1^3D_2)\rho$ & $\chi_{c0}\pi\,\,\{^3P_1\}$ & $\chi_{c1}\pi\,\,\{^3P_1\}$ & $\chi_{c2}\pi\,\,\{^3P_1\}$\\

     $|35_{cs},1_c,1\rangle_{(21,\bar{21})}$ & $-\frac{1}{3}H_{\rho}^{01}-\frac{2\sqrt{2}}{3}\tilde{H}_{\rho}^{01}$ & $\frac{1}{6}H_{\rho}^{21}+\frac{\sqrt{2}}{3}\tilde{H}_{\rho}^{21}$ & $-\frac{\sqrt{3}}{6}H_{\rho}^{21}-\frac{\sqrt{6}}{3}\tilde{H}_{\rho}^{21}$ & $\frac{\sqrt{3}}{9}H_{\pi}^{11}+\frac{2}{3}\tilde{H}_{\pi}^{11}$ & $-\frac{1}{6}H_{\pi}^{11}-\frac{\sqrt{2}}{3}\tilde{H}_{\pi}^{11}$ & $-\frac{\sqrt{15}}{18}H_{\pi}^{11}-\frac{\sqrt{5}}{3}\tilde{H}_{\pi}^{11}$ \\

     $|35_{cs},1_c,1\rangle_{(15,\bar{15})}$ & $-\frac{1}{3}H_{\rho}^{01}-\frac{2\sqrt{2}}{3}\tilde{H}_{\rho}^{01}$ & $\frac{1}{6}H_{\rho}^{21}+\frac{\sqrt{2}}{3}\tilde{H}_{\rho}^{21}$ & $-\frac{\sqrt{3}}{6}H_{\rho}^{21}-\frac{\sqrt{6}}{3}\tilde{H}_{\rho}^{21}$ & $\frac{\sqrt{3}}{9}H_{\pi}^{11}+\frac{2}{3}\tilde{H}_{\pi}^{11}$ & $-\frac{1}{6}H_{\pi}^{11}-\frac{\sqrt{2}}{3}\tilde{H}_{\pi}^{11}$ & $-\frac{\sqrt{15}}{18}H_{\pi}^{11}-\frac{\sqrt{5}}{3}\tilde{H}_{\pi}^{11}$ \\

     $|280_{cs},1_c,1\rangle_{(21,\bar{15})}$ & $\frac{2\sqrt{2}}{3}H_{\rho}^{01}-\frac{1}{3}\tilde{H}_{\rho}^{01}$ & $-\frac{\sqrt{2}}{3}H_{\rho}^{21}+\frac{1}{6}\tilde{H}_{\rho}^{21}$ & $\frac{\sqrt{6}}{3}H_{\rho}^{21}-\frac{\sqrt{3}}{6}\tilde{H}_{\rho}^{21}$ & $-\frac{2}{3}H_{\pi}^{11}+\frac{\sqrt{3}}{9}\tilde{H}_{\pi}^{11}$ & $\frac{\sqrt{2}}{3}H_{\pi}^{11}-\frac{1}{6}\tilde{H}_{\pi}^{11}$ & $\frac{\sqrt{5}}{3}H_{\pi}^{11}-\frac{\sqrt{15}}{18}\tilde{H}_{\pi}^{11}$ \\

     $|280_{cs},1_c,1\rangle_{(15,\bar{21})}$ & $\frac{2\sqrt{2}}{3}H_{\rho}^{01}-\frac{1}{3}\tilde{H}_{\rho}^{01}$ & $-\frac{\sqrt{2}}{3}H_{\rho}^{21}+\frac{1}{6}\tilde{H}_{\rho}^{21}$ & $\frac{\sqrt{6}}{3}H_{\rho}^{21}-\frac{\sqrt{3}}{6}\tilde{H}_{\rho}^{21}$ & $-\frac{2}{3}H_{\pi}^{11}+\frac{\sqrt{3}}{9}\tilde{H}_{\pi}^{11}$ & $\frac{\sqrt{2}}{3}H_{\pi}^{11}-\frac{1}{6}\tilde{H}_{\pi}^{11}$ & $\frac{\sqrt{5}}{3}H_{\pi}^{11}-\frac{\sqrt{15}}{18}\tilde{H}_{\pi}^{11}$ \\

     \midrule[1pt]

       {Initial state} & \multicolumn{5}{c}{Final state} & \\\midrule[1pt]
       $0^-(1^{+-})$ &  $h_c\sigma$ & $J/\psi(1^3S_1)\eta$ & $\psi(1^3D_1)\eta$ & $\eta_c\omega$ &  $\eta_{c2}\omega$ \\
      %\midrule[1pt]
       $|35_{cs},1_c,1\rangle_{(21,\bar{21})}$          &$-\frac{\sqrt{2}}{2}H_{\sigma}^{11}$    &$\frac{\sqrt{2}}{2}H_{\eta}^{00}$    &$0$ &$-\frac{\sqrt{2}}{2}H_{\omega}^{01}$     &$-\frac{\sqrt{2}}{2}H_{\omega}^{21}$      \\

       $|35_{cs},1_c,1\rangle_{(15,\bar{15})}$          &$-\frac{\sqrt{2}}{2}H_{\sigma}^{11}$    &$\frac{\sqrt{2}}{2}H_{\eta}^{00}$    &$0$ &$-\frac{\sqrt{2}}{2}H_{\omega}^{01}$     &$-\frac{\sqrt{2}}{2}H_{\omega}^{21}$      \\

      $|35_{cs},1_c,1\rangle_{(21,\bar{15})}$   & $\frac{\sqrt{3}}{3}H_{\sigma}^{11}+\frac{\sqrt{6}}{6}\tilde{H}_{\sigma}^{11}$   &$\frac{\sqrt{3}}{3}H_{\eta}^{00}+\frac{\sqrt{6}}{6}\tilde{H}_{\eta}^{00}$    &$0$      &$\frac{\sqrt{3}}{3}H_{\omega}^{01}+\frac{\sqrt{6}}{6}\tilde{H}_{\omega}^{01}$     &$\frac{\sqrt{3}}{3}H_{\omega}^{21}+\frac{\sqrt{6}}{6}\tilde{H}_{\omega}^{21}$  \\

       $|35_{cs},1_c,1\rangle_{(15,\bar{21})}$  & $\frac{\sqrt{6}}{6}H_{\sigma}^{11}-\frac{\sqrt{3}}{3}\tilde{H}_{\sigma}^{11}$     &$\frac{\sqrt{6}}{6}H_{\eta}^{00}-\frac{\sqrt{3}}{3}\tilde{H}_{\eta}^{00}$    &$0$      &$\frac{\sqrt{6}}{6}H_{\omega}^{01}-\frac{\sqrt{3}}{3}\tilde{H}_{\omega}^{01}$     &$\frac{\sqrt{6}}{6}H_{\omega}^{21}-\frac{\sqrt{3}}{3}\tilde{H}_{\omega}^{21}$ \\

      $|280_{cs},1_c,1\rangle_{(21,\bar{15})}$          &$\frac{\sqrt{2}}{2}\tilde{H}_{\sigma}^{11}$    &$-\frac{\sqrt{2}}{2}\tilde{H}_{\eta}^{00}$    &$0$ &$\frac{\sqrt{2}}{2}\tilde{H}_{\omega}^{01}$     &$\frac{\sqrt{2}}{2}\tilde{H}_{\omega}^{21}$      \\

       $|280_{cs},1_c,1\rangle_{(15,\bar{21})}$          &$\frac{\sqrt{2}}{2}\tilde{H}_{\sigma}^{11}$    &$-\frac{\sqrt{2}}{2}\tilde{H}_{\eta}^{00}$    &$0$ &$\frac{\sqrt{2}}{2}\tilde{H}_{\omega}^{01}$     &$\frac{\sqrt{2}}{2}\tilde{H}_{\omega}^{21}$      \\

      \midrule[1pt]

        {Initial state} & \multicolumn{7}{c}{Final state} \\\midrule[1pt]

        $0^+(1^{++})$ & $\chi_{c1}\sigma$ & $J/\psi(1^3S_1)\omega$ & $\psi(1^3D_1)\omega$  & $\psi(1^3D_2)\omega$ & $\chi_{c0}\eta\,\,\{^3P_1\}$ & $\chi_{c1}\eta\,\,\{^3P_1\}$ & $\chi_{c2}\eta\,\,\{^3P_1\}$ \\

      $|35_{cs},1_c,1\rangle_{(21,\bar{21})}$ & $-\frac{1}{3}H_{\sigma}^{01}-\frac{2\sqrt{2}}{3}\tilde{H}_{\sigma}^{01}$ & $-\frac{1}{3}H_{\omega}^{01}-\frac{2\sqrt{2}}{3}\tilde{H}_{\omega}^{01}$ & $\frac{1}{6}H_{\omega}^{21}+\frac{\sqrt{2}}{3}\tilde{H}_{\omega}^{21}$ & $-\frac{\sqrt{3}}{6}H_{\omega}^{21}-\frac{\sqrt{6}}{3}\tilde{H}_{\omega}^{21}$ & $\frac{\sqrt{3}}{9}H_{\eta}^{11}+\frac{2}{3}\tilde{H}_{\eta}^{11}$ & $-\frac{1}{6}H_{\eta}^{11}-\frac{\sqrt{2}}{3}\tilde{H}_{\eta}^{11}$ & $-\frac{\sqrt{15}}{18}H_{\eta}^{11}-\frac{\sqrt{5}}{3}\tilde{H}_{\eta}^{11}$ \\

     $|35_{cs},1_c,1\rangle_{(15,\bar{15})}$ & $-\frac{1}{3}H_{\sigma}^{01}-\frac{2\sqrt{2}}{3}\tilde{H}_{\sigma}^{01}$ & $-\frac{1}{3}H_{\omega}^{01}-\frac{2\sqrt{2}}{3}\tilde{H}_{\omega}^{01}$ & $\frac{1}{6}H_{\omega}^{21}+\frac{\sqrt{2}}{3}\tilde{H}_{\omega}^{21}$ & $-\frac{\sqrt{3}}{6}H_{\omega}^{21}-\frac{\sqrt{6}}{3}\tilde{H}_{\omega}^{21}$ & $\frac{\sqrt{3}}{9}H_{\eta}^{11}+\frac{2}{3}\tilde{H}_{\eta}^{11}$ & $-\frac{1}{6}H_{\eta}^{11}-\frac{\sqrt{2}}{3}\tilde{H}_{\eta}^{11}$ & $-\frac{\sqrt{15}}{18}H_{\eta}^{11}-\frac{\sqrt{5}}{3}\tilde{H}_{\eta}^{11}$ \\

     $|280_{cs},1_c,1\rangle_{(21,\bar{15})}$ & $\frac{2\sqrt{2}}{3}H_{\sigma}^{01}-\frac{1}{3}\tilde{H}_{\sigma}^{01}$& $\frac{2\sqrt{2}}{3}H_{\omega}^{01}-\frac{1}{3}\tilde{H}_{\omega}^{01}$ & $-\frac{\sqrt{2}}{3}H_{\omega}^{21}+\frac{1}{6}\tilde{H}_{\omega}^{21}$ & $\frac{\sqrt{6}}{3}H_{\omega}^{21}-\frac{\sqrt{3}}{6}\tilde{H}_{\omega}^{21}$ & $-\frac{2}{3}H_{\eta}^{11}+\frac{\sqrt{3}}{9}\tilde{H}_{\eta}^{11}$ & $\frac{\sqrt{2}}{3}H_{\eta}^{11}-\frac{1}{6}\tilde{H}_{\eta}^{11}$ & $\frac{\sqrt{5}}{3}H_{\eta}^{11}-\frac{\sqrt{15}}{18}\tilde{H}_{\eta}^{11}$ \\

     $|280_{cs},1_c,1\rangle_{(15,\bar{21})}$ & $\frac{2\sqrt{2}}{3}H_{\sigma}^{01}-\frac{1}{3}\tilde{H}_{\sigma}^{01}$& $\frac{2\sqrt{2}}{3}H_{\omega}^{01}-\frac{1}{3}\tilde{H}_{\omega}^{01}$ & $-\frac{\sqrt{2}}{3}H_{\omega}^{21}+\frac{1}{6}\tilde{H}_{\omega}^{21}$ & $\frac{\sqrt{6}}{3}H_{\omega}^{21}-\frac{\sqrt{3}}{6}\tilde{H}_{\omega}^{21}$ & $-\frac{2}{3}H_{\eta}^{11}+\frac{\sqrt{3}}{9}\tilde{H}_{\eta}^{11}$ & $\frac{\sqrt{2}}{3}H_{\eta}^{11}-\frac{1}{6}\tilde{H}_{\eta}^{11}$ & $\frac{\sqrt{5}}{3}H_{\eta}^{11}-\frac{\sqrt{15}}{18}\tilde{H}_{\eta}^{11}$ \\

     \midrule[1pt]

  {Initial state} & \multicolumn{3}{c}{Final state} \\\midrule[1pt]
      $1^-(0^{++})$ & $\eta_c\pi$ & $J/\psi(1^3S_1)\rho$ & $\psi(1^3D_1)\rho$ & $\chi_{c1}\pi\,\,\{^3P_0\}$ \\
      %\midrule[1pt]
       $|1_{cs},1_c,0\rangle_{(21,\bar{21})}$     &$\frac{\sqrt{21}}{6}H_{\pi}^{00}+\frac{\sqrt{42}}{21}\tilde{H}_{\pi}^{00}$    &$-\frac{\sqrt{7}}{14}H_{\rho}^{01}-\frac{\sqrt{14}}{7}\tilde{H}_{\rho}^{01}$      & $-\frac{\sqrt{7}}{14}H_{\rho}^{21}-\frac{\sqrt{14}}{7}\tilde{H}_{\rho}^{21}$     &$-\frac{\sqrt{7}}{14}H_{\pi}^{11}-\frac{\sqrt{14}}{7}\tilde{H}_{\pi}^{11}$   \\

      $|405_{cs},1_c,0\rangle_{(21,\bar{21})}$     &$\frac{3\sqrt{7}}{14}\tilde{H}_{\pi}^{00}$    &$-\frac{2\sqrt{42}}{21}H_{\rho}^{01}+\frac{5\sqrt{21}}{42}\tilde{H}_{\rho}^{01}$      & $-\frac{2\sqrt{42}}{21}H_{\rho}^{21}+\frac{5\sqrt{21}}{42}\tilde{H}_{\rho}^{21}$   &$-\frac{2\sqrt{42}}{21}H_{\pi}^{11}+\frac{5\sqrt{21}}{42}\tilde{H}_{\pi}^{11}$ \\

       $|1_{cs},1_c,0\rangle_{(15,\bar{15})}$     &$\frac{\sqrt{15}}{6}H_{\pi}^{00}-\frac{\sqrt{30}}{15}\tilde{H}_{\pi}^{00}$    &$\frac{\sqrt{5}}{10}H_{\rho}^{01}+\frac{\sqrt{10}}{5}\tilde{H}_{\rho}^{01}$      & $\frac{\sqrt{5}}{10}H_{\rho}^{21}+\frac{\sqrt{10}}{5}\tilde{H}_{\rho}^{21}$     &$\frac{\sqrt{5}}{10}H_{\pi}^{11}+\frac{\sqrt{10}}{5}\tilde{H}_{\pi}^{11}$ \\

      $|189_{cs},1_c,0\rangle_{(15,\bar{15})}$     &$-\frac{3\sqrt{5}}{10}\tilde{H}_{\pi}^{00}$    &$-\frac{2\sqrt{30}}{15}H_{\rho}^{01}-\frac{\sqrt{15}}{30}\tilde{H}_{\rho}^{01}$      & $-\frac{2\sqrt{30}}{15}H_{\rho}^{21}-\frac{\sqrt{15}}{30}\tilde{H}_{\rho}^{21}$   &$-\frac{2\sqrt{30}}{15}H_{\pi}^{11}-\frac{\sqrt{15}}{30}\tilde{H}_{\pi}^{11}$ \\

      \midrule[1pt]

 {Initial state} & \multicolumn{5}{c}{Final state} \\\midrule[1pt]
       $0^+(0^{++})$ & $\chi_{c0}\sigma$ & $\eta_c\eta$ & $J/\psi(1^3S_1)\omega$ & $\psi(1^3D_1)\omega$ & $\chi_{c1}\eta\,\,\{^3P_0\}$  \\
      %\midrule[1pt]
        $|1_{cs},1_c,0\rangle_{(21,\bar{21})}$ &$-\frac{\sqrt{7}}{14}H_{\sigma}^{11}-\frac{\sqrt{14}}{7}\tilde{H}_{\sigma}^{11}$      &$\frac{\sqrt{21}}{6}H_{\eta}^{00}+\frac{\sqrt{42}}{21}\tilde{H}_{\eta}^{00}$    &$-\frac{\sqrt{7}}{14}H_{\omega}^{01}-\frac{\sqrt{14}}{7}\tilde{H}_{\omega}^{01}$      & $-\frac{\sqrt{7}}{14}H_{\omega}^{21}-\frac{\sqrt{14}}{7}\tilde{H}_{\omega}^{21}$     &$-\frac{\sqrt{7}}{14}H_{\eta}^{11}-\frac{\sqrt{14}}{7}\tilde{H}_{\eta}^{11}$   \\

       $|405_{cs},1_c,0\rangle_{(21,\bar{21})}$     &$-\frac{2\sqrt{42}}{21}H_{\sigma}^{01}+\frac{5\sqrt{21}}{42}\tilde{H}_{\sigma}^{01}$  &$\frac{3\sqrt{7}}{14}\tilde{H}_{\eta}^{00}$    &$-\frac{2\sqrt{42}}{21}H_{\omega}^{01}+\frac{5\sqrt{21}}{42}\tilde{H}_{\omega}^{01}$      & $-\frac{2\sqrt{42}}{21}H_{\omega}^{21}+\frac{5\sqrt{21}}{42}\tilde{H}_{\omega}^{21}$   &$-\frac{2\sqrt{42}}{21}H_{\eta}^{11}+\frac{5\sqrt{21}}{42}\tilde{H}_{\eta}^{11}$ \\

       $|1_{cs},1_c,0\rangle_{(15,\bar{15})}$   &$\frac{\sqrt{5}}{10}H_{\sigma}^{01}+\frac{\sqrt{10}}{5}\tilde{H}_{\sigma}^{01}$   &$\frac{\sqrt{15}}{6}H_{\eta}^{00}-\frac{\sqrt{30}}{15}\tilde{H}_{\eta}^{00}$    &$\frac{\sqrt{5}}{10}H_{\omega}^{01}+\frac{\sqrt{10}}{5}\tilde{H}_{\omega}^{01}$      & $\frac{\sqrt{5}}{10}H_{\omega}^{21}+\frac{\sqrt{10}}{5}\tilde{H}_{\omega}^{21}$     &$\frac{\sqrt{5}}{10}H_{\eta}^{11}+\frac{\sqrt{10}}{5}\tilde{H}_{\eta}^{11}$ \\

       $|189_{cs},1_c,0\rangle_{(15,\bar{15})}$   &$-\frac{2\sqrt{30}}{15}H_{\sigma}^{01}-\frac{\sqrt{15}}{30}\tilde{H}_{\sigma}^{01}$  &$-\frac{3\sqrt{5}}{10}\tilde{H}_{\eta}^{00}$    &$-\frac{2\sqrt{30}}{15}H_{\omega}^{01}-\frac{\sqrt{15}}{30}\tilde{H}_{\omega}^{01}$      & $-\frac{2\sqrt{30}}{15}H_{\omega}^{21}-\frac{\sqrt{15}}{30}\tilde{H}_{\omega}^{21}$   &$-\frac{2\sqrt{30}}{15}H_{\eta}^{11}-\frac{\sqrt{15}}{30}\tilde{H}_{\eta}^{11}$ \\

      \midrule[1pt]

      \end{tabular}
\end{center}
\end{table*}

\renewcommand{\arraystretch}{1.5}
\begin{table*}[htbp]
\scriptsize
\begin{center}
\caption{\label{tab:1}  The decay matrix elements of the S-wave
tetraquarks with the constraint of the color-spin $SU(6)_{cs}$
symmetry. The reduced matrix elements $H_{\alpha}^{ij}\propto\langle
Q,i\|H_{eff}(\alpha)\|j\rangle$, where the indices $i$ and $j$
denote the light spin of the final and initial hadron respectively,
and $Q$ is the angular momentum of the final light meson. The
quantum numbers in the bracelets represent the total angular
momentum configurations of the final state particles.}\label{tab:2}
   \begin{tabular}{c l ccccccccccc} \toprule[1pt]
          $I^G(J^{pc})$ & {Initial state} & \multicolumn{5}{c}{Final state} \\\midrule[1pt]
      & & $J/\psi(1^3S_1)\pi$ & $\psi(1^3D_1)\pi$ & $\eta_c\rho$ &  $\eta_{c2}\rho$ & $h_c\pi\,\,\{^3P_1\}$  \\
      %\midrule[1pt]
      \multirow{6}{*}{$1^+(1^{+-})$} & $|35_{cs},1_c,1\rangle_{(21,\bar{21})}$     &$\frac{\sqrt{2}}{2}H_{\pi}^{00}$    &$0$      &$-\frac{\sqrt{2}}{2}H_{\rho}^{01}$     &$-\frac{\sqrt{2}}{2}H_{\rho}^{21}$ & $\epsilon$  \\

      & $|35_{cs},1_c,1\rangle_{(15,\bar{15})}$     &$\frac{\sqrt{2}}{2}H_{\pi}^{00}$    &$0$      &$-\frac{\sqrt{2}}{2}H_{\rho}^{01}$     &$-\frac{\sqrt{2}}{2}H_{\rho}^{21}$ & $\epsilon$   \\

     & $|35_{cs},1_c,1\rangle_{(21,\bar{15})}$     &$\frac{\sqrt{3}}{3}H_{\pi}^{00}+\tilde{\epsilon}$    &$0$      &$\frac{\sqrt{3}}{3}H_{\rho}^{01}+\tilde{\epsilon}$     &$\frac{\sqrt{3}}{3}H_{\rho}^{21}+\tilde{\epsilon}$ & $\epsilon+\tilde{\epsilon}$  \\

      & $|35_{cs},1_c,1\rangle_{(15,\bar{21})}$     &$\frac{\sqrt{6}}{6}H_{\pi}^{00}-\tilde{\epsilon}$    &$0$      &$\frac{\sqrt{6}}{6}H_{\rho}^{01}-\tilde{\epsilon}$     &$\frac{\sqrt{6}}{6}H_{\rho}^{21}-\tilde{\epsilon}$ & $\epsilon-\tilde{\epsilon}$  \\

      & $|280_{cs},1_c,1\rangle_{(21,\bar{15})}$     &$\tilde{\epsilon}$    &$0$      &$\tilde{\epsilon}$     &$\tilde{\epsilon}$ & $\tilde{\epsilon}$   \\

      & $|280_{cs},1_c,1\rangle_{(15,\bar{21})}$     &$\tilde{\epsilon}$    &$0$      &$\tilde{\epsilon}$     &$\tilde{\epsilon}$ & $\tilde{\epsilon}$   \\

      \midrule[1pt]

     $I^G(J^{pc})$ & {Initial state} & \multicolumn{6}{c}{Final state} \\\midrule[1pt]

     & & $J/\psi(1^3S_1)\rho$ & $\psi(1^3D_1)\rho$  & $\psi(1^3D_2)\rho$ & $\chi_{c0}\pi\,\,\{^3P_1\}$ & $\chi_{c1}\pi\,\,\{^3P_1\}$ & $\chi_{c2}\pi\,\,\{^3P_1\}$\\

     \multirow{4}{*}{$1^-(1^{++})$}     &$|35_{cs},1_c,1\rangle_{(21,\bar{21})}$ & $-\frac{1}{3}H_{\rho}^{01}-\frac{2\sqrt{2}}{3}\tilde{H}_{\rho}^{01}$ & $\frac{1}{6}H_{\rho}^{21}+\frac{\sqrt{2}}{3}\tilde{H}_{\rho}^{21}$ & $-\frac{\sqrt{3}}{6}H_{\rho}^{21}-\frac{\sqrt{6}}{3}\tilde{H}_{\rho}^{21}$ & $\epsilon+\tilde{\epsilon}$ & $-\epsilon-\tilde{\epsilon}$ & $-\epsilon-\tilde{\epsilon}$ \\

     &$|35_{cs},1_c,1\rangle_{(15,\bar{15})}$ & $-\frac{1}{3}H_{\rho}^{01}-\frac{2\sqrt{2}}{3}\tilde{H}_{\rho}^{01}$ & $\frac{1}{6}H_{\rho}^{21}+\frac{\sqrt{2}}{3}\tilde{H}_{\rho}^{21}$ & $-\frac{\sqrt{3}}{6}H_{\rho}^{21}-\frac{\sqrt{6}}{3}\tilde{H}_{\rho}^{21}$ & $\epsilon+\tilde{\epsilon}$ & $-\epsilon-\tilde{\epsilon}$ & $-\epsilon-\tilde{\epsilon}$ \\

     &$|280_{cs},1_c,1\rangle_{(21,\bar{15})}$ & $\frac{2\sqrt{2}}{3}H_{\rho}^{01}-\frac{1}{3}\tilde{H}_{\rho}^{01}$ & $-\frac{\sqrt{2}}{3}H_{\rho}^{21}+\frac{1}{6}\tilde{H}_{\rho}^{21}$ & $\frac{\sqrt{6}}{3}H_{\rho}^{21}-\frac{\sqrt{3}}{6}\tilde{H}_{\rho}^{21}$ & $-\epsilon+\tilde{\epsilon}$ & $\epsilon-\tilde{\epsilon}$ & $\epsilon-\tilde{\epsilon}$ \\

     &$|280_{cs},1_c,1\rangle_{(15,\bar{21})}$ & $\frac{2\sqrt{2}}{3}H_{\rho}^{01}-\frac{1}{3}\tilde{H}_{\rho}^{01}$ & $-\frac{\sqrt{2}}{3}H_{\rho}^{21}+\frac{1}{6}\tilde{H}_{\rho}^{21}$ & $\frac{\sqrt{6}}{3}H_{\rho}^{21}-\frac{\sqrt{3}}{6}\tilde{H}_{\rho}^{21}$ & $-\epsilon+\tilde{\epsilon}$ & $\epsilon-\tilde{\epsilon}$ & $\epsilon-\tilde{\epsilon}$ \\

     \midrule[1pt]

      $I^G(J^{pc})$ & {Initial state} & \multicolumn{5}{c}{Final state} & \\\midrule[1pt]
      & &  $h_c\sigma$ & $J/\psi(1^3S_1)\eta$ & $\psi(1^3D_1)\eta$ & $\eta_c\omega$ &  $\eta_{c2}\omega$ \\
      %\midrule[1pt]
      \multirow{6}{*}{$0^-(1^{+-})$} & $|35_{cs},1_c,1\rangle_{(21,\bar{21})}$          &$-\epsilon$    &$\frac{\sqrt{2}}{2}H_{\eta}^{00}$    &$0$ &$-\frac{\sqrt{2}}{2}H_{\omega}^{01}$     &$-\frac{\sqrt{2}}{2}H_{\omega}^{21}$      \\

      & $|35_{cs},1_c,1\rangle_{(15,\bar{15})}$          &$-\epsilon$    &$\frac{\sqrt{2}}{2}H_{\eta}^{00}$    &$0$ &$-\frac{\sqrt{2}}{2}H_{\omega}^{01}$     &$-\frac{\sqrt{2}}{2}H_{\omega}^{21}$      \\

     & $|35_{cs},1_c,1\rangle_{(21,\bar{15})}$   & $\epsilon+\tilde{\epsilon}$   &$\frac{\sqrt{3}}{3}H_{\eta}^{00}+\tilde{\epsilon}$    &$0$      &$\frac{\sqrt{3}}{3}H_{\omega}^{01}+\tilde{\epsilon}$     &$\frac{\sqrt{3}}{3}H_{\omega}^{21}+\tilde{\epsilon}$  \\

      & $|35_{cs},1_c,1\rangle_{(15,\bar{21})}$  & $\epsilon-\tilde{\epsilon}$     &$\frac{\sqrt{6}}{6}H_{\eta}^{00}-\tilde{\epsilon}$    &$0$      &$\frac{\sqrt{6}}{6}H_{\omega}^{01}-\tilde{\epsilon}$     &$\frac{\sqrt{6}}{6}H_{\omega}^{21}-\tilde{\epsilon}$ \\

     & $|280_{cs},1_c,1\rangle_{(21,\bar{15})}$          &$\tilde{\epsilon}$    &$-\tilde{\epsilon}$    &$0$ &$\tilde{\epsilon}$     &$\tilde{\epsilon}$      \\

      & $|280_{cs},1_c,1\rangle_{(15,\bar{21})}$          &$\tilde{\epsilon}$    &$-\tilde{\epsilon}$    &$0$ &$\tilde{\epsilon}$     &$\tilde{\epsilon}$      \\

      \midrule[1pt]

       $I^G(J^{pc})$ & {Initial state} & \multicolumn{7}{c}{Final state} \\\midrule[1pt]

       & & $\chi_{c1}\sigma$ & $J/\psi(1^3S_1)\omega$ & $\psi(1^3D_1)\omega$  & $\psi(1^3D_2)\omega$ & $\chi_{c0}\eta\,\,\{^3P_1\}$ & $\chi_{c1}\eta\,\,\{^3P_1\}$ & $\chi_{c2}\eta\,\,\{^3P_1\}$ \\

       \multirow{4}{*}{$0^+(1^{++})$}     &$|35_{cs},1_c,1\rangle_{(21,\bar{21})}$ & $-\epsilon-\tilde{\epsilon}$ & $-\frac{1}{3}H_{\omega}^{01}-\frac{2\sqrt{2}}{3}\tilde{H}_{\omega}^{01}$ & $\frac{1}{6}H_{\omega}^{21}+\frac{\sqrt{2}}{3}\tilde{H}_{\omega}^{21}$ & $-\frac{\sqrt{3}}{6}H_{\omega}^{21}-\frac{\sqrt{6}}{3}\tilde{H}_{\omega}^{21}$ & $\epsilon+\tilde{\epsilon}$ & $-\epsilon-\tilde{\epsilon}$ & $-\epsilon-\tilde{\epsilon}$ \\

     &$|35_{cs},1_c,1\rangle_{(15,\bar{15})}$ & $-\epsilon-\tilde{\epsilon}$ & $-\frac{1}{3}H_{\omega}^{01}-\frac{2\sqrt{2}}{3}\tilde{H}_{\omega}^{01}$ & $\frac{1}{6}H_{\omega}^{21}+\frac{\sqrt{2}}{3}\tilde{H}_{\omega}^{21}$ & $-\frac{\sqrt{3}}{6}H_{\omega}^{21}-\frac{\sqrt{6}}{3}\tilde{H}_{\omega}^{21}$ & $\epsilon+\tilde{\epsilon}$ & $-\epsilon-\tilde{\epsilon}$ & $-\epsilon-\tilde{\epsilon}$ \\

     &$|280_{cs},1_c,1\rangle_{(21,\bar{15})}$ & $\epsilon-\tilde{\epsilon}$& $\frac{2\sqrt{2}}{3}H_{\omega}^{01}-\frac{1}{3}\tilde{H}_{\omega}^{01}$ & $-\frac{\sqrt{2}}{3}H_{\omega}^{21}+\frac{1}{6}\tilde{H}_{\omega}^{21}$ & $\frac{\sqrt{6}}{3}H_{\omega}^{21}-\frac{\sqrt{3}}{6}\tilde{H}_{\omega}^{21}$ & $-\epsilon+\tilde{\epsilon}$ & $\epsilon-\tilde{\epsilon}$ & $\epsilon-\tilde{\epsilon}$ \\

     &$|280_{cs},1_c,1\rangle_{(15,\bar{21})}$ & $\epsilon-\tilde{\epsilon}$& $\frac{2\sqrt{2}}{3}H_{\omega}^{01}-\frac{1}{3}\tilde{H}_{\omega}^{01}$ & $-\frac{\sqrt{2}}{3}H_{\omega}^{21}+\frac{1}{6}\tilde{H}_{\omega}^{21}$ & $\frac{\sqrt{6}}{3}H_{\omega}^{21}-\frac{\sqrt{3}}{6}\tilde{H}_{\omega}^{21}$ & $-\epsilon+\tilde{\epsilon}$ & $\epsilon-\tilde{\epsilon}$ & $\epsilon-\tilde{\epsilon}$ \\

     \midrule[1pt]

 $I^G(J^{pc})$ & {Initial state} & \multicolumn{3}{c}{Final state} \\\midrule[1pt]
      & & $\eta_c\pi$ & $J/\psi(1^3S_1)\rho$ & $\psi(1^3D_1)\rho$ & $\chi_{c1}\pi\,\,\{^3P_0\}$ \\
      %\midrule[1pt]
      \multirow{4}{*}{$1^-(0^{++})$} & $|1_{cs},1_c,0\rangle_{(21,\bar{21})}$     &$\frac{\sqrt{21}}{6}H_{\pi}^{00}+\tilde{\epsilon}$    &$-\frac{\sqrt{7}}{14}H_{\rho}^{01}-\frac{\sqrt{14}}{7}\tilde{H}_{\rho}^{01}$      & $-\frac{\sqrt{7}}{14}H_{\rho}^{21}-\frac{\sqrt{14}}{7}\tilde{H}_{\rho}^{21}$     &$-\epsilon-\tilde{\epsilon}$   \\

      & $|405_{cs},1_c,0\rangle_{(21,\bar{21})}$     &$\tilde{\epsilon}$    &$-\frac{2\sqrt{42}}{21}H_{\rho}^{01}+\frac{5\sqrt{21}}{42}\tilde{H}_{\rho}^{01}$      & $-\frac{2\sqrt{42}}{21}H_{\rho}^{21}+\frac{5\sqrt{21}}{42}\tilde{H}_{\rho}^{21}$   &$-\epsilon+\tilde{\epsilon}$ \\

      & $|1_{cs},1_c,0\rangle_{(15,\bar{15})}$     &$\frac{\sqrt{15}}{6}H_{\pi}^{00}-\tilde{\epsilon}$    &$\frac{\sqrt{5}}{10}H_{\rho}^{01}+\frac{\sqrt{10}}{5}\tilde{H}_{\rho}^{01}$      & $\frac{\sqrt{5}}{10}H_{\rho}^{21}+\frac{\sqrt{10}}{5}\tilde{H}_{\rho}^{21}$     &$\epsilon+\tilde{\epsilon}$ \\

      & $|189_{cs},1_c,0\rangle_{(15,\bar{15})}$     &$-\tilde{\epsilon}$    &$-\frac{2\sqrt{30}}{15}H_{\rho}^{01}-\frac{\sqrt{15}}{30}\tilde{H}_{\rho}^{01}$      & $-\frac{2\sqrt{30}}{15}H_{\rho}^{21}-\frac{\sqrt{15}}{30}\tilde{H}_{\rho}^{21}$   &$-\epsilon-\tilde{\epsilon}$ \\

      \midrule[1pt]

$I^G(J^{pc})$ & {Initial state} & \multicolumn{5}{c}{Final state}
\\\midrule[1pt]
      & & $\chi_{c0}\sigma$ & $\eta_c\eta$ & $J/\psi(1^3S_1)\omega$ & $\psi(1^3D_1)\omega$ & $\chi_{c1}\eta\,\,\{^3P_0\}$  \\
      %\midrule[1pt]
       \multirow{4}{*}{$0^+(0^{++})$} & $|1_{cs},1_c,0\rangle_{(21,\bar{21})}$ &$-\epsilon-\tilde{\epsilon}$      &$\frac{\sqrt{21}}{6}H_{\eta}^{00}+\tilde{\epsilon}$    &$-\frac{\sqrt{7}}{14}H_{\omega}^{01}-\frac{\sqrt{14}}{7}\tilde{H}_{\omega}^{01}$      & $-\frac{\sqrt{7}}{14}H_{\omega}^{21}-\frac{\sqrt{14}}{7}\tilde{H}_{\omega}^{21}$     &$-\epsilon-\tilde{\epsilon}$   \\

      & $|405_{cs},1_c,0\rangle_{(21,\bar{21})}$     &$-\epsilon+\tilde{\epsilon}$  &$\tilde{\epsilon}$    &$-\frac{2\sqrt{42}}{21}H_{\omega}^{01}+\frac{5\sqrt{21}}{42}\tilde{H}_{\omega}^{01}$      & $-\frac{2\sqrt{42}}{21}H_{\omega}^{21}+\frac{5\sqrt{21}}{42}\tilde{H}_{\omega}^{21}$   &$-\epsilon+\tilde{\epsilon}$ \\

      & $|1_{cs},1_c,0\rangle_{(15,\bar{15})}$   &$\epsilon+\tilde{\epsilon}$   &$\frac{\sqrt{15}}{6}H_{\eta}^{00}-\tilde{\epsilon}$    &$\frac{\sqrt{5}}{10}H_{\omega}^{01}+\frac{\sqrt{10}}{5}\tilde{H}_{\omega}^{01}$      & $\frac{\sqrt{5}}{10}H_{\omega}^{21}+\frac{\sqrt{10}}{5}\tilde{H}_{\omega}^{21}$     &$\epsilon+\tilde{\epsilon}$ \\

      & $|189_{cs},1_c,0\rangle_{(15,\bar{15})}$   &$-\epsilon-\tilde{\epsilon}$  &$-\tilde{\epsilon}$    &$-\frac{2\sqrt{30}}{15}H_{\omega}^{01}-\frac{\sqrt{15}}{30}\tilde{H}_{\omega}^{01}$      & $-\frac{2\sqrt{30}}{15}H_{\omega}^{21}-\frac{\sqrt{15}}{30}\tilde{H}_{\omega}^{21}$   &$-\epsilon-\tilde{\epsilon}$ \\

      \midrule[1pt]

      \end{tabular}
\end{center}
\end{table*}

With the help of the heavy quark symmetry, we are ready to discuss
the strong decay behavior of the S-wave tetraquarks. In the decay
processes, the heavy spin, light spin, parity, C-parity, G-parity
and isospin should be conserved. We collect some typical decay
matrix elements  in Table \ref{tab:1}.

All six tetraquark states with $I^G(J^{PC})=1^+(1^{+-})$ can decay
into $J/\psi\pi$, $\eta_c\rho$, $\eta_{c2}\rho$ through S-wave and
decay into $h_c\pi\,\,\{^3P_1\}$ through P-wave. The decay modes of
the states $|35_{cs},1_c,1\rangle_{(21,\bar{21})}$ and
$|35_{cs},1_c,1\rangle_{(15,\bar{15})}$ are dominated by their color
singlet, while the decay modes of the states
$|280_{cs},1_c,1\rangle_{(21,\bar{15})}$ and
$|280_{cs},1_c,1\rangle_{(15,\bar{21})}$ are governed by the color
octet configurations. The six $1^+(1^{+-})$ states decay into
$J/\psi\pi$ through the spin configuration $(1_H\otimes0_l)_1^{+-}$.
They decay into $\eta_c\rho$ and $\eta_{c2}\rho$ through the spin
configuration $(0_H\otimes1_l)_1^{+-}$. However, none of the six
states can decay into $\psi(1^3D_1)\pi$ since the $\psi(1^3D_1)\pi$
is governed by the spin configuration $(1_H\otimes2_l)_1^{+-}$,
which does not appear in the color-spin wave functions of the six
$1^+(1^{+-})$ tetraquark states. Therefore their decay into
$\psi(1^3D_1)\pi$ is strongly suppressed in the heavy quark limit.
Similar suppression was derived for the molecular structures with
$I^G(J^{PC})=1^+(1^{+-})$ \cite{Ma:2014ofa,Ma:2014zva}.

For the isoscalar $1^{+-}$ tetraquark states, the allowed decay
modes are $h_c\sigma$, $J/\psi\eta$, $\eta_c\omega$ and
$\eta_{c2}\omega$. The decay mode $\psi(1^3D_1)\eta$ is suppressed
due to the non-conservation of the light spin under heavy quark
symmetry.

The isovector $1^{++}$ systems have four kinds of tetraquark states.
All of them can decay into $J/\psi\rho$, $\psi(1^3D_1)\rho$ and
$\psi(1^3D_2)\rho$ via S-wave, and decay into $\pi
\chi_{cJ}\,\,(J=0,1,2)$ via P-wave. It's interesting to note that
the hidden-charm molecular states with the above quantum numbers
have the same decay patterns. And the decay modes are all dominated
by the spin configuration $(1_H\otimes1_l)^{++}$.

The $0^+(1^{++})$ tetraquark states have the S-wave decay modes such
as $\chi_{c1}\sigma$, $J/\psi\omega$, $\psi(1^3D_1)\omega$ and
$\psi(1^3D_2)\omega$. They also have the P-wave decay modes
$\chi_{cJ}\eta\,\,(J=0,1,2)$. All the decays occur through the spin
configuration $(1_H\otimes1_l)_0^{++}$. The color octet
configurations also contribute to the decay.

The allowed decay modes of the $1^-(0^{++})$ tetraquark states are
$\eta_c\pi$, $J/\psi\rho$ and $\psi(1^3D_1)\rho$. The P-wave decay
mode $\chi_{c1}\pi$ is also allowed in the heavy quark limit. The
decay mode $\eta_c\pi$ is governed by the spin configuration
$(0_H\otimes0_l)_0^{++}$. The S-wave decay modes $J/\psi\rho$,
$\psi(1^3D_1)\rho$ and the P-wave decay mode $\chi_{c1}\pi$ are all
dominated by the spin configuration $(1_H\otimes1_l)_0^{++}$. As
listed in Table \ref{tab:1}, the decay mode $\eta_c\pi$ for the
states $|1_{cs},1_c,0\rangle_{(21,\bar{21})}$ and
$|1_{cs},1_c,0\rangle_{(15,\bar{15})}$ arises from both the color
singlet and color octet configurations. For the states
$|405_{cs},1_c,0\rangle_{(21,\bar{21})}$ and
$|189_{cs},1_c,0\rangle_{(15,\bar{15})}$, only the color singlet
contributes to the decay.

The isoscalar $0^{++}$ tetraquark states have the S-wave decay modes
$\chi_{c0}\sigma$, $\eta_c\eta$, $J/\psi\omega$,
$\psi(1^3D_1)\omega$ and the P-wave decay mode $\chi_{c1}\eta$.
Similar to their isovector partners, the decay mode $\eta_c\eta$ for
the states $|1_{cs},1_c,0\rangle_{(21,\bar{21})}$ and
$|1_{cs},1_c,0\rangle_{(15,\bar{15})}$ are from the contributions of
both the color singlet and color octet configurations. The states
$|405_{cs},1_c,0\rangle_{(21,\bar{21})}$ and
$|189_{cs},1_c,0\rangle_{(15,\bar{15})}$ decay via the color singlet
only.

The above discussions are based on the heavy quark symmetry without
considering any decay dynamics. In fact, the decay patterns shown in
Table \ref{tab:1} are the same as the decay patterns within the
molecule framework listed in Refs. \cite{Ma:2014ofa,Ma:2014zva} for
a hidden-charm system with $I^G(J^{PC})=1^+(1^{+-})$.

%%%%%%%%%%%%%%%%%%%%%%%%%%
\subsection{Further analysis of the decay patterns}\label{subsec2.2}
%%%%%%%%%%%%%%%%%%%%%%%%%%%

In the above subsection, we discussed the strong decay patterns of
the S-wave hidden-charm tetraquarks under the heavy quark symmetry.
We notice that some decays depend on the transition matrix elements
between the initial and final states which either have different
color configurations or belong to different color-spin $SU(6)_{cs}$
representations. In this subsection, we discuss two kinds of
suppressions.

Suppose the strong decays are induced by the general interaction
which includes the Coulomb interaction, the linear confinement and
the color-magnetic interaction,
\begin{equation}\label{VVV}
V_{eff}=-\sum_{i>j}\frac{a_{ij}}{|r_i-r_j|}+\sum_{i>j}b_{ij}|r_i-r_j|-\sum_{i>j}c_{ij}
\vec{\lambda}_i\cdot\vec{\lambda}_j\vec{\sigma}_i\cdot\vec{\sigma}_j,
\end{equation}
where $a_{ij}$ and $b_{ij}$ are coefficients depending on
$\vec{\lambda}_i\cdot\vec{\lambda}_j $. Their specific values are
not important for the following discussions. Generally, the
coefficient $c_{ij}\propto1/(m_im_j)$. In the heavy quark limit,
$m_c\rightarrow\infty$, the color-magnetic interaction between two
light quarks is dominant. Since both $\vec{\lambda}_i $ and
$\vec{\lambda}_i\vec{\sigma}_i$ are generators of the $SU(6)_{cs}$
color-spin group, these operators will not change the $SU(6)_{cs}$
representations. In other words, the strong decay Hamiltonian is
invariant under the color-spin $SU(6)_{cs}$ transformation, which
requires that the color-spin $SU(6)_{cs}$ representations between
the two light quarks be conserved in the strong decays. Therefore,
the transition between different $SU(6)_{cs}$ representations is
strongly suppressed in the heavy quark symmetry limit.

Similarly, the transition between different color configurations is
also suppressed since such a process involves the exchange of the
soft gluons. This suppression may not be so strong as the
suppression due to the violation of the color-spin symmetry.

With the above analysis, we can simplify the decay matrix elements
of the S-wave hidden charm tetraquarks further, which are shown in
Table \ref{tab:2}. We use $\epsilon$ to denote the suppressed decay
matrix due to the non-conservation of the color-spin $SU(6)_{cs}$
representations. We use $\tilde{\epsilon}$ to denote the dualy
suppressed decay matrix due to the non-conservation of the color
configuration and color-spin $SU(6)_{cs}$ representation.

From Table \ref{tab:2}, the P-wave decay mode $h_c\pi$ for all the
$1^+(1^{+-})$ tetraquarks is suppressed. The final state
$h_c\pi\,\,\{^3P_1\}$ is governed by the spin configuration
$(0_H\otimes1_l)_1^{+-}$, and $\pi$ belongs to the color-spin
$SU(6)_{cs}$ singlet. In contrast, the spin configuration
$(0_H\otimes1_l)_1^{+-}$ in the initial states arises only from the
color-spin configuration $(1,1,0)_H^{-+}\otimes(35,1,1)_l^{--}$ or
$(35,8,0)_H^{-+}\otimes(35,8,1)_l^{--}$, both of which contain the
color-spin $35_{cs}$ representation only. Thus, the color-spin
$SU(6)_{cs}$ representations between two light quarks are not the
same for the decay mode $h_c\pi\,\,\{^3P_1\}$. Therefore, their
$h_c\pi$ decay width is suppressed in the heavy quark limit. We want
to emphasize that the decay mode $h_c\pi\,\,\{^3P_1\}$ of the
$1^+(1^{+-})$ molecular states is not suppressed under the heavy
quark symmetry as shown in Refs. \cite{Ma:2014ofa,Ma:2014zva}.

As shown in Table \ref{tab:2}, the decay modes $J/\psi\pi$,
$\psi(1^3D_1)\pi$, $\eta_c\rho$, $\eta_{c2}\rho$ and
$h_c\pi\,\,\{^3P_1\}$ of the $1^+(1^{+-})$ tetraquarks with the
color-spin representations $280_{cs}$ are all suppressed. The states
$|280_{cs},1_c,1\rangle_{(21,\bar{15})}$ and
$|280_{cs},1_c,1\rangle_{(15,\bar{21})}$ may have narrower widths
than the states with the color-spin representations $35_{cs}$, which
provides an effective way to distinguish the color-spin
representation of the $1^+(1^{+-})$ tetraquarks.

The P-wave decay modes $\chi_{cJ}\pi\,\,\{^3P_1\}\,\,(J=0,1,2)$ are
not allowed for the four kinds of $1^-(1^{++})$ tetraquarks due to
the non-conservation of the color-spin $SU(6)_{cs}$ representations.

The decay modes $J/\psi\eta$, $\psi(1^3D_1)\eta$, $\eta_c\omega$,
$\eta_{c2}\omega$ and $h_c\sigma$ of the $0^-(1^{+-})$ tetraquarks
with color-spin representations $280_{cs}$ are all suppressed, which
is similar to the decays of its isovector partners into $J/\psi\pi$,
$\psi(1^3D_1)\pi$, $\eta_c\rho$, $\eta_{c2}\rho$ and
$h_c\pi\,\,\{^3P_1\}$. The final state $h_c\sigma$ is governed by
the spin configuration $(0_H\otimes1_l)_1^{+-}$ with the color-spin
$SU(6)_{cs}$ representation $1_{cs}$, which does not appear in the
color-spin wave functions of all the $0^-(1^{+-})$ tetraquarks.
Therefore, their decay into $h_c\sigma$ is suppressed in the heavy
quark limit. However, the $h_c\sigma$ mode of the $0^-(1^{+-})$
molecular states is not suppressed \cite{Ma:2014ofa,Ma:2014zva}.

Since the isoscalar $1^{++}$ tetraquarks contain the color-spin
configurations $(35,8,1)_H^{--}\otimes(35,8,1)_l^{--}$ and
$(35,1,1)_H^{--}\otimes(35,1,1)_l^{--}$ only, their decays into
$\chi_{c1}\sigma$ and $\chi_{cJ}\eta\,\,\{^3P_1\}\,\,(J=0,1,2)$ are
suppressed, which are dominated by the color-spin configuration
$(35,1,1)_H^{--}\otimes(1,1,1)_l^{--}$. In contrast, these decay
modes of the molecular states with the same quantum numbers are not
suppressed. This discrepancy may be used to distinguish the inner
structure of the $0^+(1^{++})$ hidden-charm states.

From the Table \ref{tab:2}, we notice that all the $1^-(0^{++})$
tetraquarks are not allowed to decay into $\chi_{c1} \pi$ via
P-wave. $\chi_{c1}\pi\,\,\{^3P_0\}$ is governed by the color-spin
configuration $(35,1,1)_H^{--}\otimes(1,1,1)_l^{--}$ while the
color-spin representation of all the initial states with the light
spin $S_l=1$ is $35_{cs}$. Therefore, these decays are suppressed in
the heavy quark limit. We also notice that $\eta_c\pi$ is an allowed
decay mode for the $1^-(0^{++})$ tetraquarks with the color-spin
$SU(6)_{cs}$ representation $1_{cs}$, while this decay mode is not
allowed for the $1^-(0^{++})$ tetraquarks with the color-spin
$SU(6)_{cs}$ representation $405_{cs}$ or $189_{cs}$.

The $\eta_c\pi$ decay mode of the states
$|1_{cs},1_c,0\rangle_{(21,\bar{21})}$ and
$|1_{cs},1_c,0\rangle_{(15,\bar{15})}$ arises from both the color
singlet and color octet components. The contribution from the color
octet is suppressed due to the non-conservation of the color-spin
representations. The $\eta_c\pi$ decay mode of the initial states
$|405_{cs},1_c,0\rangle_{(21,\bar{21})}$ and
$|189_{cs},1_c,0\rangle_{(15,\bar{15})}$ contain the contribution
from the color octet only. Thus, their decay into $\eta_c\pi$ is
suppressed, which provides a way to distinguish
$|405_{cs},1_c,0\rangle_{(21,\bar{21})}$ and
$|189_{cs},1_c,0\rangle_{(15,\bar{15})}$.

For the $0^+(0^{++})$ tetraquarks, the decay mode $\eta_c\eta$ is
allowed for the states $|1_{cs},1_c,0\rangle_{(21,\bar{21})}$ and
$|1_{cs},1_c,0\rangle_{(15,\bar{15})}$. Both the color singlet and
color octet contribute to the $\eta_c\eta$ decay width. The decay
mode $\eta_c\eta$ is not allowed for the states
$|405_{cs},1_c,0\rangle_{(21,\bar{21})}$ and
$|189_{cs},1_c,0\rangle_{(15,\bar{15})}$. Only the color octet
contributes to this decay. This is similar to the decay patterns of
their isovector partners. All the four $0^+(0^{++})$ states are not
allowed to decay into $\chi_{c0}\sigma$ and
$\chi_{c1}\eta\,\,\{^3P_0\}$. The color-spin configuration
$(35,1,1)_H^{--}\otimes(1,1,1)_l^{--}$ is dominant for these decay
modes. This color-spin configuration does not appear in all the four
initial states. Again, this feature is different from the decay
patterns of their molecular counterparts
\cite{Ma:2014ofa,Ma:2014zva}.

%%%%%%%%%%
%%%%%%%%%%%%%%%%%%%%%%%%%%%%
\section{Decay patterns of the P-wave tetraquarks}\label{sec4}
%%%%%%%%%%%%%%%%%%%%%%%%%%%%
%%%%%%%%%%

%%%%%%%%%%%%%%%%%%%%%
\subsection{Color-spin structures of the P-wave tetraquarks}\label{subsec1}
%%%%%%%%%%%%%%%%%%%%%%

The P-wave excitation can exist either between the diquark and
anti-diquark, or inside the diquark or anti-diquark. Here we use
"type \uppercase\expandafter{\romannumeral1}" to denote the case
where the P-wave excitation is between the diquark and anti-diquark
pair, and "type \uppercase\expandafter{\romannumeral2}" to denote
the case where the P-wave excitation is either inside the diquark or
anti-diquark. We categorize the two kinds of situations into two
subsections.

%%%%%%%%%%%%%%%%%%
\subsubsection{The type \uppercase\expandafter{\romannumeral1} P-wave tetraquarks }
%%%%%%%%%%%%%%%%%%

According to the discussion in Sec. \ref{sec2}, if the P-wave
excitation exists between the diquark and anti-diquark pair, the
P-wave will contribute to the C-parity of the system under C
transformation. We have listed the color-spin wave functions of the
P-wave tetraquarks in Appendix B. The type
\uppercase\expandafter{\romannumeral1} P-wave tetraquarks with
$J^{PC}=1^{--}$ have three kinds of configurations, which are
$1^{--}(^1P_1)$, $1^{--}(^3P_1)$ and $1^{--}(^5P_1)$. The notation
$(^1P_1)$ means the spin of all the four quarks is $0$, and the
total spin couples with the P-wave into the total angular momentum
$1$.

There are four $1^{--}(^1P_1)$ tetraquark states. Since the isospin
wave functions do not affect the results of the color-spin
rearrangement, their isoscalar and isovector states have the same
color-spin wave functions. All the four states contain the color
$SU(3)_c$ singlet and octet, which is different from the color-spin
wave functions of their molecular counterparts. The color octet
terms in the color-spin wave functions of the states
$|1_{cs},1_c,0\rangle_{(21,\bar{21})}$ and
$|1_{cs},1_c,0\rangle_{(15,\bar{15})}$ have both the heavy spin
singlet and heavy spin triplet. The color octet configurations of
the states $|405_{cs},1_c,0\rangle_{(21,\bar{21})}$ and
$|189_{cs},1_c,0\rangle_{(15,\bar{15})}$ have the heavy spin singlet
only.

From the color-spin wave functions of the $1^{--}(^3P_1)$ states, we
notice that the states $|35_{cs},1_c,1\rangle_{(21,\bar{21})}$ and
$|35_{cs},1_c,1\rangle_{(15,\bar{15})}$ have the same color-spin
wave functions. All the four tetraquarks have spin configurations
such as $(1_H\otimes0_l)_1^{--}$, $(1_H\otimes1_l)_1^{--}$ and
$(1_H\otimes2_l)_1^{--}$. Each of these spin configurations includes
the color singlet and octet. Nevertheless, none of the states
contains the spin configuration $(0_H\otimes1_l)_1^{--}$. This
feature is different from the color-spin wave functions of the
$1^{--}(^1P_1)$ tetraquarks.

There are only two $1^{--}(^5P_1)$ tetraquark states. Their
color-spin wave functions are similar to those of $1^{--}(^3P_1)$
states, which contain the spin configurations like
$(1_H\otimes0_l)_1^{--}$, $(1_H\otimes1_l)_1^{--}$ and
$(1_H\otimes2_l)_1^{--}$, but do not have the spin configuration
$(0_H\otimes1_l)_1^{--}$.

The P-wave $1^{-+}$ tetraquarks of type
\uppercase\expandafter{\romannumeral1} have six kinds of states. As
shown in Appendix B, the states
$|35_{cs},1_c,1\rangle_{(21,\bar{21})}$ and
$|35_{cs},1_c,1\rangle_{(15,\bar{15})}$ have the same color-spin
wave functions, which contain the color singlet configuration only.
The states $|280_{cs},1_c,1\rangle_{(21,\bar{15})}$ and
$|280_{cs},1_c,1\rangle_{(15,\bar{21})}$ also have the same
color-spin wave functions which contain the color octet
configuration only.

There are four tetraquark states with $J^{PC}=0^{--}$ of type
\uppercase\expandafter{\romannumeral1}. All the four states contain
the spin configuration $(1_H\otimes1_l)_0^{--}$ only. But none of
them has the spin configuration $(0_H\otimes0_l)_0^{--}$. Their spin
configuration $(1_H\otimes1_l)_0^{--}$ arises from both the color
singlet and color octet. The states
$|35_{cs},1_c,1\rangle_{(21,\bar{21})}$ and
$|35_{cs},1_c,1\rangle_{(15,\bar{15})}$ have the same color-spin
wave functions. They have similar strong decay patterns without
considering their phase space difference.

There are six $0^{-+}$ P-wave tetraquarks of type
\uppercase\expandafter{\romannumeral1}. The state
$|35_{cs},1_c,1\rangle_{(21,\bar{21})}$ has the same re-coupling
coefficients with the state $|35_{cs},1_c,1\rangle_{(15,\bar{15})}$.
Their spin configurations are from the color singlet only. The state
$|280_{cs},1_c,1\rangle_{(21,\bar{15})}$ also has the same
re-coupling coefficients with the state
$|280_{cs},1_c,1\rangle_{(15,\bar{21})}$. Both of them have the
color octet configuration only.

%%%%%%%%%%%%%%%%%%
\subsubsection{The type \uppercase\expandafter{\romannumeral2} P-wave tetraquarks}
%%%%%%%%%%%%%%%%%%

We have listed the color-spin wave functions of the type
\uppercase\expandafter{\romannumeral2} P-wave tetraquarks in
Appendix B. The P-wave tetraquarks with $J^{PC}=1^{--}$ have only
one configuration $1^{--}(^3P_1)$, which have six kinds of states.
The states $|35_{cs},1_c,1\rangle_{(21,\bar{21})}$ and
$|35_{cs},1_c,1\rangle_{(15,\bar{15})}$ have the same re-coupling
coefficients. The state $|280_{cs},1_c,1\rangle_{(21,\bar{15})}$ and
$|280_{cs},1_c,1\rangle_{(15,\bar{21})}$ also have the same
re-coupling coefficients. Moreover, the states
$|35_{cs},1_c,1\rangle_{(21,\bar{21})}$ and
$|35_{cs},1_c,1\rangle_{(15,\bar{15})}$ have the color octet
configuration only, while the states
$|280_{cs},1_c,1\rangle_{(21,\bar{15})}$ and
$|280_{cs},1_c,1\rangle_{(15,\bar{21})}$ contain the color singlet
configuration only. The states
$|35_{cs},1_c,1\rangle_{(21,\bar{15})}$ and
$|35_{cs},1_c,1\rangle_{(15,\bar{21})}$ contain not only the color
singlet but also the color octet. However, none of them has the spin
configuration $(1_H\otimes0_l)_1^{--}$, which differs from the
$1^{--}$ P-wave tetraquarks of type
\uppercase\expandafter{\romannumeral1}.

The $1^{-+}$ tetraquarks have three kinds of configurations, which
are the $1^{-+}(^1P_1)$, $1^{-+}(^3P_1)$ and $1^{-+}(^5P_1)$. The
$1^{-+}(^1P_1)$ states have four kinds of states. Their color-spin
wave functions include both the color singlet and color octet. The
spin configuration $(0_H\otimes1_l)_1^{-+}$ of the states
$|1_{cs},1_c,0\rangle_{(21,\bar{21})}$ and
$|1_{cs},1_c,0\rangle_{(15,\bar{15})}$ comes from not only the color
singlet terms but also from the color octet terms. The spin
configuration $(0_H\otimes1_l)_1^{-+}$ of the states
$|405_{cs},1_c,0\rangle_{(21,\bar{21})}$ and
$|189_{cs},1_c,0\rangle_{(15,\bar{15})}$ comes from the color octet
terms only. The obvious difference between the $1^{-+}(^3P_1)$
states and $1^{-+}(^1P_1)$ states is that none of the
$1^{-+}(^3P_1)$ states contains the spin configuration
$(0_H\otimes1_l)_1^{-+}$. We also notice that the states
$|35_{cs},1_c,1\rangle_{(21,\bar{21})}$ and
$|35_{cs},1_c,1\rangle_{(15,\bar{15})}$ have the same color-spin
wave functions, and the states
$|280_{cs},1_c,1\rangle_{(21,\bar{15})}$ and
$|280_{cs},1_c,1\rangle_{(15,\bar{21})}$ also have the same
color-spin wave functions.

There are only two $1^{-+}(^5P_1)$ tetraquark states. Neither of
them contains the spin configuration $(0_H\otimes1_l)_1^{-+}$, which
is similar to the $1^{-+}(^3P_1)$ states.

The P-wave $0^{--}$ tetraquarks of type
\uppercase\expandafter{\romannumeral1} include six kinds of
different states. The state $|35_{cs},1_c,1\rangle_{(21,\bar{21})}$
has the same color-spin structures with the state
$|35_{cs},1_c,1\rangle_{(15,\bar{15})}$. Their spin configurations
arise from the color singlet only. The state
$|280_{cs},1_c,1\rangle_{(21,\bar{15})}$ also has the same
color-spin structures with the state
$|280_{cs},1_c,1\rangle_{(15,\bar{21})}$. Their spin configurations
come from the color octet configuration only.

There are four tetraquark states with $J^{PC}=0^{-+}$ of type
\uppercase\expandafter{\romannumeral1}. All of them contain the spin
configuration $(1_H\otimes1_l)_0^{-+}$. Nevertheless, none of them
has the spin configuration $(0_H\otimes0_l)_0^{-+}$. Their spin
configuration $(1_H\otimes1_l)_0^{-+}$ comes not only from the color
singlet but also from the color octet. The states
$|35_{cs},1_c,1\rangle_{(21,\bar{21})}$ and
$|35_{cs},1_c,1\rangle_{(15,\bar{15})}$ also have the same
color-spin wave functions.

%%%%%%%%%%%%%%%%%%%%%%%%%%
\subsection{The strong decay matrix for P-wave tetraquarks}\label{subsec2}
%%%%%%%%%%%%%%%%%%%%%%%%%%%

\renewcommand{\arraystretch}{1.2}
\begin{table*}[htbp]
\scriptsize
\begin{center}
\caption{The decay matrix elements of the tetraquarks with
$J^{PC}=1^{--}$. The reduced matrix elements $H_{\alpha}^{ij}\propto\langle
Q,i\|H_{eff}(\alpha)\|j\rangle$, where the indices $i$ and $j$
denote the light spin of the final and initial hadron respectively,
and $Q$ is the angular momentum of the final light meson. The
quantum numbers in the bracelets represent the total angular
momentum configurations of the final state particles.}\label{tab:3}
   \begin{tabular}{p{1.6cm} ccccccccccc} \toprule[1pt]

    {$1^+(1^{--})\{^1P_1\}$} & \multicolumn{7}{c}{Final state} \\\midrule[1pt]
       {Type \uppercase\expandafter{\romannumeral1}} & $h_c\pi$ & $\chi_{c0}\rho$ & $\chi_{c1}\rho$ & $\chi_{c2}\rho$ & $J/\psi\pi\,\,\{^3P_1\}$ & $J/\psi a_0(980)$ & $\psi(1^3D_1)\pi/\psi(1^3D_2)\pi$  \\
      %\midrule[1pt]
       $|1_{cs},1_c,1\rangle_{(21,\bar{21})}$     &$\frac{\sqrt{21}}{6}H_{\pi}^{11}+\frac{\sqrt{42}}{21}\tilde{H}_{\pi}^{11}$    & $A_1$      &$A_2$     &$A_3$ &$\frac{\sqrt{21}}{42}H_{\pi}^{01}+\frac{\sqrt{21}}{42}\tilde{H}_{\pi}^{01}$ & $-\frac{\sqrt{7}}{42}H_{a_0}^{00}-\frac{\sqrt{7}}{42}\tilde{H}_{a_0}^{00}$ &$B_1/B_2$  \\

      $|405_{cs},1_c,1\rangle_{(21,\bar{21})}$  &   $\frac{3\sqrt{7}}{14}\tilde{H}_{\pi}^{11}$    & $A_4$      &$A_5$     &$A_6$ &$\frac{2\sqrt{14}}{21}H_{\pi}^{01}-\frac{5\sqrt{7}}{42}\tilde{H}_{\pi}^{01}$ & $-\frac{2\sqrt{42}}{63}H_{a_0}^{00}-\frac{5\sqrt{21}}{126}\tilde{H}_{a_0}^{00}$ &$B_3/B_4$  \\

      $|1_{cs},1_c,1\rangle_{(15,\bar{15})}$  &$\frac{\sqrt{15}}{6}H_{\pi}^{11}-\frac{\sqrt{30}}{15}\tilde{H}_{\pi}^{11}$    &$A_7$    & $A_8$     & $A_9$  &$-\frac{\sqrt{15}}{30}H_{\pi}^{01}-\frac{\sqrt{30}}{15}\tilde{H}_{\pi}^{01}$ & $-\frac{\sqrt{5}}{30}H_{a_0}^{00}+\frac{\sqrt{10}}{15}\tilde{H}_{a_0}^{00}$ &$B_5/B_6$ \\

      $|189_{cs},1_c,1\rangle_{(15,\bar{15})}$ & $-\frac{3\sqrt{5}}{10}\tilde{H}_{\pi}^{11}$    &$A_{10}$    & $A_{11}$      &$A_{12}$  &$\frac{2\sqrt{10}}{15}H_{\pi}^{01}+\frac{\sqrt{5}}{30}\tilde{H}_{\pi}^{01}$ & $-\frac{2\sqrt{30}}{45}H_{a_0}^{00}-\frac{\sqrt{15}}{90}\tilde{H}_{a_0}^{00}$ & $B_7/B_8$ \\

      \midrule[1pt]

    {$1^+(1^{--})\{^3P_1\}$} & \multicolumn{7}{c}{Final state} \\\midrule[1pt]
       {Type \uppercase\expandafter{\romannumeral1}} & $h_c\pi$ & $\chi_{c0}\rho$ & $\chi_{c1}\rho$ & $\chi_{c2}\rho$ & $J/\psi\pi\,\,\{^3P_1\}$ & $J/\psi a_0(980)$ & $\psi(1^3D_1)\pi/\psi(1^3D_2)\pi$ \\
      %\midrule[1pt]
       $|35_{cs},1_c,1\rangle_{(21,\bar{21})}$     &$0$    & $A_{13}$      &$A_{14}$     &$A_{15}$ &$-\frac{1}{6}H_{\pi}^{01}-\frac{\sqrt{2}}{3}\tilde{H}_{\pi}^{01}$ & $\frac{\sqrt{3}}{9}H_{a_0}^{00}+\frac{2\sqrt{6}}{9}\tilde{H}_{a_0}^{00}$ &$B_9/B_{10}$ \\

      $|35_{cs},1_c,1\rangle_{(15,\bar{15})}$     &$0$    & $A_{13}$      &$A_{14}$     &$A_{15}$ &$-\frac{1}{6}H_{\pi}^{01}-\frac{\sqrt{2}}{3}\tilde{H}_{\pi}^{01}$ & $\frac{\sqrt{3}}{9}H_{a_0}^{00}+\frac{2\sqrt{6}}{9}\tilde{H}_{a_0}^{00}$ &$B_9/B_{10}$  \\

      $|280_{cs},1_c,1\rangle_{(21,\bar{15})}$  &$0$    &$A_{16}$    & $A_{17}$     & $A_{18}$  &$\frac{\sqrt{2}}{3}H_{\pi}^{01}-\frac{1}{6}\tilde{H}_{\pi}^{01}$ & $-\frac{2\sqrt{6}}{9}H_{a_0}^{00}+\frac{\sqrt{3}}{9}\tilde{H}_{a_0}^{00}$ &$B_{11}/B_{12}$ \\

      $|280_{cs},1_c,1\rangle_{(15,\bar{21})}$ &$0$    &$A_{16}$    & $A_{17}$     & $A_{18}$  &$\frac{\sqrt{2}}{3}H_{\pi}^{01}-\frac{1}{6}\tilde{H}_{\pi}^{01}$ & $-\frac{2\sqrt{6}}{9}H_{a_0}^{00}+\frac{\sqrt{3}}{9}\tilde{H}_{a_0}^{00}$  &$B_{11}/B_{12}$ \\

      \midrule[1pt]

      {$1^+(1^{--})\{^5P_1\}$} & \multicolumn{7}{c}{Final state} \\\midrule[1pt]
       {Type \uppercase\expandafter{\romannumeral1}} & $h_c\pi$ & $\chi_{c0}\rho$ & $\chi_{c1}\rho$ & $\chi_{c2}\rho$ & $J/\psi\pi\,\,\{^3P_1\}$ & $J/\psi a_0(980)$ & $\psi(1^3D_1)\pi/\psi(1^3D_2)\pi$ \\
      %\midrule[1pt]
       $|405_{cs},1_c,1\rangle_{(21,\bar{21})}$     &$0$    & $A_{19}$      &$A_{20}$     &$A_{21}$ &$\frac{\sqrt{10}}{6}H_{\pi}^{01}+\frac{\sqrt{5}}{6}\tilde{H}_{\pi}^{01}$ & $\frac{\sqrt{30}}{9}H_{a_0}^{00}+\frac{\sqrt{15}}{9}\tilde{H}_{a_0}^{00}$ &$B_{13}/B_{14}$ \\

      $|189_{cs},1_c,1\rangle_{(15,\bar{15})}$     &$0$    & $A_{22}$      &$A_{23}$     &$A_{24}$ &$-\frac{\sqrt{5}}{6}H_{\pi}^{01}+\frac{\sqrt{10}}{6}\tilde{H}_{\pi}^{01}$ & $-\frac{\sqrt{15}}{9}H_{a_0}^{00}+\frac{\sqrt{30}}{9}\tilde{H}_{a_0}^{00}$ &$B_{15}/B_{16}$  \\

      \midrule[1pt]

     {$1^+(1^{--})\{^3P_1\}$} & \multicolumn{7}{c}{Final state} \\\midrule[1pt]
       {Type \uppercase\expandafter{\romannumeral2}} & $h_c\pi$ & $\chi_{c0}\rho$ & $\chi_{c1}\rho$ & $\chi_{c2}\rho$ & $J/\psi\pi\,\,\{^3P_1\}$ & $J/\psi a_0(980)$ & $\psi(1^3D_1)\pi/\psi(1^3D_2)\pi$  \\
      %\midrule[1pt]
       $|35_{cs},1_c,1\rangle_{(21,\bar{21})}$     &$-\frac{\sqrt{2}}{2}H_{\pi}^{11}$    & $-\frac{\sqrt{6}}{6}H_{\rho}^{11}$      &$\frac{\sqrt{2}}{4}H_{\rho}^{11}$     &$\frac{\sqrt{30}}{12}H_{\rho}^{11}$ &$\frac{\sqrt{2}}{2}H_{\pi}^{01}$& $0$ &$B_{17}/B_{18}$  \\

       $|35_{cs},1_c,1\rangle_{(15,\bar{15})}$     &$-\frac{\sqrt{2}}{2}H_{\pi}^{11}$    & $-\frac{\sqrt{6}}{6}H_{\rho}^{11}$      &$\frac{\sqrt{2}}{4}H_{\rho}^{11}$     &$\frac{\sqrt{30}}{12}H_{\rho}^{11}$ &$\frac{\sqrt{2}}{2}H_{\pi}^{01}$ & $0$ &$B_{17}/B_{18}$  \\

       $|35_{cs},1_c,1\rangle_{(21,\bar{15})}$     &$\frac{\sqrt{3}}{3}H_{\pi}^{11}+\frac{\sqrt{6}}{6}\tilde{H}_{\pi}^{11}$    & $-\frac{1}{3}H_{\rho}^{11}-\frac{\sqrt{2}}{6}\tilde{H}_{\rho}^{11}$      &$\frac{\sqrt{3}}{6}H_{\rho}^{11}+\frac{\sqrt{6}}{12}\tilde{H}_{\rho}^{11}$     &$\frac{\sqrt{5}}{6}H_{\rho}^{11}+\frac{\sqrt{10}}{12}\tilde{H}_{\rho}^{11}$ &$\frac{\sqrt{3}}{3}H_{\pi}^{01}+\frac{\sqrt{6}}{6}\tilde{H}_{\pi}^{01}$ & $0$  &$B_{19}/B_{20}$ \\

       $|35_{cs},1_c,1\rangle_{(15,\bar{21})}$     &$\frac{\sqrt{6}}{6}H_{\pi}^{11}-\frac{\sqrt{3}}{3}\tilde{H}_{\pi}^{11}$     & $-\frac{\sqrt{2}}{6}H_{\rho}^{11}+\frac{1}{3}\tilde{H}_{\rho}^{11}$      &$\frac{\sqrt{6}}{12}H_{\rho}^{11}-\frac{\sqrt{3}}{6}\tilde{H}_{\rho}^{11}$     &$\frac{\sqrt{10}}{12}H_{\rho}^{11}-\frac{\sqrt{5}}{6}\tilde{H}_{\rho}^{11}$ &$\frac{\sqrt{6}}{6}H_{\pi}^{01}-\frac{\sqrt{3}}{3}\tilde{H}_{\pi}^{01}$ & $0$  &$B_{21}/B_{22}$ \\

      $|280_{cs},1_c,1\rangle_{(21,\bar{15})}$     &$\frac{\sqrt{2}}{2}\tilde{H}_{\pi}^{11}$    & $-\frac{\sqrt{6}}{6}\tilde{H}_{\rho}^{11}$      &$\frac{\sqrt{2}}{4}\tilde{H}_{\rho}^{11}$     &$\frac{\sqrt{30}}{12}\tilde{H}_{\rho}^{11}$ &$-\frac{\sqrt{2}}{2}\tilde{H}_{\pi}^{01}$ & $0$ &$B_{23}/B_{24}$  \\

      $|280_{cs},1_c,1\rangle_{(15,\bar{21})}$    &$\frac{\sqrt{2}}{2}\tilde{H}_{\pi}^{11}$    & $-\frac{\sqrt{6}}{6}\tilde{H}_{\rho}^{11}$      &$\frac{\sqrt{2}}{4}\tilde{H}_{\rho}^{11}$     &$\frac{\sqrt{30}}{12}\tilde{H}_{\rho}^{11}$ &$-\frac{\sqrt{2}}{2}\tilde{H}_{\pi}^{01}$  & $0$ &$B_{23}/B_{24}$  \\

      \midrule[1pt]

     {$0^-(1^{--})\{^1P_1\}$} & \multicolumn{7}{c}{Final state} \\\midrule[1pt]
       {Type \uppercase\expandafter{\romannumeral1}} & $J/\psi\sigma$ & $\psi(1^3D_1)\sigma$ & $h_c\eta$ & $\chi_{c0}\omega$ & $\chi_{c1}\omega$ & $\chi_{c2}\omega$   & $\psi(1^3D_1)\eta/\psi(1^3D_2)\eta$ & $J/\psi\eta\,\,\{^3P_1\}$ \\
      %\midrule[1pt]
       $|1_{cs},1_c,1\rangle_{(21,\bar{21})}$      &$-\frac{\sqrt{7}}{42}H_{\sigma}^{01}-\frac{\sqrt{7}}{42}\tilde{H}_{\sigma}^{01}$    &$-\frac{\sqrt{35}}{42}H_{\sigma}^{21}-\frac{\sqrt{35}}{42}\tilde{H}_{\sigma}^{21}$   &$\frac{\sqrt{21}}{6}H_{\eta}^{11}+\frac{\sqrt{42}}{21}\tilde{H}_{\eta}^{11}$   & $A_1$      &$A_2$     &$A_3$  &$B_1/B_2$ &$\frac{\sqrt{21}}{42}H_{\eta}^{01}+\frac{\sqrt{21}}{42}\tilde{H}_{\eta}^{01}$\\

      $|405_{cs},1_c,1\rangle_{(21,\bar{21})}$  &$-\frac{2\sqrt{42}}{63}H_{\sigma}^{01}+\frac{5\sqrt{21}}{126}\tilde{H}_{\sigma}^{01}$    &$-\frac{2\sqrt{210}}{63}H_{\sigma}^{21}+\frac{5\sqrt{105}}{126}\tilde{H}_{\sigma}^{21}$  &   $\frac{3\sqrt{7}}{14}\tilde{H}_{\eta}^{11}$  & $A_4$      &$A_5$     &$A_6$  &$B_3/B_4$ &$\frac{2\sqrt{14}}{21}H_{\eta}^{01}-\frac{5\sqrt{7}}{42}\tilde{H}_{\eta}^{01}$\\

      $|1_{cs},1_c,1\rangle_{(15,\bar{15})}$  & $\frac{\sqrt{5}}{30}H_{\sigma}^{01}+\frac{\sqrt{10}}{15}\tilde{H}_{\sigma}^{01}$    &$\frac{1}{6}H_{\sigma}^{21}+\frac{\sqrt{2}}{3}\tilde{H}_{\sigma}^{21}$  &$\frac{\sqrt{15}}{6}H_{\eta}^{11}-\frac{\sqrt{30}}{15}\tilde{H}_{\eta}^{11}$   & $A_7$      &$A_8$     &$A_9$  &$B_5/B_6$ &$-\frac{\sqrt{15}}{30}H_{\eta}^{01}-\frac{\sqrt{30}}{15}\tilde{H}_{\eta}^{01}$\\

      $|189_{cs},1_c,1\rangle_{(15,\bar{15})}$  & $-\frac{2\sqrt{30}}{45}H_{\sigma}^{01}-\frac{\sqrt{15}}{90}\tilde{H}_{\sigma}^{01}$    &$-\frac{2\sqrt{6}}{9}H_{\sigma}^{21}-\frac{\sqrt{3}}{18}\tilde{H}_{\sigma}^{21}$  & $-\frac{3\sqrt{5}}{10}\tilde{H}_{\eta}^{11}$  & $A_{10}$      &$A_{11}$     &$A_{12}$  & $B_7/B_8$&$\frac{2\sqrt{10}}{15}H_{\eta}^{01}+\frac{\sqrt{5}}{30}\tilde{H}_{\eta}^{01}$\\

      \midrule[1pt]

     {$0^-(1^{--})\{^3P_1\}$} & \multicolumn{7}{c}{Final state} \\\midrule[1pt]
       {Type \uppercase\expandafter{\romannumeral1}} & $J/\psi\sigma$ & $\psi(1^3D_1)\sigma$ & $h_c\eta$ & $\chi_{c0}\omega$ & $\chi_{c1}\omega$ & $\chi_{c2}\omega$   & $\psi(1^3D_1)\eta/\psi(1^3D_2)\eta$ & $J/\psi\eta\,\,\{^3P_1\}$\\
      %\midrule[1pt]
       $|35_{cs},1_c,1\rangle_{(21,\bar{21})}$      &$\frac{\sqrt{3}}{9}H_{\sigma}^{01}+\frac{2\sqrt{6}}{9}\tilde{H}_{\sigma}^{01}$    &$-\frac{\sqrt{15}}{18}H_{\sigma}^{21}-\frac{\sqrt{30}}{9}\tilde{H}_{\sigma}^{21}$   &$0$   & $A_{13}$      &$A_{14}$     &$A_{15}$  &$B_{9}/B_{10}$ &$-\frac{1}{6}H_{\eta}^{01}-\frac{\sqrt{2}}{3}\tilde{H}_{\eta}^{01}$\\

      $|35_{cs},1_c,1\rangle_{(21,\bar{21})}$  &$\frac{\sqrt{3}}{9}H_{\sigma}^{01}+\frac{2\sqrt{6}}{9}\tilde{H}_{\sigma}^{01}$    &$-\frac{\sqrt{15}}{18}H_{\sigma}^{21}-\frac{\sqrt{30}}{9}\tilde{H}_{\sigma}^{21}$   &$0$   & $A_{13}$      &$A_{14}$     &$A_{15}$  &$B_{9}/B_{10}$ &$-\frac{1}{6}H_{\eta}^{01}-\frac{\sqrt{2}}{3}\tilde{H}_{\eta}^{01}$ \\

      $|280_{cs},1_c,1\rangle_{(21,\bar{15})}$  & $-\frac{2\sqrt{6}}{9}H_{\sigma}^{01}+\frac{\sqrt{3}}{9}\tilde{H}_{\sigma}^{01}$    &$\frac{\sqrt{30}}{9}H_{\sigma}^{21}-\frac{\sqrt{15}}{18}\tilde{H}_{\sigma}^{21}$  & $0$  & $A_{16}$      &$A_{17}$     &$A_{18}$  &$B_{11}/B_{12}$ &$\frac{\sqrt{2}}{3}H_{\eta}^{01}-\frac{1}{6}\tilde{H}_{\eta}^{01}$\\

      $|280_{cs},1_c,1\rangle_{(15,\bar{21})}$  & $-\frac{2\sqrt{6}}{9}H_{\sigma}^{01}+\frac{\sqrt{3}}{9}\tilde{H}_{\sigma}^{01}$    &$\frac{\sqrt{30}}{9}H_{\sigma}^{21}-\frac{\sqrt{15}}{18}\tilde{H}_{\sigma}^{21}$  & $0$  & $A_{16}$      &$A_{17}$     &$A_{18}$  &$B_{11}/B_{12}$&$\frac{\sqrt{2}}{3}H_{\eta}^{01}-\frac{1}{6}\tilde{H}_{\eta}^{01}$\\

      \midrule[1pt]

      {$0^-(1^{--})\{^5P_1\}$} & \multicolumn{7}{c}{Final state} \\\midrule[1pt]
       {Type \uppercase\expandafter{\romannumeral1}} & $J/\psi\sigma$ & $\psi(1^3D_1)\sigma$ & $h_c\eta$ & $\chi_{c0}\omega$ & $\chi_{c1}\omega$ & $\chi_{c2}\omega$   & $\psi(1^3D_1)\eta/\psi(1^3D_2)\eta$ & $J/\psi\eta\,\,\{^3P_1\}$\\
      %\midrule[1pt]

      $|405_{cs},1_c,1\rangle_{(21,\bar{21})}$  & $\frac{\sqrt{30}}{9}H_{\sigma}^{01}+\frac{\sqrt{15}}{9}\tilde{H}_{\sigma}^{01}$    &$\frac{\sqrt{6}}{18}H_{\sigma}^{21}+\frac{\sqrt{3}}{18}\tilde{H}_{\sigma}^{21}$  & $0$  & $A_{19}$      &$A_{20}$     &$A_{21}$ &$B_{13}/B_{14}$ &$\frac{\sqrt{10}}{6}H_{\eta}^{01}+\frac{\sqrt{5}}{6}\tilde{H}_{\eta}^{01}$ \\

      $|189_{cs},1_c,1\rangle_{(15,\bar{15})}$  & $-\frac{\sqrt{15}}{9}H_{\sigma}^{01}+\frac{\sqrt{30}}{9}\tilde{H}_{\sigma}^{01}$    &$-\frac{\sqrt{3}}{18}H_{\sigma}^{21}+\frac{\sqrt{6}}{18}\tilde{H}_{\sigma}^{21}$  & $0$  & $A_{22}$      &$A_{23}$     &$A_{24}$  &$B_{15}/B_{16}$&$-\frac{\sqrt{5}}{6}H_{\eta}^{01}+\frac{\sqrt{10}}{6}\tilde{H}_{\eta}^{01}$\\

      \midrule[1pt]

      {$0^-(1^{--})\{^3P_1\}$} & \multicolumn{7}{c}{Final state} \\\midrule[1pt]
       {Type \uppercase\expandafter{\romannumeral2}} & $J/\psi\sigma$ & $\psi(1^3D_1)\sigma$ & $h_c\eta$ & $\chi_{c0}\omega$ & $\chi_{c1}\omega$ & $\chi_{c2}\omega$   & $\psi(1^3D_1)\eta/\psi(1^3D_2)\eta$ & $J/\psi\eta\,\,\{^3P_1\}$\\
      %\midrule[1pt]

      $|35_{cs},1_c,1\rangle_{(21,\bar{21})}$  & $0$    &$0$  & $-\frac{\sqrt{2}}{2}H_{\eta}^{11}$  & $-\frac{\sqrt{6}}{6}H_{\omega}^{11}$      &$\frac{\sqrt{2}}{4}H_{\omega}^{11}$     &$\frac{\sqrt{30}}{12}H_{\omega}^{11}$  &$B_{17}/B_{18}$ &$\frac{\sqrt{2}}{2}H_{\eta}^{01}$\\

      $|35_{cs},1_c,1\rangle_{(15,\bar{15})}$  & $0$    &$0$  & $-\frac{\sqrt{2}}{2}H_{\eta}^{11}$  & $-\frac{\sqrt{6}}{6}H_{\omega}^{11}$      &$\frac{\sqrt{2}}{4}H_{\omega}^{11}$     &$\frac{\sqrt{30}}{12}H_{\omega}^{11}$ &$B_{17}/B_{18}$&$\frac{\sqrt{2}}{2}H_{\eta}^{01}$ \\

     $|35_{cs},1_c,1\rangle_{(21,\bar{15})}$  & $0$    &$0$  & $\frac{\sqrt{3}}{3}H_{\eta}^{11}+\frac{\sqrt{6}}{6}\tilde{H}_{\eta}^{11}$  & $-\frac{1}{3}H_{\omega}^{11}-\frac{\sqrt{2}}{6}\tilde{H}_{\omega}^{11}$      &$\frac{\sqrt{3}}{6}H_{\omega}^{11}+\frac{\sqrt{6}}{12}\tilde{H}_{\omega}^{11}$     &$\frac{\sqrt{5}}{6}H_{\omega}^{11}+\frac{\sqrt{10}}{12}\tilde{H}_{\omega}^{11}$  &$B_{19}/B_{20}$ &$\frac{\sqrt{3}}{3}H_{\eta}^{01}+\frac{\sqrt{6}}{6}\tilde{H}_{\eta}^{01}$\\

     $|35_{cs},1_c,1\rangle_{(15,\bar{21})}$  & $0$    &$0$ & $\frac{\sqrt{6}}{6}H_{\eta}^{11}-\frac{\sqrt{3}}{3}\tilde{H}_{\eta}^{11}$  & $-\frac{\sqrt{2}}{6}H_{\omega}^{11}+\frac{1}{3}\tilde{H}_{\omega}^{11}$      &$\frac{\sqrt{6}}{12}H_{\omega}^{11}-\frac{\sqrt{3}}{6}\tilde{H}_{\omega}^{11}$     &$\frac{\sqrt{10}}{12}H_{\omega}^{11}-\frac{\sqrt{5}}{6}\tilde{H}_{\omega}^{11}$   &$B_{21}/B_{22}$ &$\frac{\sqrt{6}}{6}H_{\eta}^{01}-\frac{\sqrt{3}}{3}\tilde{H}_{\eta}^{01}$\\

     $|280_{cs},1_c,1\rangle_{(21,\bar{15})}$  & $0$    &$0$ & $\frac{\sqrt{2}}{2}\tilde{H}_{\eta}^{11}$  & $-\frac{\sqrt{6}}{6}\tilde{H}_{\omega}^{11}$      &$\frac{\sqrt{2}}{4}\tilde{H}_{\omega}^{11}$     &$\frac{\sqrt{30}}{12}\tilde{H}_{\omega}^{11}$  &$B_{23}/B_{24}$ &$-\frac{\sqrt{2}}{2}\tilde{H}_{\eta}^{01}$\\

     $|280_{cs},1_c,1\rangle_{(15,\bar{21})}$  & $0$    &$0$ & $\frac{\sqrt{2}}{2}\tilde{H}_{\eta}^{11}$  & $-\frac{\sqrt{6}}{6}\tilde{H}_{\omega}^{11}$      &$\frac{\sqrt{2}}{4}\tilde{H}_{\omega}^{11}$     &$\frac{\sqrt{30}}{12}\tilde{H}_{\omega}^{11}$ &$B_{23}/B_{24}$ &$-\frac{\sqrt{2}}{2}\tilde{H}_{\eta}^{01}$\\

      \midrule[1pt]

      \end{tabular}
\end{center}
\end{table*}

With the preparations in Sec. \ref{sec2}, we are ready to discuss
the strong decay behavior of the P-wave tetraquarks. The strong
decay matrix elements of the P-wave tetraquarks are listed in Table
\ref{tab:3}, where the parameters $A_{(1-24)}$ and $B_{(1-24)}$ are
defined in Appendix C. The P-wave tetraquarks of type
\uppercase\expandafter{\romannumeral1} with $J^{PC}=1^{--}$ have
three kinds of configurations, which are $1^{--}(^1P_1)$,
$1^{--}(^3P_1)$ and $1^{--}(^5P_1)$.

All the four $1^{--}(^1P_1)$ states of type
\uppercase\expandafter{\romannumeral1} have the S-wave decay modes
$h_c\pi$, $\chi_{cJ}\rho\,\,(J=0,1,2)$, $J/\psi a_0(980)$ and the
P-wave decay modes $J/\psi\pi\,\,\{^3P_1\}$,
$\psi(1^3D_1)\pi\,\,\{^3P_1\}$, $\psi(1^3D_2)\pi\,\,\{^5P_1\}$. The
decay mode $h_c\pi$ is dominated by the spin configuration
$(0_H\otimes1_l)_1^{--}$. The spin configuration
$(0_H\otimes1_l)_0^{--}$ of the initial states
$|405_{cs},1_c,1\rangle_{(21,\bar{21})}$ and
$|189_{cs},1_c,1\rangle_{(15,\bar{15})}$ arises from the
contributions of the color octet only. The initial states
$|1_{cs},1_c,1\rangle_{(21,\bar{21})}$ and
$|1_{cs},1_c,1\rangle_{(15,\bar{15})}$ decay into $h_c\pi$ through
both the color singlet and color octet. All the four $1^{--}(^1P_1)$
states can decay into $J/\psi\pi $ via P-wave, which is governed by
the spin configuration $(1_H\otimes1_l)_1^{--}$. The decay mode
$J/\psi a_0(980)$ is dominated by the spin configuration
$(1_H\otimes0_l)_1^{--}$. Their decays into $J/\psi a_0(980)$ have
the contributions from both the color singlet and color octet. The
decay modes $\chi_{cJ}\rho\,\,(J=0,1,2)$ depend on the spin
configurations $(1_H\otimes0_l)_0^{--}$, $(1_H\otimes1_l)_0^{--}$
and $(1_H\otimes2_l)_0^{--}$. The P-wave decay modes
$\psi(1^3D_1)\pi\,\,\{^3P_1\}$ and $\psi(1^3D_2)\pi\,\,\{^5P_1\}$
both depend on the spin configurations $(1_H\otimes1_l)_0^{--}$ and
$(1_H\otimes2_l)_0^{--}$. There is obvious disfference between
tetraquarks with configurations $1^{--}(^1P_1)$ and $1^{--}(^3P_1)$
of type \uppercase\expandafter{\romannumeral1}. All the
$1^{--}(^1P_1)$ states can decay into $h_c \pi$ while all the
$1^{--}(^3P_1)$ of type \uppercase\expandafter{\romannumeral1} can
not. The $1^{--}(^1P_1)$ and $1^{--}(^3P_1)$ tetraquarks of type
\uppercase\expandafter{\romannumeral1} have the common decay modes
$\chi_{cJ}\rho\,\,(J=0,1,2)$, $J/\psi a_0(980)$,
$J/\psi\pi\,\,\{^3P_1\}$, $\psi(1^3D_1)\pi\,\,\{^3P_1\}$ and
$\psi(1^3D_2)\pi\,\,\{^5P_1\}$.

The $1^{--}(^5P_1)$ states of type
\uppercase\expandafter{\romannumeral1} can not decay into $h_c\pi$
in the heavy quark limit, which provides a way to distinguish the
$1^{--}(^1P_1)$ states from $1^{--}(^3P_1)$ and $1^{--}(^5P_1)$. The
$1^{--}(^5P_1)$ states also have the S-wave decay modes
$\chi_{cJ}\rho\,\,(J=0,1,2)$, $J/\psi a_0(980)$, and the P-wave
decay modes $J/\psi\pi\,\,\{^3P_1\}$, $\psi(1^3D_1)\pi\,\,\{^3P_1\}$
and $\psi(1^3D_2)\pi\,\,\{^5P_1\}$. The $1^{--}(^3P_1)$ states of
type \uppercase\expandafter{\romannumeral2} have the S-wave decay
mode $h_c\pi$ and the P-wave decay mode $J/\psi\pi$, where the
decays from the states $|35_{cs},1_c,1\rangle_{(21,\bar{21})}$ and
$|35_{cs},1_c,1\rangle_{(15,\bar{15})}$ are dominated by the color
singlet, while the decays from the states
$|280_{cs},1_c,1\rangle_{(21,\bar{15})}$ and
$|280_{cs},1_c,1\rangle_{(15,\bar{21})}$ are dominated by the color
octet. The above decays from the states
$|35_{cs},1_c,1\rangle_{(21,\bar{15})}$ and
$|35_{cs},1_c,1\rangle_{(15,\bar{21})}$ are governed not only by the
color singlet but also the color octet. They can also decay into
$\chi_{cJ}\rho\,\,(J=0,1,2)$, $\psi(1^3D_1)\pi\,\,\{^3P_1\}$ and
$\psi(1^3D_2)\pi\,\,\{^5P_1\}$, where the spin configurations
$(1_H\otimes1_l)_1^{--}$ and $(1_H\otimes2_l)_1^{--}$ are dominant.
Their decay mode $J/\psi a_0(980)$ is suppressed. The final state
$J/\psi a_0(980)$ contains the spin configuration
$(1_H\otimes0_l)_1^{--}$ only, which does not appear in all the
initial $1^{--}(^3P_1)$ states of type
\uppercase\expandafter{\romannumeral2}. Therefore, their decays into
$J/\psi a_0(980)$ are strongly suppressed under the heavy quark
symmetry.

If the P-wave excitation is between the diquark and anti-diquark
pair, the $0^-(1^{--})$ tetraquarks and their isovector partners
have three kinds of configurations $1^{--}(^1P_1)$, $1^{--}(^3P_1)$
and $1^{--}(^5P_1)$. $J/\psi\sigma$, $\psi(1^3D_1)\sigma$,
$h_c\eta$, $\chi_{cJ}\omega\,\,(J=0,1,2)$,
$J/\psi\eta\,\,\{^3P_1\}$, $\psi(1^3D_1)\eta\,\,\{^3P_1\}$ and
$\psi(1^3D_2)\eta\,\,\{^5P_1\}$ are the allowed decay modes of the
$1^{--}(^1P_1)$ tetraquarks of type
\uppercase\expandafter{\romannumeral1}. The $J/\psi\sigma$ is
governed by the spin configuration $(1_H\otimes0_l)_1^{--}$. The
decay mode $\psi(1^3D_1)\sigma$ is governed by the spin
configuration $(1_H\otimes2_l)_1^{--}$. The P-wave decay mode
$J/\psi\eta$ is dominated by the spin configuration
$(1_H\otimes1_l)_1^{--}$. The spin configuration
$(0_H\otimes1_l)_1^{--}$ is dominant for the decay mode $h_c\eta$.
The spin configuration $(0_H\otimes1_l)_0^{--}$ of the initial
states $|405_{cs},1_c,1\rangle_{(21,\bar{21})}$ and
$|189_{cs},1_c,1\rangle_{(15,\bar{15})}$ arises from the color octet
only. The initial states $|1_{cs},1_c,1\rangle_{(21,\bar{21})}$ and
$|1_{cs},1_c,1\rangle_{(15,\bar{15})}$ decay into $h_c\pi$ via both
the color singlet and color octet. The decay modes
$\chi_{cJ}\omega\,\,(J=0,1,2)$ depend on the spin configurations
$(1_H\otimes0_l)_0^{--}$, $(1_H\otimes1_l)_0^{--}$ and
$(1_H\otimes2_l)_0^{--}$. The isoscalar $1^{--}(^3P_1)$ and
$1^{--}(^5P_1)$ tetraquarks of type
\uppercase\expandafter{\romannumeral1} also have the decay modes
$J/\psi\sigma$, $\psi(1^3D_1)\sigma$,
$\chi_{cJ}\omega\,\,(J=0,1,2)$, $J/\psi\eta\,\,\{^3P_1\}$,
$\psi(1^3D_1)\eta\,\,\{^3P_1\}$ and $\psi(1^3D_2)\eta\,\,\{^5P_1\}$,
where the spin configurations which contribute to their decays are
similar to those of the $1^{--}(^1P_1)$ tetraquarks of type
\uppercase\expandafter{\romannumeral1}. We notice that the decay
mode $h_c\eta$ is suppressed for the $1^{--}(^3P_1)$ and
$1^{--}(^5P_1)$ tetraquarks, while it is allowed for the
$1^{--}(^1P_1)$ tetraquarks of type
\uppercase\expandafter{\romannumeral1}. This feature provides a way
to distinguish the $1^{--}(^1P_1)$ tetraquarks from the
$1^{--}(^3P_1)$ and $1^{--}(^5P_1)$ tetraquarks.

The $1^{--}(^3P_1)$ tetraquarks of type
\uppercase\expandafter{\romannumeral2} can also  decay into
$\chi_{cJ}\omega\,\,(J=0,1,2)$, $\psi(1^3D_1)\eta\,\,\{^3P_1\}$ and
$\psi(1^3D_2)\eta\,\,\{^5P_1\}$, where the spin configurations
$(1_H\otimes1_l)_0^{--}$ and $(1_H\otimes2_l)_0^{--}$ are dominant.
Similar to the case of the $1^{--}(^1P_1)$ tetraquarks of type
\uppercase\expandafter{\romannumeral1}, the $1^{--}(^3P_1)$
tetraquarks of type \uppercase\expandafter{\romannumeral2} also have
the S-wave decay mode $h_c\eta$ and the P-wave decay mode
$J/\psi\eta\,\,\{^3P_1\}$, which are governed by the spin
configurations $(0_H\otimes1_l)_0^{--}$ and $(1_H\otimes1_l)_0^{--}$
respectively. The states $|35_{cs},1_c,1\rangle_{(21,\bar{21})}$ and
$|35_{cs},1_c,1\rangle_{(15,\bar{15})}$ contain the color singlet
only while the states $|280_{cs},1_c,1\rangle_{(21,\bar{15})}$ and
$|280_{cs},1_c,1\rangle_{(15,\bar{21})}$ contain the color octet
only. Their decays into $h_c\eta$ and $J/\psi\eta\,\,\{^3P_1\}$
depend on the color singlet and the color octet respectively.

In Table \ref{tab:3}, we notice that the $1^{--}$ tetraquarks of
type \uppercase\expandafter{\romannumeral1} differ from the $1^{--}$
tetraquarks of type \uppercase\expandafter{\romannumeral2} greatly
in the decay modes $J/\psi\sigma$ and $\psi(1^3D_1)\sigma$. The
isoscalar $1^{--}$ tetraquarks of type
\uppercase\expandafter{\romannumeral1} can decay into $J/\psi\sigma$
and $\psi(1^3D_1)\sigma$. However, the isoscalar $1^{--}$
tetraquarks of type \uppercase\expandafter{\romannumeral2} do not
decay into these modes under the heavy quark symmetry.

\renewcommand{\arraystretch}{1.2}
\begin{table*}[htbp]
\scriptsize
\begin{center}
\caption{The decay matrix elements of the tetraquarks with
$J^{PC}=1^{-+}$. The reduced matrix elements $H_{\alpha}^{ij}\propto\langle
Q,i\|H_{eff}(\alpha)\|j\rangle$, where the indices $i$ and $j$
denote the light spin of the final and initial hadron respectively,
and $Q$ is the angular momentum of the final light meson. The
quantum numbers in the bracelets represent the total angular
momentum configurations of the final state particles.}\label{tab:4}
   \begin{tabular}{c cccccc} \toprule[1pt]

$I^G(J^{pc})$ & {Initial state} & \multicolumn{4}{c}{Final state}
\\\midrule[1pt]
      & {Type \uppercase\expandafter{\romannumeral1}} & $\chi_{c1}\pi$ & $h_c\rho$  & $\eta_c\pi\,\,\{^1P_1\}$ & $\eta_{c2}\pi\,\,\{^5P_1\}$ \\
      %\midrule[1pt]
      \multirow{6}{*}{$1^-(1^{-+})$} & $|35_{cs},1_c,1\rangle_{(21,\bar{21})}$     &$\frac{\sqrt{2}}{2}H_{\pi}^{11}$    &$-\frac{\sqrt{2}}{2}H_{\rho}^{11}$  &$-\frac{\sqrt{2}}{2}H_{\pi}^{01}$  &$-\frac{\sqrt{2}}{2}H_{\pi}^{21}$  \\

      &$|35_{cs},1_c,1\rangle_{(15,\bar{15})}$ &$\frac{\sqrt{2}}{2}H_{\pi}^{11}$    &$-\frac{\sqrt{2}}{2}H_{\rho}^{11}$  &$-\frac{\sqrt{2}}{2}H_{\pi}^{01}$  &$-\frac{\sqrt{2}}{2}H_{\pi}^{21}$   \\

      &$|35_{cs},1_c,1\rangle_{(21,\bar{15})}$ &$\frac{\sqrt{3}}{3}H_{\pi}^{11}+\frac{\sqrt{6}}{6}\tilde{H}_{\pi}^{11}$    &$\frac{\sqrt{3}}{3}H_{\rho}^{11}+\frac{\sqrt{6}}{6}\tilde{H}_{\rho}^{11}$   & $\frac{\sqrt{3}}{3}H_{\pi}^{01}+\frac{\sqrt{6}}{6}\tilde{H}_{\pi}^{01}$  & $\frac{\sqrt{3}}{3}H_{\pi}^{21}+\frac{\sqrt{6}}{6}\tilde{H}_{\pi}^{21}$ \\

      &$|35_{cs},1_c,1\rangle_{(15,\bar{21})}$ &$\frac{\sqrt{6}}{6}H_{\pi}^{11}-\frac{\sqrt{3}}{3}\tilde{H}_{\pi}^{11}$    &$\frac{\sqrt{6}}{6}H_{\rho}^{11}-\frac{\sqrt{3}}{3}\tilde{H}_{\rho}^{11}$   & $\frac{\sqrt{6}}{6}H_{\pi}^{01}-\frac{\sqrt{3}}{3}\tilde{H}_{\pi}^{01}$  & $\frac{\sqrt{6}}{6}H_{\pi}^{21}-\frac{\sqrt{3}}{3}\tilde{H}_{\pi}^{21}$  \\

      &$|280_{cs},1_c,1\rangle_{(21,\bar{15})}$ & $-\frac{\sqrt{2}}{2}\tilde{H}_{\pi}^{11}$    &$\frac{\sqrt{2}}{2}\tilde{H}_{\rho}^{11}$   & $\frac{\sqrt{2}}{2}\tilde{H}_{\pi}^{01}$  & $\frac{\sqrt{2}}{2}\tilde{H}_{\pi}^{21}$ \\

      &$|280_{cs},1_c,1\rangle_{(15,\bar{21})}$ & $-\frac{\sqrt{2}}{2}\tilde{H}_{\pi}^{11}$    &$\frac{\sqrt{2}}{2}\tilde{H}_{\rho}^{11}$   & $\frac{\sqrt{2}}{2}\tilde{H}_{\pi}^{01}$  & $\frac{\sqrt{2}}{2}\tilde{H}_{\pi}^{21}$ \\

      \midrule[1pt]

$I^G(J^{pc})$ & {Initial state} & \multicolumn{4}{c}{Final state}
\\\midrule[1pt]
      & {Type \uppercase\expandafter{\romannumeral2}} & $\chi_{c1}\pi$ & $h_c\rho$  & $\eta_c\pi\,\,\{^1P_1\}$ & $\eta_{c2}\pi\,\,\{^5P_1\}$ \\
      %\midrule[1pt]
      \multirow{4}{*}{$1^-(1^{-+})\,\,\{^1P_1\}$} & $|1_{cs},1_c,1\rangle_{(21,\bar{21})}$     &$\frac{\sqrt{21}}{42}H_{\pi}^{11}+\frac{\sqrt{21}}{42}\tilde{H}_{\pi}^{11}$    &$\frac{\sqrt{21}}{6}H_{\rho}^{11}+\frac{\sqrt{42}}{21}\tilde{H}_{\rho}^{11}$  &$\frac{\sqrt{21}}{6}H_{\pi}^{01}+\frac{\sqrt{42}}{21}\tilde{H}_{\pi}^{01}$  &$\frac{\sqrt{21}}{6}H_{\pi}^{21}+\frac{\sqrt{42}}{21}\tilde{H}_{\pi}^{21}$  \\

      &$|405_{cs},1_c,1\rangle_{(21,\bar{21})}$ &$\frac{2\sqrt{14}}{21}H_{\pi}^{11}-\frac{5\sqrt{7}}{42}\tilde{H}_{\pi}^{11}$    &$\frac{3\sqrt{7}}{14}\tilde{H}_{\rho}^{11}$  &$\frac{3\sqrt{7}}{14}\tilde{H}_{\pi}^{01}$  &$\frac{3\sqrt{7}}{14}\tilde{H}_{\pi}^{21}$ \\

      &$|1_{cs},1_c,1\rangle_{(15,\bar{15})}$ &$-\frac{\sqrt{15}}{30}H_{\pi}^{11}-\frac{\sqrt{30}}{15}\tilde{H}_{\pi}^{11}$    &$\frac{\sqrt{15}}{6}H_{\rho}^{11}-\frac{\sqrt{30}}{15}\tilde{H}_{\rho}^{11}$  &$\frac{\sqrt{15}}{6}H_{\pi}^{01}-\frac{\sqrt{30}}{15}\tilde{H}_{\pi}^{01}$  &$\frac{\sqrt{15}}{6}H_{\pi}^{21}-\frac{\sqrt{30}}{15}\tilde{H}_{\pi}^{21}$   \\

      &$|189_{cs},1_c,1\rangle_{(15,\bar{15})}$ &$\frac{2\sqrt{10}}{15}H_{\pi}^{11}+\frac{\sqrt{5}}{30}\tilde{H}_{\pi}^{11}$    &$-\frac{3\sqrt{5}}{10}\tilde{H}_{\rho}^{11}$  &$-\frac{3\sqrt{5}}{10}\tilde{H}_{\pi}^{01}$  &$-\frac{3\sqrt{5}}{10}\tilde{H}_{\pi}^{21}$ \\

      \midrule[1pt]

$I^G(J^{pc})$ & {Initial state} & \multicolumn{4}{c}{Final state}
\\\midrule[1pt]
      & {Type \uppercase\expandafter{\romannumeral2}} & $\chi_{c1}\pi$ & $h_c\rho$  & $\eta_c\pi\,\,\{^1P_1\}$ & $\eta_{c2}\pi\,\,\{^5P_1\}$ \\
      %\midrule[1pt]
      \multirow{4}{*}{$1^-(1^{-+})\,\,\{^3P_1\}$} & $|35_{cs},1_c,1\rangle_{(21,\bar{21})}$     &$-\frac{1}{6}H_{\pi}^{11}-\frac{\sqrt{2}}{3}\tilde{H}_{\pi}^{11}$    &$0$  &$0$  &$0$  \\

      &$|35_{cs},1_c,1\rangle_{(15,\bar{15})}$ &$-\frac{1}{6}H_{\pi}^{11}-\frac{\sqrt{2}}{3}\tilde{H}_{\pi}^{11}$    &$0$  &$0$  &$0$  \\

      &$|280_{cs},1_c,1\rangle_{(21,\bar{15})}$ &$\frac{\sqrt{2}}{3}H_{\pi}^{11}-\frac{1}{6}\tilde{H}_{\pi}^{11}$    &$0$  &$0$  &$0$  \\

      &$|280_{cs},1_c,1\rangle_{(15,\bar{21})}$ &$\frac{\sqrt{2}}{3}H_{\pi}^{11}-\frac{1}{6}\tilde{H}_{\pi}^{11}$    &$0$  &$0$  &$0$  \\

      \midrule[1pt]

$I^G(J^{pc})$ & {Initial state} & \multicolumn{4}{c}{Final state}
\\\midrule[1pt]
      & {Type \uppercase\expandafter{\romannumeral2}} & $\chi_{c1}\pi$ & $h_c\rho$  & $\eta_c\pi\,\,\{^1P_1\}$ & $\eta_{c2}\pi\,\,\{^5P_1\}$ \\
      %\midrule[1pt]
      \multirow{2}{*}{$1^-(1^{-+})\,\,\{^5P_1\}$} & $|405_{cs},1_c,1\rangle_{(21,\bar{21})}$      &$-\frac{\sqrt{10}}{6}H_{\pi}^{11}+\frac{\sqrt{5}}{6}\tilde{H}_{\pi}^{11}$    &$0$  &$0$  &$0$ \\

      &$|189_{cs},1_c,1\rangle_{(15,\bar{15})}$  &$-\frac{\sqrt{5}}{6}H_{\pi}^{11}+\frac{\sqrt{10}}{6}\tilde{H}_{\pi}^{11}$    &$0$  &$0$  &$0$  \\

      \midrule[1pt]

$I^G(J^{pc})$ & {Initial state} & \multicolumn{4}{c}{Final state}
\\\midrule[1pt]
      & {Type \uppercase\expandafter{\romannumeral1}} & $\chi_{c1}\eta$ & $h_c\omega$  & $\eta_c\eta\,\,\{^1P_1\}$ & $\eta_{c2}\eta\,\,\{^5P_1\}$ \\
      %\midrule[1pt]
      \multirow{6}{*}{$0^+(1^{-+})$} & $|35_{cs},1_c,1\rangle_{(21,\bar{21})}$     &$\frac{\sqrt{2}}{2}H_{\eta}^{11}$    &$-\frac{\sqrt{2}}{2}H_{\omega}^{11}$  &$-\frac{\sqrt{2}}{2}H_{\eta}^{01}$  &$-\frac{\sqrt{2}}{2}H_{\eta}^{21}$  \\

      &$|35_{cs},1_c,1\rangle_{(15,\bar{15})}$ &$\frac{\sqrt{2}}{2}H_{\eta}^{11}$    &$-\frac{\sqrt{2}}{2}H_{\omega}^{11}$  &$-\frac{\sqrt{2}}{2}H_{\eta}^{01}$  &$-\frac{\sqrt{2}}{2}H_{\eta}^{21}$  \\

      &$|35_{cs},1_c,1\rangle_{(21,\bar{15})}$ &$\frac{\sqrt{3}}{3}H_{\eta}^{11}+\frac{\sqrt{6}}{6}\tilde{H}_{\eta}^{11}$    &$\frac{\sqrt{3}}{3}H_{\omega}^{11}+\frac{\sqrt{6}}{6}\tilde{H}_{\omega}^{11}$   & $\frac{\sqrt{3}}{3}H_{\eta}^{01}+\frac{\sqrt{6}}{6}\tilde{H}_{\eta}^{01}$  & $\frac{\sqrt{3}}{3}H_{\eta}^{21}+\frac{\sqrt{6}}{6}\tilde{H}_{\eta}^{21}$  \\

      &$|35_{cs},1_c,1\rangle_{(15,\bar{21})}$ &$\frac{\sqrt{6}}{6}H_{\eta}^{11}-\frac{\sqrt{3}}{3}\tilde{H}_{\eta}^{11}$    &$\frac{\sqrt{6}}{6}H_{\omega}^{11}-\frac{\sqrt{3}}{3}\tilde{H}_{\omega}^{11}$   & $\frac{\sqrt{6}}{6}H_{\eta}^{01}-\frac{\sqrt{3}}{3}\tilde{H}_{\eta}^{01}$  & $\frac{\sqrt{6}}{6}H_{\eta}^{21}-\frac{\sqrt{3}}{3}\tilde{H}_{\eta}^{21}$  \\

      &$|280_{cs},1_c,1\rangle_{(21,\bar{15})}$ & $-\frac{\sqrt{2}}{2}\tilde{H}_{\eta}^{11}$    &$\frac{\sqrt{2}}{2}\tilde{H}_{\omega}^{11}$   & $\frac{\sqrt{2}}{2}\tilde{H}_{\eta}^{01}$  & $\frac{\sqrt{2}}{2}\tilde{H}_{\eta}^{21}$ \\

      &$|280_{cs},1_c,1\rangle_{(15,\bar{21})}$ & $-\frac{\sqrt{2}}{2}\tilde{H}_{\eta}^{11}$    &$\frac{\sqrt{2}}{2}\tilde{H}_{\omega}^{11}$   & $\frac{\sqrt{2}}{2}\tilde{H}_{\eta}^{01}$  & $\frac{\sqrt{2}}{2}\tilde{H}_{\eta}^{21}$ \\

      \midrule[1pt]

$I^G(J^{pc})$ & {Initial state} & \multicolumn{4}{c}{Final state}
\\\midrule[1pt]
      & {Type \uppercase\expandafter{\romannumeral2}} & $\chi_{c1}\eta$ & $h_c\omega$  & $\eta_c\eta\,\,\{^1P_1\}$ & $\eta_{c2}\eta\,\,\{^5P_1\}$ \\
      %\midrule[1pt]
      \multirow{4}{*}{$0^+(1^{-+})\,\,\{^1P_1\}$} & $|1_{cs},1_c,1\rangle_{(21,\bar{21})}$     &$\frac{\sqrt{21}}{42}H_{\eta}^{11}+\frac{\sqrt{21}}{42}\tilde{H}_{\eta}^{11}$    &$\frac{\sqrt{21}}{6}H_{\omega}^{11}+\frac{\sqrt{42}}{21}\tilde{H}_{\omega}^{11}$  &$\frac{\sqrt{21}}{6}H_{\eta}^{01}+\frac{\sqrt{42}}{21}\tilde{H}_{\eta}^{01}$  &$\frac{\sqrt{21}}{6}H_{\eta}^{21}+\frac{\sqrt{42}}{21}\tilde{H}_{\eta}^{21}$  \\

      &$|405_{cs},1_c,1\rangle_{(21,\bar{21})}$ &$\frac{2\sqrt{14}}{21}H_{\eta}^{11}-\frac{5\sqrt{7}}{42}\tilde{H}_{\eta}^{11}$    &$\frac{3\sqrt{7}}{14}\tilde{H}_{\omega}^{11}$  &$\frac{3\sqrt{7}}{14}\tilde{H}_{\eta}^{01}$  &$\frac{3\sqrt{7}}{14}\tilde{H}_{\eta}^{21}$ \\

      &$|1_{cs},1_c,1\rangle_{(15,\bar{15})}$ &$-\frac{\sqrt{15}}{30}H_{\eta}^{11}-\frac{\sqrt{30}}{15}\tilde{H}_{\eta}^{11}$    &$\frac{\sqrt{15}}{6}H_{\omega}^{11}-\frac{\sqrt{30}}{15}\tilde{H}_{\omega}^{11}$  &$\frac{\sqrt{15}}{6}H_{\eta}^{01}-\frac{\sqrt{30}}{15}\tilde{H}_{\eta}^{01}$  &$\frac{\sqrt{15}}{6}H_{\eta}^{21}-\frac{\sqrt{30}}{15}\tilde{H}_{\eta}^{21}$  \\

      &$|189_{cs},1_c,1\rangle_{(15,\bar{15})}$ &$\frac{2\sqrt{10}}{15}H_{\eta}^{11}+\frac{\sqrt{5}}{30}\tilde{H}_{\eta}^{11}$    &$-\frac{3\sqrt{5}}{10}\tilde{H}_{\omega}^{11}$  &$-\frac{3\sqrt{5}}{10}\tilde{H}_{\eta}^{01}$  &$-\frac{3\sqrt{5}}{10}\tilde{H}_{\eta}^{21}$ \\

      \midrule[1pt]

$I^G(J^{pc})$ & {Initial state} & \multicolumn{4}{c}{Final state}
\\\midrule[1pt]
      & {Type \uppercase\expandafter{\romannumeral2}} & $\chi_{c1}\eta$ & $h_c\omega$  & $\eta_c\eta\,\,\{^1P_1\}$ & $\eta_{c2}\eta\,\,\{^5P_1\}$ \\
      %\midrule[1pt]
      \multirow{4}{*}{$0^+(1^{-+})\,\,\{^3P_1\}$} & $|35_{cs},1_c,1\rangle_{(21,\bar{21})}$     &$-\frac{1}{6}H_{\eta}^{11}-\frac{\sqrt{2}}{3}\tilde{H}_{\eta}^{11}$    &$0$  &$0$  &$0$ \\

      &$|35_{cs},1_c,1\rangle_{(15,\bar{15})}$ &$-\frac{1}{6}H_{\eta}^{11}-\frac{\sqrt{2}}{3}\tilde{H}_{\eta}^{11}$    &$0$  &$0$  &$0$  \\

      &$|280_{cs},1_c,1\rangle_{(21,\bar{15})}$ &$\frac{\sqrt{2}}{3}H_{\eta}^{11}-\frac{1}{6}\tilde{H}_{\eta}^{11}$    &$0$  &$0$  &$0$  \\

      &$|280_{cs},1_c,1\rangle_{(15,\bar{21})}$ &$\frac{\sqrt{2}}{3}H_{\eta}^{11}-\frac{1}{6}\tilde{H}_{\eta}^{11}$    &$0$  &$0$  &$0$  \\

      \midrule[1pt]

$I^G(J^{pc})$ & {Initial state} & \multicolumn{4}{c}{Final state}
\\\midrule[1pt]
      & {Type \uppercase\expandafter{\romannumeral2}} & $\chi_{c1}\eta$ & $h_c\omega$  & $\eta_c\eta\,\,\{^1P_1\}$ & $\eta_{c2}\eta\,\,\{^5P_1\}$ \\
      %\midrule[1pt]
      \multirow{2}{*}{$0^+(1^{-+})\,\,\{^5P_1\}$} & $|405_{cs},1_c,1\rangle_{(21,\bar{21})}$      &$-\frac{\sqrt{10}}{6}H_{\eta}^{11}+\frac{\sqrt{5}}{6}\tilde{H}_{\eta}^{11}$    &$0$  &$0$  &$0$  \\

      &$|189_{cs},1_c,1\rangle_{(15,\bar{15})}$  &$-\frac{\sqrt{5}}{6}H_{\eta}^{11}+\frac{\sqrt{10}}{6}\tilde{H}_{\eta}^{11}$    &$0$  &$0$  &$0$  \\

      \midrule[1pt]

      \end{tabular}
\end{center}
\end{table*}

There are two types of $1^{-+}$ states. For the isovector $1^{-+}$
states of type \uppercase\expandafter{\romannumeral1}, the allowed
S-wave decay modes are $\chi_{c1}\pi$ and $h_c\rho$, where the spin
configurations $(1_H\otimes1_l)_0^{-+}$ and $(0_H\otimes1_l)_0^{-+}$
are dominant respectively. The allowed P-wave decay modes are
$\eta_c\pi$ and $\eta_{c2}\pi$, where the spin configuration
$(0_H\otimes1_l)_0^{-+}$ is dominant. The allowed decay modes from
$|35_{cs},1_c,1\rangle_{(21,\bar{21})}$ and
$|35_{cs},1_c,1\rangle_{(15,\bar{15})}$ are governed by the color
singlet, while the decay modes from
$|280_{cs},1_c,1\rangle_{(21,\bar{15})}$ and
$|280_{cs},1_c,1\rangle_{(15,\bar{21})}$ are governed by the color
octet. The mode $\chi_{c1}\pi$ is allowed for the isovector
$1^{-+}\,\,\{^1P_1\}$, $1^{-+}\,\,\{^3P_1\}$ and
$1^{-+}\,\,\{^5P_1\}$ states of type
\uppercase\expandafter{\romannumeral2}.

For the $1^{-+}\,\,\{^1P_1\}$, $1^{-+}\,\,\{^3P_1\}$ and
$1^{-+}\,\,\{^5P_1\}$ states of type
\uppercase\expandafter{\romannumeral2}, the $1^{-+}\,\,\{^1P_1\}$
can decay into $h_c\rho$, $\eta_c\pi\,\,\{^1P_1\}$ and
$\eta_{c2}\pi\,\,\{^5P_1\}$. These decay modes are suppressed for
the states $1^{-+}\,\,\{^3P_1\}$ and $1^{-+}\,\,\{^5P_1\}$. The
re-coupled color-spin wave functions of the $1^{-+}\,\,\{^3P_1\}$
and $1^{-+}\,\,\{^5P_1\}$ states of type
\uppercase\expandafter{\romannumeral2} do not contain the spin
configuration $(0_H\otimes1_l)_0^{-+}$. Therefore, their decays into
$h_c\rho$, $\eta_c\pi\,\,\{^1P_1\}$ and $\eta_{c2}\pi\,\,\{^5P_1\}$
are strongly suppressed in the heavy quark limit. The decay modes
$h_c\rho$, $\eta_c\pi\,\,\{^1P_1\}$ and $\eta_{c2}\pi\,\,\{^5P_1\}$
of the $1^-(1^{-+})\,\,\{^1P_1\}$ states
$|405_{cs},1_c,1\rangle_{(21,\bar{21})}$ and
$|189_{cs},1_c,1\rangle_{(15,\bar{15})}$ of type
\uppercase\expandafter{\romannumeral2} are governed by their color
octet and the spin configuration $(0_H\otimes1_l)_0^{-+}$.

The isoscalar $1^{-+}$ states of type
\uppercase\expandafter{\romannumeral1} can decay into
$\chi_{c1}\eta$, $h_c\omega$, $\eta_c\eta\,\,\{^1P_1\}$ and
$\eta_{c2}\eta\,\,\{^5P_1\}$. These decay modes are also allowed for
the $1^{-+}\,\,\{^1P_1\}$ states of Type
\uppercase\expandafter{\romannumeral2}. However, the states
$1^{-+}\,\,\{^3P_1\}$ and $1^{-+}\,\,\{^5P_1\}$ of type
\uppercase\expandafter{\romannumeral2} do not decay into
$h_c\omega$, $\eta_c\eta\,\,\{^1P_1\}$ and
$\eta_{c2}\eta\,\,\{^5P_1\}$. These decay modes are dominated by the
the spin configuration $(0_H\otimes1_l)_0^{-+}$ which does not
appear in the states $1^{-+}\,\,\{^3P_1\}$ and $1^{-+}\,\,\{^5P_1\}$
of type \uppercase\expandafter{\romannumeral2}. The decay mode
$\chi_{c1}\eta$ is allowed for all the isoscalar $1^{-+}$ states.

\begin{table*}[htbp]
\scriptsize
\begin{center}
\caption{The decay matrix elements of the tetraquarks with
$J^{PC}=0^{-+},\,\,0^{--}$. The reduced matrix elements $H_{\alpha}^{ij}\propto\langle
Q,i\|H_{eff}(\alpha)\|j\rangle$, where the indices $i$ and $j$
denote the light spin of the final and initial hadron respectively,
and $Q$ is the angular momentum of the final light meson. The
quantum numbers in the bracelets represent the total angular
momentum configurations of the final state particles.}\label{tab:5}
   \begin{tabular}{c cccc|cccccccccc} \toprule[1pt]

 $I^G(J^{pc})$ & {Initial state} & \multicolumn{3}{c}{Final state} &  $I^G(J^{pc})$ & {Initial state} & \multicolumn{4}{c}{Final state} \\\midrule[1pt]
      & {Type \uppercase\expandafter{\romannumeral1}} & $\chi_{c0}\pi$ & $h_c\rho$ & $\eta_c a_0(980)$ & & {Type \uppercase\expandafter{\romannumeral2}} & $\chi_{c1}\rho$ & $h_c a_0(980)\,\,\{^3P_0\}$ & $\eta_c\rho\,\,\{^3P_0\}$ & $J/\psi\pi\,\,\{^3P_0\}$  \\
      %\midrule[1pt]
      \multirow{6}{*}{$1^-(0^{-+})$} & $|35_{cs},1_c,0\rangle_{(21,\bar{21})}$     &$\frac{\sqrt{2}}{2}H_{\pi}^{11}$    &$-\frac{\sqrt{2}}{2}H_{\rho}^{10}$ &$-\frac{\sqrt{2}}{2}H_{a_0}^{00}$ &\multirow{6}{*}{$1^+(0^{--})$ }    &$|35_{cs},1_c,0\rangle_{(21,\bar{21})}$ & $\frac{\sqrt{2}}{2}H_{\rho}^{11}$ & $-\frac{\sqrt{2}}{2}H_{a_0}^{11}$ & $-\frac{\sqrt{2}}{2}H_{\rho}^{01}$ & $\frac{\sqrt{2}}{2}H_{\pi}^{01}$ \\

      & $|35_{cs},1_c,0\rangle_{(15,\bar{15})}$     &$\frac{\sqrt{2}}{2}H_{\pi}^{11}$    &$-\frac{\sqrt{2}}{2}H_{\rho}^{10}$ &$-\frac{\sqrt{2}}{2}H_{a_0}^{00}$ &   &$|35_{cs},1_c,0\rangle_{(15,\bar{15})}$ & $\frac{\sqrt{2}}{2}H_{\rho}^{11}$ & $-\frac{\sqrt{2}}{2}H_{a_0}^{10}$ & $-\frac{\sqrt{2}}{2}H_{\rho}^{10}$ & $\frac{\sqrt{2}}{2}H_{\pi}^{01}$ \\

      & $|35_{cs},1_c,0\rangle_{(21,\bar{15})}$     &$\frac{\sqrt{3}}{3}H_{\pi}^{11}+\frac{\sqrt{6}}{6}\tilde{H}_{\pi}^{11}$    &$\frac{\sqrt{3}}{3}H_{\rho}^{10}+\frac{\sqrt{6}}{6}\tilde{H}_{\rho}^{10}$ & $\frac{\sqrt{3}}{3}H_{a_0}^{00}+\frac{\sqrt{6}}{6}\tilde{H}_{a_0}^{00}$ &    &$|35_{cs},1_c,0\rangle_{(21,\bar{15})}$ &$\frac{\sqrt{3}}{3}H_{\rho}^{11}+\frac{\sqrt{6}}{6}\tilde{H}_{\rho}^{11}$ &$\frac{\sqrt{3}}{3}H_{a_0}^{10}+\frac{\sqrt{6}}{6}\tilde{H}_{a_0}^{10}$ &$\frac{\sqrt{3}}{3}H_{\rho}^{10}+\frac{\sqrt{6}}{6}\tilde{H}_{\rho}^{10}$ &$\frac{\sqrt{3}}{3}H_{\pi}^{00}+\frac{\sqrt{6}}{6}\tilde{H}_{\pi}^{00}$ \\

     & $|35_{cs},1_c,0\rangle_{(15,\bar{21})}$     &$\frac{\sqrt{6}}{6}H_{\pi}^{11}-\frac{\sqrt{3}}{3}\tilde{H}_{\pi}^{11}$    &$\frac{\sqrt{6}}{6}H_{\rho}^{10}-\frac{\sqrt{3}}{3}\tilde{H}_{\rho}^{10}$ & $\frac{\sqrt{6}}{6}H_{a_0}^{00}-\frac{\sqrt{3}}{3}\tilde{H}_{a_0}^{00}$ &    &$|35_{cs},1_c,0\rangle_{(15,\bar{21})}$ &$\frac{\sqrt{6}}{6}H_{\rho}^{11}-\frac{\sqrt{3}}{3}\tilde{H}_{\rho}^{11}$ &$\frac{\sqrt{6}}{6}H_{a_0}^{10}-\frac{\sqrt{3}}{3}\tilde{H}_{a_0}^{10}$ &$\frac{\sqrt{6}}{6}H_{\rho}^{10}-\frac{\sqrt{3}}{3}\tilde{H}_{\rho}^{10}$ &$\frac{\sqrt{6}}{6}H_{\pi}^{00}-\frac{\sqrt{3}}{3}\tilde{H}_{\pi}^{00}$ \\

    & $|280_{cs},1_c,0\rangle_{(21,\bar{15})}$     &$-\frac{\sqrt{2}}{2}\tilde{H}_{\pi}^{11}$    &$\frac{\sqrt{2}}{2}\tilde{H}_{\rho}^{10}$ &$\frac{\sqrt{2}}{2}\tilde{H}_{a_0}^{00}$ &   &$|280_{cs},1_c,0\rangle_{(21,\bar{15})}$ & $-\frac{\sqrt{2}}{2}\tilde{H}_{\rho}^{11}$ & $\frac{\sqrt{2}}{2}\tilde{H}_{a_0}^{10}$ & $\frac{\sqrt{2}}{2}\tilde{H}_{\rho}^{10}$ & $-\frac{\sqrt{2}}{2}\tilde{H}_{\pi}^{01}$ \\

    & $|280_{cs},1_c,0\rangle_{(15,\bar{21})}$     &$-\frac{\sqrt{2}}{2}\tilde{H}_{\pi}^{11}$    &$\frac{\sqrt{2}}{2}\tilde{H}_{\rho}^{10}$ &$\frac{\sqrt{2}}{2}\tilde{H}_{a_0}^{00}$ &   &$|280_{cs},1_c,0\rangle_{(15,\bar{21})}$ & $-\frac{\sqrt{2}}{2}\tilde{H}_{\rho}^{11}$ & $\frac{\sqrt{2}}{2}\tilde{H}_{a_0}^{10}$ & $\frac{\sqrt{2}}{2}\tilde{H}_{\rho}^{10}$ & $-\frac{\sqrt{2}}{2}\tilde{H}_{\pi}^{01}$ \\

      \midrule[1pt]

$I^G(J^{pc})$ & {Initial state} & \multicolumn{3}{c}{Final state} &
$I^G(J^{pc})$ & {Initial state} & \multicolumn{3}{c}{Final state}
\\\midrule[1pt]
      & {Type \uppercase\expandafter{\romannumeral2}} & $\chi_{c0}\pi$ & $h_c\rho$ & $\eta_c a_0(980)$ & & {Type \uppercase\expandafter{\romannumeral1}} & $\chi_{c1}\rho$ & $h_c a_0(980)\,\,\{^3P_0\}$ & $\eta_c\rho\,\,\{^3P_0\}$ & $J/\psi\pi\,\,\{^3P_0\}$  \\
      %\midrule[1pt]
      \multirow{4}{*}{$1^-(0^{-+})$} & $|35_{cs},1_c,0\rangle_{(21,\bar{21})}$     &$-\frac{1}{3}H_{\pi}^{11}-\frac{2\sqrt{2}}{3}\tilde{H}_{\pi}^{11}$   &$0$ & $0$ &\multirow{4}{*}{$1^+(0^{--})$ }    &$|35_{cs},1_c,0\rangle_{(21,\bar{21})}$ &$-\frac{1}{3}H_{\rho}^{11}-\frac{2\sqrt{2}}{3}\tilde{H}_{\rho}^{11}$   &$0$ &$0$ & $-\frac{1}{3}H_{\pi}^{01}-\frac{2\sqrt{2}}{3}\tilde{H}_{\pi}^{01}$ \\

      & $|35_{cs},1_c,0\rangle_{(15,\bar{15})}$     &$-\frac{1}{3}H_{\pi}^{11}-\frac{2\sqrt{2}}{3}\tilde{H}_{\pi}^{11}$   &$0$ & $0$ &    &$|35_{cs},1_c,0\rangle_{(15,\bar{15})}$ &$-\frac{1}{3}H_{\rho}^{11}-\frac{2\sqrt{2}}{3}\tilde{H}_{\rho}^{11}$   &$0$&$0$ & $-\frac{1}{3}H_{\pi}^{01}-\frac{2\sqrt{2}}{3}\tilde{H}_{\pi}^{01}$ \\

    & $|280_{cs},1_c,0\rangle_{(21,\bar{15})}$     &$\frac{2\sqrt{2}}{3}H_{\pi}^{11}-\frac{1}{3}\tilde{H}_{\pi}^{11}$    &$0$ & $0$ &   &$|280_{cs},1_c,0\rangle_{(21,\bar{15})}$ &$\frac{2\sqrt{2}}{3}H_{\rho}^{11}-\frac{1}{3}\tilde{H}_{\rho}^{11}$    &$0$&$0$ & $\frac{2\sqrt{2}}{3}H_{\pi}^{01}-\frac{1}{3}\tilde{H}_{\pi}^{01}$\\

    & $|280_{cs},1_c,0\rangle_{(15,\bar{21})}$     &$\frac{2\sqrt{2}}{3}H_{\pi}^{11}-\frac{1}{3}\tilde{H}_{\pi}^{11}$    &$0$ & $0$ &   &$|280_{cs},1_c,0\rangle_{(15,\bar{21})}$ &$\frac{2\sqrt{2}}{3}H_{\rho}^{11}-\frac{1}{3}\tilde{H}_{\rho}^{11}$    &$0$ &$0$ & $\frac{2\sqrt{2}}{3}H_{\pi}^{01}-\frac{1}{3}\tilde{H}_{\pi}^{01}$\\

      \midrule[1pt]

$I^G(J^{pc})$ & {Initial state} & \multicolumn{3}{c}{Final state} &
$I^G(J^{pc})$ & {Initial state} & \multicolumn{3}{c}{Final state}
\\\midrule[1pt]
      & {Type \uppercase\expandafter{\romannumeral1}} & $\chi_{c0}\eta$ & $h_c\omega$ & $\eta_c\sigma$ & & {Type \uppercase\expandafter{\romannumeral2}} & $\chi_{c1}\omega$ & $h_c\sigma\,\,\{^3P_0\}$ & $\eta_c\omega\,\,\{^3P_0\}$ & $J/\psi\eta\,\,\{^3P_0\}$  \\
      %\midrule[1pt]
      \multirow{6}{*}{$0^+(0^{-+})$} & $|35_{cs},1_c,0\rangle_{(21,\bar{21})}$     &$\frac{\sqrt{2}}{2}H_{\eta}^{11}$    &$-\frac{\sqrt{2}}{2}H_{\omega}^{10}$ &$-\frac{\sqrt{2}}{2}H_{\sigma}^{00}$ &\multirow{6}{*}{$0^-(0^{--})$ }    &$|35_{cs},1_c,0\rangle_{(21,\bar{21})}$ & $\frac{\sqrt{2}}{2}H_{\omega}^{11}$ & $-\frac{\sqrt{2}}{2}H_{\sigma}^{10}$ & $-\frac{\sqrt{2}}{2}H_{\omega}^{10}$ & $\frac{\sqrt{2}}{2}H_{\eta}^{01}$ \\

      & $|35_{cs},1_c,0\rangle_{(15,\bar{15})}$     &$\frac{\sqrt{2}}{2}H_{\eta}^{11}$    &$-\frac{\sqrt{2}}{2}H_{\omega}^{10}$ &$-\frac{\sqrt{2}}{2}H_{\sigma}^{00}$ &   &$|35_{cs},1_c,0\rangle_{(15,\bar{15})}$ & $\frac{\sqrt{2}}{2}H_{\omega}^{11}$ & $-\frac{\sqrt{2}}{2}H_{\sigma}^{10}$ & $-\frac{\sqrt{2}}{2}H_{\omega}^{10}$ & $\frac{\sqrt{2}}{2}H_{\eta}^{01}$ \\

      & $|35_{cs},1_c,0\rangle_{(21,\bar{15})}$     &$\frac{\sqrt{3}}{3}H_{\eta}^{11}+\frac{\sqrt{6}}{6}\tilde{H}_{\eta}^{11}$    &$\frac{\sqrt{3}}{3}H_{\omega}^{10}+\frac{\sqrt{6}}{6}\tilde{H}_{\omega}^{10}$ & $\frac{\sqrt{3}}{3}H_{\sigma}^{00}+\frac{\sqrt{6}}{6}\tilde{H}_{\sigma}^{00}$ &    &$|35_{cs},1_c,0\rangle_{(21,\bar{15})}$ &$\frac{\sqrt{3}}{3}H_{\omega}^{11}+\frac{\sqrt{6}}{6}\tilde{H}_{\omega}^{11}$ &$\frac{\sqrt{3}}{3}H_{\sigma}^{10}+\frac{\sqrt{6}}{6}\tilde{H}_{\sigma}^{10}$ &$\frac{\sqrt{3}}{3}H_{\omega}^{10}+\frac{\sqrt{6}}{6}\tilde{H}_{\omega}^{10}$ &$\frac{\sqrt{3}}{3}H_{\sigma}^{00}+\frac{\sqrt{6}}{6}\tilde{H}_{\sigma}^{00}$ \\

     & $|35_{cs},1_c,0\rangle_{(15,\bar{21})}$     &$\frac{\sqrt{6}}{6}H_{\eta}^{11}-\frac{\sqrt{3}}{3}\tilde{H}_{\eta}^{11}$    &$\frac{\sqrt{6}}{6}H_{\omega}^{10}-\frac{\sqrt{3}}{3}\tilde{H}_{\omega}^{10}$ & $\frac{\sqrt{6}}{6}H_{\sigma}^{00}-\frac{\sqrt{3}}{3}\tilde{H}_{\sigma}^{00}$ &    &$|35_{cs},1_c,0\rangle_{(15,\bar{21})}$ &$\frac{\sqrt{6}}{6}H_{\omega}^{11}-\frac{\sqrt{3}}{3}\tilde{H}_{\omega}^{11}$ &$\frac{\sqrt{6}}{6}H_{\sigma}^{10}-\frac{\sqrt{3}}{3}\tilde{H}_{\sigma}^{10}$ &$\frac{\sqrt{6}}{6}H_{\omega}^{10}-\frac{\sqrt{3}}{3}\tilde{H}_{\omega}^{10}$ &$\frac{\sqrt{6}}{6}H_{\sigma}^{00}-\frac{\sqrt{3}}{3}\tilde{H}_{\sigma}^{00}$ \\

    & $|280_{cs},1_c,0\rangle_{(21,\bar{15})}$     &$-\frac{\sqrt{2}}{2}\tilde{H}_{\eta}^{11}$    &$\frac{\sqrt{2}}{2}\tilde{H}_{\omega}^{10}$ &$\frac{\sqrt{2}}{2}\tilde{H}_{\sigma}^{00}$ &   &$|280_{cs},1_c,0\rangle_{(21,\bar{15})}$ & $-\frac{\sqrt{2}}{2}\tilde{H}_{\omega}^{11}$ & $\frac{\sqrt{2}}{2}\tilde{H}_{\sigma}^{10}$ & $\frac{\sqrt{2}}{2}\tilde{H}_{\omega}^{10}$ & $-\frac{\sqrt{2}}{2}\tilde{H}_{\eta}^{01}$ \\

    & $|280_{cs},1_c,0\rangle_{(15,\bar{21})}$     &$-\frac{\sqrt{2}}{2}\tilde{H}_{\eta}^{11}$    &$\frac{\sqrt{2}}{2}\tilde{H}_{\omega}^{10}$ &$\frac{\sqrt{2}}{2}\tilde{H}_{\sigma}^{00}$ &   &$|280_{cs},1_c,0\rangle_{(15,\bar{21})}$ & $-\frac{\sqrt{2}}{2}\tilde{H}_{\omega}^{11}$ & $\frac{\sqrt{2}}{2}\tilde{H}_{\sigma}^{10}$ & $\frac{\sqrt{2}}{2}\tilde{H}_{\omega}^{10}$ & $-\frac{\sqrt{2}}{2}\tilde{H}_{\eta}^{01}$ \\

      \midrule[1pt]

$I^G(J^{pc})$ & {Initial state} & \multicolumn{3}{c}{Final state} &
$I^G(J^{pc})$ & {Initial state} & \multicolumn{3}{c}{Final state}
\\\midrule[1pt]
      & {Type \uppercase\expandafter{\romannumeral2}} & $\chi_{c0}\eta$ & $h_c\omega$ & $\eta_c\sigma$ & & {Type \uppercase\expandafter{\romannumeral1}} & $\chi_{c1}\omega$ & $h_c\sigma\,\,\{^3P_0\}$ & $\eta_c\omega\,\,\{^3P_0\}$ & $J/\psi\eta\,\,\{^3P_0\}$  \\
      %\midrule[1pt]
      \multirow{4}{*}{$0^+(0^{-+})$} & $|35_{cs},1_c,0\rangle_{(21,\bar{21})}$     &$-\frac{1}{3}H_{\eta}^{11}-\frac{2\sqrt{2}}{3}\tilde{H}_{\eta}^{11}$   &$0$ & $0$ &\multirow{4}{*}{$0^-(0^{--})$ }    &$|35_{cs},1_c,0\rangle_{(21,\bar{21})}$ &$-\frac{1}{3}H_{\omega}^{11}-\frac{2\sqrt{2}}{3}\tilde{H}_{\omega}^{11}$   &$0$ &$0$ & $-\frac{1}{3}H_{\eta}^{01}-\frac{2\sqrt{2}}{3}\tilde{H}_{\eta}^{01}$ \\

      & $|35_{cs},1_c,0\rangle_{(15,\bar{15})}$     &$-\frac{1}{3}H_{\eta}^{11}-\frac{2\sqrt{2}}{3}\tilde{H}_{\eta}^{11}$   &$0$ & $0$ &    &$|35_{cs},1_c,0\rangle_{(15,\bar{15})}$ &$-\frac{1}{3}H_{\omega}^{11}-\frac{2\sqrt{2}}{3}\tilde{H}_{\omega}^{11}$   &$0$ &$0$ & $-\frac{1}{3}H_{\eta}^{01}-\frac{2\sqrt{2}}{3}\tilde{H}_{\eta}^{01}$ \\

    & $|280_{cs},1_c,0\rangle_{(21,\bar{15})}$     &$\frac{2\sqrt{2}}{3}H_{\eta}^{11}-\frac{1}{3}\tilde{H}_{\eta}^{11}$    &$0$ & $0$ &   &$|280_{cs},1_c,0\rangle_{(21,\bar{15})}$ &$\frac{2\sqrt{2}}{3}H_{\omega}^{11}-\frac{1}{3}\tilde{H}_{\omega}^{11}$    &$0$ &$0$ & $\frac{2\sqrt{2}}{3}H_{\eta}^{01}-\frac{1}{3}\tilde{H}_{\eta}^{01}$\\

    & $|280_{cs},1_c,0\rangle_{(15,\bar{21})}$     &$\frac{2\sqrt{2}}{3}H_{\eta}^{11}-\frac{1}{3}\tilde{H}_{\eta}^{11}$    &$0$ & $0$ &   &$|280_{cs},1_c,0\rangle_{(15,\bar{21})}$ &$\frac{2\sqrt{2}}{3}H_{\omega}^{11}-\frac{1}{3}\tilde{H}_{\omega}^{11}$    &$0$ &$0$ & $\frac{2\sqrt{2}}{3}H_{\eta}^{01}-\frac{1}{3}\tilde{H}_{\eta}^{01}$\\

      \midrule[1pt]

      \end{tabular}
\end{center}
\end{table*}

Both of the $0^{-+}$ and the $0^{--}$ tetraquark states have two
types. In Table \ref{tab:5}, we notice that all the isovector
$0^{-+}$ states can decay into $\chi_{c0}\pi$, which is governed by
the spin configuration $(1_H\otimes1_l)_0^{-+}$. The difference
between the $1^-(0^{-+})$ states of type
\uppercase\expandafter{\romannumeral1} and the $1^-(0^{-+})$ states
of type \uppercase\expandafter{\romannumeral2} is reflected in their
decay modes. The decay modes $h_c\rho$ and $\eta_c a_0(980)$ are
dominated by the spin configuration $(0_H\otimes1_l)_0^{-+}$, which
exists in the $1^-(0^{-+})$ states of type
\uppercase\expandafter{\romannumeral1} and is absent in the
$1^-(0^{-+})$ states of type \uppercase\expandafter{\romannumeral2}.
Thus, these decay modes from the $1^-(0^{-+})$ states of type
\uppercase\expandafter{\romannumeral2} are suppressed.

The same situation occurs for the isoscalar $0^{-+}$ states of type
\uppercase\expandafter{\romannumeral1} and type
\uppercase\expandafter{\romannumeral2}. The decay modes $h_c\omega$
and $\eta_c\sigma$ are governed by the spin configuration
$(0_H\otimes1_l)_0^{-+}$ as well. Nevertheless, this spin
configuration $(0_H\otimes1_l)_0^{-+}$ is absent in the
$0^+(0^{-+})$ states of type \uppercase\expandafter{\romannumeral2}.
The heavy spin and light spin are not conserved for the decay modes
$h_c\omega$ and $\eta_c\sigma$. These decay widths are strongly
suppressed in the heavy quark limit. The decay modes $h_c\omega$ and
$\eta_c\sigma$ are allowed for the $0^+(0^{-+})$ states of type
\uppercase\expandafter{\romannumeral1}. Both types of states can
decays into $\chi_{c0}\eta$.

The $\chi_{c1}\rho$, $h_c a_0(980)\,\,\{^3P_0\}$,
$\eta_c\rho\,\,\{^3P_0\}$ and $J/\psi\pi\,\,\{^3P_0\}$ are the
allowed decay modes for the $1^+(0^{--})$ states of type
\uppercase\expandafter{\romannumeral2}. The $\chi_{c1}\rho$ and
$J/\psi\pi\,\,\{^3P_0\}$ modes are governed by the spin
configuration $(1_H\otimes1_l)_0^{--}$ while the $h_c
a_0(980)\,\,\{^3P_0\}$ and the $\eta_c\rho\,\,\{^3P_0\}$ depend on
the spin configuration $(0_H\otimes1_l)_0^{--}$. As shown in
Appendix B, the $1^+(0^{--})$ states of type
\uppercase\expandafter{\romannumeral1} contain the spin
configuration $(1_H\otimes1_l)_0^{--}$ only. They can decay into
$\chi_{c1}\rho$ and $J/\psi\pi\,\,\{^3P_0\}$. But they can not decay
into $h_c a_0(980)\,\,\{^3P_0\}$ and $\eta_c\rho\,\,\{^3P_0\}$ in
the heavy quark limit.

The $0^-(0^{--})$ states of type
\uppercase\expandafter{\romannumeral2} decay into $\chi_{c1}\omega$,
$h_c\sigma\,\,\{^3P_0\}$, $\eta_c\omega\,\,\{^3P_0\}$ and
$J/\psi\eta\,\,\{^3P_0\}$. However, the modes
$h_c\sigma\,\,\{^3P_0\}$ and $\eta_c\omega\,\,\{^3P_0\}$ of the
$0^-(0^{--})$ states of type \uppercase\expandafter{\romannumeral1}
are sppressed.

%%%%%%%%%%%%%%%%%%%%%%%%%%%%%%%%%%%%%%%%%%%
\section{Decay patterns of $Y(4260)$}\label{sec5}
%%%%%%%%%%%%%%%%%%%%%%%%%%%%%%%%%%%%%%%%%%%

The charmoniumlike state $Y(4260)$ with $I^G(J^{PC})=0^-(1^{--})$
was first discovered by the BaBar Collaboration in the
$e^+e^-\rightarrow J/\psi\pi^+\pi^-$ process \cite{Aubert:2005rm}.
There have been extensive experimental and theoretical
investigations of this puzzling state
\cite{Maiani:2005pe,Zhu:2005hp,LlanesEstrada:2005hz,Eichten:2005ga,Segovia:2008zz,Li:2009zu,Ebert:2008wm,Ding:2008gr,Wang:2013cya,Li:2013yka,Close:2010wq,Liu:2005ay,Yuan:2005dr,Qiao:2005av,MartinezTorres:2009xb,Chen:2010nv,Wang:2013kra,Cleven:2013mka}.
Some authors proposed that $Y(4260)$ could be reproduced by the
interference of $e^+e^-\rightarrow \psi(4160)/\psi(4415)\rightarrow
J/\psi\pi^+\pi^-$ and the background contribution
\cite{Chen:2010nv}. $Y(4260)$ may be an exotic hybrid charmonium
state or a conventional charmonium \cite{Zhu:2005hp}. In this work,
we compare the decay patterns under various resonant assumptions.

From the conservation of the total angular momentum, parity, C
parity and G parity, the allowed decay modes of $Y(4260)$ are
$J/\psi\sigma$, $\psi(1^3D_1)\sigma$, $h_c\eta$,
$\chi_{cJ}\omega\,\,(J=0,1,2)$, $J/\psi\eta\,\,\{^3P_1\}$,
$\psi(1^3D_1)\eta\,\,\{^3P_1\}$ and $\psi(1^3D_2)\eta\,\,\{^5P_1\}$.
We collect the current experimental information on the decay modes
of $Y(4260)$ in Table \ref{experiment}.

\renewcommand{\arraystretch}{1.5}
\begin{table}[htbp]
\caption{Experimental information on the decay modes of
$Y(4260)$.}\label{experiment}
\begin{center}
   \begin{tabular}{ c |c } \toprule[1pt]
      %\multicolumn{2}{|c|}
      {$Y(4260)$ decay modes} & {Observation results}   \\
      \midrule[1pt]

      $J/\psi\pi^+\pi^-$;\,\, $J/\psi\pi^0\pi^0$;\,\, $J/\psi K^+K^-$ & {seen} \\
      $X(3872)\gamma$ & {seen} \\
      $J/\psi\eta$;\,\, $J/\psi\eta'$;\,\, $J/\psi\eta\eta$;\,\, $J/\psi\pi^0$;\,\, $J/\psi\pi^+\pi^-\pi^0$ & {not seen} \\
      $\psi(2S)\pi^+\pi^-$;\,\, $\psi(2S)\eta$ & {not seen} \\
      $\chi_{c0}\omega$;\,\, $\chi_{c1}\gamma$;\,\, $\chi_{c2}\gamma$;\,\, $\chi_{c1}\pi^+\pi^-\pi^0$;\,\, $\chi_{c2}\pi^+\pi^-\pi^0$ & {not seen} \\
      $h_c(1P)\pi^+\pi^-$;\,\, $\phi\pi^+\pi^-$ & {not seen} \\
      $D\bar{D}$;\,\, $D^{\ast}\bar{D}+c.c.$;\,\, $D^{\ast}\bar{D}^{\ast}$ & {not seen} \\
      $D^{\ast}\bar{D}^{\ast}\pi$;\,\, $D^0D^{\ast -}+c.c.$;\,\, $D\bar{D}^{\ast}+c.c.$ & {not seen} \\
      $D_s^+D_s^-$;\,\, $D_s^{\ast +}D_s^-+c.c$;\,\, $D_s^{\ast +}D_s^{\ast -}$  & {not seen} \\
      $J/\psi\eta\pi^0$  & {not seen} \\
      \midrule[1pt]
      \end{tabular}
  \end{center}
\end{table}

%%%%%%%%%%%%%%%%%%
\subsection{Decay patterns of $Y(4260)$ as a conventional
charmonium}
%%%%%%%%%%%%%%%%%%%

$Y(4260)$ could be a conventional charmonium the $\psi(4S)$ or the
mixture of 4S and 3D vector charmonia $Y(4260)$
\cite{Zhu:2005hp,LlanesEstrada:2005hz,Eichten:2005ga,Li:2009zu}.
Recall that the spin structures of $\psi(4S)$ and $\psi(3D)$ are
$(1_H\otimes0_l)_1^{--}$ and $(1_H\otimes2_l)_1^{--}$ respectively.
If $Y(4260)$ is the mixture of $\psi(4S)$ and $\psi(3D)$, it can
easily decay into $J/\psi\sigma$, $\psi(1^3D_1)\sigma$ and $J/\psi
f_0(980)$. However, the strong decay mode $h_c\eta$ and radiative
decay modes $\eta_c\gamma(M1)$ and $\eta_{c2}\gamma(M1)$ are all
dominated by the spin configuration $(0_H\otimes1_l)_1^{--}$ and are
suppressed in this case.

For comparison, the spin configuration $(0_H\otimes1_l)_1^{--}$
appears in the color-spin wave functions of the
$0^-(1^{--})\,\,\{^1P_1\}$ tetraquarks of type
\uppercase\expandafter{\romannumeral1}. All the $h_c\eta$ and
$\eta_c\gamma(M1)$ and $\eta_{c2}\gamma(M1)$ modes are allowed if
$Y(4260)$ is the $0^-(1^{--})\,\,\{^1P_1\}$ of type
\uppercase\expandafter{\romannumeral1}. In other words, the strong
decay mode $h_c\eta$ and radiative decay modes $\eta_c\gamma(M1)$
and $\eta_{c2}\gamma(M1)$ can provide a way to distinguish the
tetraquark assumptions of $Y(4260)$ from the conventional charmonia
assumptions.

%%%%%%%%%%%%%%%%%%
\subsection{Decay patterns of $Y(4260)$ as a Molecule}
%%%%%%%%%%%%%%%%%%%

We made a comprehensive investigation on the decay patterns
$Y(4260)$ as a molecular state with $I^G(J^{PC})=0^-(1^{--})$ in
Refs. \cite{Ma:2014ofa,Ma:2014zva}. With the molecular assumptions,
$Y(4260)$ can decay into $h_c\eta$, $\chi_{cJ}\omega\,\,(J=0,1,2)$,
$J/\psi\eta\,\,\{^3P_1\}$, $\psi(1^3D_1)\eta\,\,\{^3P_1\}$ and
$\psi(1^3D_2)\eta\,\,\{^5P_1\}$. However, the decay modes
$J/\psi\sigma$, $\psi(1^3D_1)\sigma$ and $J/\psi f_0(980)$ are
strongly suppressed under the heavy quark symmetry. Within the
molecular scheme, $Y(4260)$ may be the mixture of the pure
$D_1\bar{D}$ and $D_1^{\prime}\bar{D}$. Considering this mixing
effect, $J/\psi\sigma$, $\psi(1^3D_1)\sigma$ and $J/\psi f_0(980)$
are also suppressed.

$Y(4260)$ was first discovered in the $J/\psi\pi^+\pi^-$ final
states where the $\pi^+\pi^-$ pair is in the S-wave, which may arise
from the intermediate $\sigma$ and $f_0(980)$ resonances. The
molecular scheme is unable to explain the discovery mode
$J/\psi\pi^+\pi^-$ of $Y(4260)$.

As the isoscalar $1^{--}$ molecule, $Y(4260)$ is composed of
$D_0\bar{D}^{\ast}$, $D'_1\bar{D}$, $D_1\bar{D}$,
$D'_1\bar{D}^{\ast}$, $D_1\bar{D}^{\ast}$ or $D_2\bar{D}^{\ast}$.
All these molecules are dominated by the spin configurations
$(0_H\otimes1_l)_1^{--}$ and $(1_H\otimes1_l)_1^{--}$. The same spin
configurations also appear in the open charm final states such as
$D\bar D$, $D^*\bar D$, $D^* \bar D^*$. In other words, $Y(4260)$
can decay into the open-charm modes easily if it's a molecular
state.

In Table \ref{experiment}, we notice that the $J/\psi\eta$ and
$J/\psi\pi^+\pi^-\pi^0$ are "not seen" experimentally. These final
states are governed by the spin configuration
$(1_H\otimes1_l)_1^{--}$, which exists in the $1^{--}$ molecular
structures. Thus these decays should be allowed for the molecule
assumption.

Moreover, the strong decay mode $h_c\eta$ and radiative decay modes
$\eta_c\gamma(M1)$ and $\eta_{c2}\gamma(M1)$ are allowed for the
molecule scheme.

The discovery mode $J/\psi\pi^+\pi^-$ and non-observation of the
open-charm decay modes $D\bar D$, $D^*\bar D$, $D^* \bar D^*$ and
hidden-charm decay modes $J/\psi\eta$ and $J/\psi\pi^+\pi^-\pi^0$
disfavor the molecule assumption of $Y(4260)$.

%%%%%%%%%%%%%%%%%%
\subsection{Decay patterns of $Y(4260)$ as a tetraquark state}
%%%%%%%%%%%%%%%%%%%

There are two types of isoscalar $1^{--}$ tetraquarks. For "type
\uppercase\expandafter{\romannumeral1}", the P-wave excitation
exists between the diquark and anti-diquark pair. For "type
\uppercase\expandafter{\romannumeral2}", the P-wave excitation
exists inside either the diquark or anti-diquark.

The total spin of the four quarks can be $0$, $1$ and $2$. The
isoscalar $1^{--}$ tetraquarks of type
\uppercase\expandafter{\romannumeral1} can be further categorized
into three kinds of structures, $0^-(1^{--})\,\,\{^1P_1\}$,
$0^-(1^{--})\,\,\{^3P_1\}$ and $0^-(1^{--})\,\,\{^5P_1\}$.

According to Table \ref{tab:3}, all the isoscalar $1^{--}$
tetraquark states of type \uppercase\expandafter{\romannumeral1} can
easily decay into $J/\psi\sigma$, $\psi(1^3D_1)\sigma$ and $J/\psi
f_0(980)$. The $J/\psi\pi^+\pi^-$ mode are allowed for all the
isoscalar $1^{--}$ tetraquarks of type
\uppercase\expandafter{\romannumeral1}.

Nevertheless, the isoscalar $1^{--}$ tetraquarks of type
\uppercase\expandafter{\romannumeral2} do not decay into
$J/\psi\sigma$, $\psi(1^3D_1)\sigma$ and $J/\psi f_0(980)$ in the
heavy quark symmetry limit. If $Y(4260)$ turns out be a tetraquark
state, its P-wave excitation exists between the diquark and
anti-diquark pair.

From Table \ref{tab:3}, we notice that the decay mode $h_c\eta$ is
allowed for the $0^-(1^{--})\,\,\{^1P_1\}$ tetraquarks of type
\uppercase\expandafter{\romannumeral1} while $h_c\eta$ is strongly
suppressed for the $0^-(1^{--})\,\,\{^3P_1\}$ and
$0^-(1^{--})\,\,\{^5P_1\}$ tetraquarks of type
\uppercase\expandafter{\romannumeral1}. The $h_c\eta$ decay mode
provides a way to distinguish the $0^-(1^{--})\,\,\{^1P_1\}$
tetraquark state from $0^-(1^{--})\,\,\{^3P_1\}$ and
$0^-(1^{--})\,\,\{^5P_1\}$.

The decay mode $\chi_{c0}\omega$ is also allowed if $Y(4260)$ is a
$0^-(1^{--})$ tetraquark of type
\uppercase\expandafter{\romannumeral1}. Unfortunately, the decay
phase space is tiny. Hence the experimental measurement of the
$\chi_{c0}\omega$ mode will be difficult.

The P-wave open charm modes $D\bar D$, $D^*\bar D$, $D^* \bar D^*$
are dominated by the spin configurations $(0_H\otimes1_l)_1^{--}$,
$(1_H\otimes0_l)_1^{--}$, $(1_H\otimes1_l)_1^{--}$ and
$(1_H\otimes2_l)_1^{--}$. The S-wave open charm final states
$D_0\bar{D}^{\ast}$, $D'_1\bar{D}$, $D_1\bar{D}$,
$D'_1\bar{D}^{\ast}$, $D_1\bar{D}^{\ast}$ or $D_2\bar{D}^{\ast}$ are
governed by the spin configurations $(0_H\otimes1_l)_1^{--}$ and
$(1_H\otimes1_l)_1^{--}$. The open charm final states $D^*\bar
D\pi$, $D^* \bar D^*\pi$ also contain the $(0_H\otimes1_l)_1^{--}$,
$(1_H\otimes0_l)_1^{--}$, $(1_H\otimes1_l)_1^{--}$ and
$(1_H\otimes2_l)_1^{--}$ configurations. The isoscalar $1^{--}$
tetraquark states of type \uppercase\expandafter{\romannumeral1}
contain the spin configuration $(1_H\otimes1_l)_1^{--}$. Therefore,
all the above open charm decay modes are allowed under the heavy
quark symmetry. The non-observation of the open charm decay modes
remains a big challenge to the tetraquark assumption of $Y(4260)$.

%%%%%%%%%%%%%%%%%%
\subsection{Decay patterns of $Y(4260)$ as a hybrid Charmonium}
%%%%%%%%%%%%%%%%%%%

\subsubsection{Scenario A}

$Y(4260)$ was proposed as a good candidate of the hybrid charmonium
\cite{Zhu:2005hp}. With Scenario A, we assume that $Y(4260)$ is
composed of a pair of $c\bar{c}$ and a gluon. We use $l_g=0,1$ to
denote the E1 and M1 gluon in the hybrid charmonium respectively.
$l_{c\bar{c}}$ is the orbital momentum of the $c\bar{c}$ pair, and
$s_{c\bar{c}}$ is the spin of $c\bar{c}$. Thus, we have $(l_g,
l_{c\bar{c}}, s_{c\bar{c}})=(0, 1, 1)$ or $(l_g, l_{c\bar{c}},
s_{c\bar{c}})=(1, 0, 0)$. The $s_{c\bar{c}}=1$ and $l_{c\bar{c}}=1$
in the $(0, 1, 1)$ can couple into $0$, $1$, $2$. We use the
superscripts $0$, $1$, $2$ to denote the three states. The spin
structures of $Y(4260)$ read
\begin{eqnarray}
Y(4260)_{(0, 1, 1)}^0 &=& \frac{1}{3}(1_H\otimes0_l)_1^{--}-\frac{\sqrt{3}}{3}(1_H\otimes1_l)_1^{--}+\frac{\sqrt{5}}{3}(1_H\otimes2_l)_1^{--},\label{hy1}\\
Y(4260)_{(0, 1, 1)}^1 &=& -\frac{\sqrt{3}}{3}(1_H\otimes0_l)_1^{--}+\frac{1}{2}(1_H\otimes1_l)_1^{--}+\frac{\sqrt{15}}{6}(1_H\otimes2_l)_1^{--},\\
Y(4260)_{(0, 1, 1)}^2 &=& \frac{\sqrt{5}}{3}(1_H\otimes0_l)_1^{--}-\frac{\sqrt{15}}{6}(1_H\otimes1_l)_1^{--}+\frac{1}{6}(1_H\otimes2_l)_1^{--},\\
Y(4260)_{(1, 0, 0)} &=& (0_H\otimes1_l)_1^{--}.\label{hy2}
\end{eqnarray}

We notice that the states $Y(4260)_{(0, 1, 1)}^0$, $Y(4260)_{(0, 1,
1)}^1$ and $Y(4260)_{(0, 1, 1)}^2$ contain the spin configuration
$(1_H\otimes0_l)_1^{--}$, while the state $Y(4260)_{(1, 0, 0)}$
contain the spin configuration $(0_H\otimes1_l)_1^{--}$ only. The
discovery mode $J/\psi\pi^+\pi^-$ is governed by the spin
configuration $(1_H\otimes0_l)_1^{--}$. Thus, $Y(4260)_{(1, 0, 0)}$
is disfavored under heavy quark symmetry.

The P-wave decay modes $D\bar{D}$, $D^{\ast}\bar{D}+c.c.$,
$D^{\ast}\bar{D}^{\ast}$ $D_s^+D_s^-$, $D_s^{\ast +}D_s^-+c.c$,
$D_s^{\ast +}D_s^{\ast -}$ and $D^{\ast}\bar{D}^{\ast}\pi$ are
dominated by the spin configurations $(0_H\otimes1_l)_1^{--}$,
$(1_H\otimes0_l)_1^{--}$, $(1_H\otimes1_l)_1^{--}$ and
$(1_H\otimes2_l)_1^{--}$. The S-wave open charm modes which contain
a P-wave charmed meson such as $D_0\bar{D}^{\ast}$, $D'_1\bar{D}$,
$D_1\bar{D}$, $D'_1\bar{D}^{\ast}$, $D_1\bar{D}^{\ast}$ or
$D_2\bar{D}^{\ast}$ are governed by the spin configurations
$(0_H\otimes1_l)_1^{--}$ and $(1_H\otimes1_l)_1^{--}$. According to
the spin structures listed in Eqs. (\ref{hy1})-(\ref{hy2}),
$Y(4260)$ with the structure $(0, 1, 1)$ should decay into the above
open charm modes easily. Such a hybrid charmonium is composed of a
$c\bar c$ pair and one M1 color-magnetic gluon.

We notice that the hybrid charmonium and hidden-charm tetraquarks
have very similar strong decay patterns. The underlying dynamics is
quite transparent. The gluon within the hybrid charmonium can easily
fluctuate into a $q\bar q$ light quark pair to form the hidden-charm
tetraquark states.

\subsubsection{Scenario B}

When the hybrid charmonium decays, the gluon may pull one or more
soft gluons from the $c\bar c$ pair. Now the vector hybrid
charmonium is effectively composed of two gluons with
$J^{PC}=0^{++}$ and a $J^{PC}=1^{--}$ $c\bar{c}$ pair with
$l_{c\bar{c}}=0$, $s_{c\bar{c}}=1$ and $l_g=0$. Under this specific
assumption, $Y(4260)$ have the spin configuration
$(1_H\otimes0_l)_1^{--}$ only and decays into $J/\psi\pi^+\pi^-$
easily under heavy quark symmetry. $Y(4260)$ also decays into the
P-wave open charm modes such as $D\bar{D}$, $D^{\ast}\bar{D}+c.c.$,
$D^{\ast}\bar{D}^{\ast}$ $D_s^+D_s^-$, $D_s^{\ast +}D_s^-+c.c$,
$D_s^{\ast +}D_s^{\ast -}$. However, $Y(4260)$ does not decay into
the open charm modes $D_0\bar{D}^{\ast}$, $D'_1\bar{D}$,
$D_1\bar{D}$, $D'_1\bar{D}^{\ast}$, $D_1\bar{D}^{\ast}$ and
$D_2\bar{D}^{\ast}$ since they are dominated by the spin
configurations $(0_H\otimes1_l)_1^{--}$ and
$(1_H\otimes1_l)_1^{--}$.

For both Scenario A and B, $Y(4260)$ will decay into
$\chi_{c0}\omega$, $\chi_{c1}\gamma$, $\chi_{c2}\gamma$,
$\chi_{c1}\pi^+\pi^-\pi^0$ and $\chi_{c2}\pi^+\pi^-\pi^0$ under
heavy quark symmetry since these modes are governed by the spin
configurations $(1_H\otimes0_l)_1^{--}$, $(1_H\otimes1_l)_1^{--}$
and $(1_H\otimes2_l)_1^{--}$. The experimental measurement of the
radiative decay modes $\chi_{c1}\gamma$, $\chi_{c2}\gamma$ will be
desirable. For both scenarios, the open charm decay modes of
$Y(4260)$ are allowed in the heavy quark symmetry limit. The
non-observation of the open charm decay modes is also very puzzling
within the hybrid charmonium framework.

%%%%%%%%%%%%%%%%%%%%%%%%%%%%%%%%%%%%%%%%%%%
\section{Decay patterns of $X(3872)$, $Z_c(3900)$, $Z_c(4025)$, $Y(4360)$ and $Z_c(4200)$}\label{sec6}
%%%%%%%%%%%%%%%%%%%%%%%%%%%%%%%%%%%%%%%%%%%

%%%%%%%%%%%%%%%
\subsection{$X(3872)$}
%%%%%%%%%%%%%%%

The charmoniumlike state $X(3872)$ was first observed by the Belle
Collaboration in the $J/\psi\pi^+\pi^-$ invariant mass spectrum of
$B\rightarrow KJ/\psi\pi^+\pi^-$ \cite{Abe:2005ix}. There exist
extensive discussions of the underlying structure of $X(3872)$. Some
plausible interpretations include the $D\bar{D}^{\ast}$ molecular
state with $J^{PC}=1^{++}$, the conventional charmonium
$\chi^\prime_{c1}$ or a mixture of $c\bar{c}$ and $D\bar{D}^{ast}$.

The $1^{++}$ molecular state and $\chi^\prime_{c1}$ can easily decay
into the S-wave decay mode $\chi_{c1}\sigma$ through the tail of the
broad $\sigma$ resonance. The potentially allowed P-wave decay modes
$\chi_{c0}\eta$, $\chi_{c1}\eta$ and $\chi_{c2}\eta$ are forbidden
by kinematics.

As a $1^{++}$ tetraquark state, the $\chi_{c1}\sigma$ mode of
$X(3872)$ is suppressed by the transition between the color octet
and color singlet which needs the exchange of soft gluons. However,
such a suppression may not be very strong. The experimental
observation of the $\chi_{c1}(\pi\pi)_{S-wave}$ mode will further
test the tetraquark assumption of $X(3872)$.

%%%%%%%%%%%%%%%
\subsection{$Z_c(3900)$ and $Z_c(4025)$}
%%%%%%%%%%%%%%%

As the first charged charmonium-like state, $Z_c(3900)$ was observed
in the $J/\psi\pi^{\pm}$ invariant mass spectrum of
$e^+e^-\rightarrow J/\psi\pi^+\pi^-$ by BESIII collaboration
\cite{Ablikim:2013mio}, and later confirmed by Belle in the same
process. $Z_c(4020)$ and $Z_c(4025)$ were observed in the
$e^+e^-\rightarrow h_c\pi^+\pi^-$ and $e^+e^-\rightarrow
(D^\ast\bar{D}^\ast)^{\pm}\pi^{\mp}$ at
$\sqrt{s}=4.26\,\,\textmd{GeV}$ respectively
\cite{Choi:2007wga,Ablikim:2013mio,Liu:2013dau,Xiao:2013iha,Ablikim:2013mio,Liu:2013dau,Ablikim:2013wzq,Yi:2010aa,Liu:2010hf,Abe:2004zs,Aaltonen:2009tz}.
We treat these two signals as the same state. Their neutral partners
have quantum numbers $I^G(J^{PC})=1^+(1^{+-})$.

$Z_c(3900)$ and $Z_c(4020)$ are speculated to be either the
isovector $D^\ast\bar{D}$ and $D^\ast\bar{D}^\ast$ molecular states
with $I^G(J^p)=1^+(1^+)$ or the $1^+$ tetraquark states.

If $Z_c(3900)$ and $Z_c(4020)$ are molecular states, their decay
modes include $J/\psi\pi$, $\psi'\pi$, $\eta_c\rho$, $\eta_{c2}\rho$
and also the P-wave decay mode $h_c\pi$. However, if $Z_c(3900)$ and
$Z_c(4020)$ are the $1^+(1^+)$ tetraquarks, the decay mode $h_c\pi$
is suppressed in the heavy quark symmetry limit. If they are the
$1^+(1^+)$ tetraquarks with the $280_{cs}$ color-spin
representation, $h_c\pi$, $J/\psi\pi$, $\psi'\pi$, $\eta_c\rho$ and
$\eta_{c2}\rho$ are all suppressed, which implies that the states
$|280_{cs},1_c,1\rangle_{(21,\bar{15})}$ and
$|280_{cs},1_c,1\rangle_{(15,\bar{21})}$ with $I^G(J^p)=1^+(1^+)$
have rather narrow widths.

No matter whether $Z_c(3900)$ and $Z_c(4020)$ are molecules or
tetraquarks, their decay into $\psi(1^3D_1)\pi$ is strongly
suppressed due to the non-conservation of the light spin.

Since $Z_c(3900)$ was observed in the $J/\psi\pi^{\pm}$ invariant
mass spectrum of $e^+e^-\rightarrow J/\psi\pi^+\pi^-$, its
possibility as a $1^+(1^+)$ tetraquark with the $280_{cs}$
color-spin representation is relatively small while the possibility
as a $1^+(1^+)$ tetraquark with the $35_{cs}$ color-spin
representation is not excluded.

Whether $Z_c(3900)$ is a good candidate of the $1^+(1^+)$ molecular
state or a tetraquark state with the $35_{cs}$ color-spin
representation can be judged by the P-wave decay mode $h_c\pi$.
Since $Z_c(4020)$ was observed in the $h_c\pi^{\pm}$ invariant mass
spectrum, the tetraquark assumption of $Z_c(4020)$ is unfavored
under the heavy quark symmetry. It is probably a
$D^\ast\bar{D}^\ast$ molecular state with $I^G(J^p)=1^+(1^+)$.

%%%%%%%%%%%%%%%
\subsection{$Y(4360)$}
%%%%%%%%%%%%%%%

The Belle Collaboration reported a charmonium-like state $Y(4360)$
in the $\psi(2S)\pi^+\pi^-$ invariant mass spectrum of the
$e^+e^-\rightarrow \psi(2S)\pi^+\pi^-$ process \cite{Wang:2007ea}.
Some authors proposed $Y(4360)$ as an isoscalar $D_1\bar{D}^{\ast}$
molecular state \cite{Chen:2013wva}. Our pervious works suggested
that the decay modes $J/\psi\sigma$,  $J/\psi f_0(980)$,
$\psi(2S)\sigma$, $\psi(1^3D_1)\sigma$ and $\psi(2S) f_0(980)$ of
the $0^-(1^{--})$ $D_1\bar{D}^{\ast}$ and
$D_1^{\prime}\bar{D}^{\ast}$ molecular states are suppressed
\cite{Ma:2014ofa,Ma:2014zva}.

$Y(4360)$ may be the mixture of the pure $D_1\bar{D}^{\ast}$ and
$D_1^{\prime}\bar{D}^{\ast}$ molecular states. Considering this
mixing effect, the decay modes $J/\psi\sigma$, $J/\psi f_0(980)$,
$\psi(2S)\sigma$, $\psi(1^3D_1)\sigma$ and $\psi(2S) f_0(980)$ are
also suppressed in heavy quark symmetry limit. Thus, the molecule
assumption of $Y(4360)$ is unfavored.

The decay modes $J/\psi\sigma$, $\psi(1^3D_1)\sigma$,
$\psi(2S)\sigma$, $J/\psi f_0(980)$ and $\psi(2S) f_0(980)$ of the
$1^{--}$ isoscalar tetraquarks of type
\uppercase\expandafter{\romannumeral2} are strongly suppressed.
Considering its discovery mode $\psi(2S)\pi^+\pi^-$, $Y(4360)$ may
be the $1^{--}$ isoscalar tetraquarks of type
\uppercase\expandafter{\romannumeral1}.

The decay mode $h_c\eta$ is allowed for the
$0^-(1^{--})\,\,\{^1P_1\}$ of type
\uppercase\expandafter{\romannumeral1}, while it is not allowed for
the $0^-(1^{--})\,\,\{^3P_1\}$ and $0^-(1^{--})\,\,\{^5P_1\}$ of
type \uppercase\expandafter{\romannumeral1}. This decay mode can be
employed to judge whether $Y(4360)$ is the tetraquark state of
$0^-(1^{--})\,\,\{^1P_1\}$, $0^-(1^{--})\,\,\{^3P_1\}$ or
$0^-(1^{--})\,\,\{^5P_1\}$.

%%%%%%%%%%%%%%%%
\subsection{$Z_c(4200)$}
%%%%%%%%%%%%%%%%

Recently, the charged charmonium-like state $Z_c(4200)^+$ was
reported by Belle Collaboration in the $Z_c(4200)^+\rightarrow
J/\psi\pi^+$ process. Its $J^P$ quantum numbers are probably $1^+$.
The neutral part of $Z_c(4200)$ have quantum numbers
$I^G(J^{PC})=1^+(1^{+-})$. Its mass and decay width are
$M=4196_{-29-13}^{+31+17}\,\,\mathrm{MeV}$ and
$\Gamma=370_{-70-132}^{+70+70}\,\,\mathrm{MeV}$. The decay width of
$Z_c(4200)$ is too large to be interpreted as a molecule. Moreover,
its mass is not close to the open charm threshold.

It was proposed as a S-wave tetraquark \cite{Chen:2015fsa}. From
Table \ref{tab:2}, the decay modes $J/\psi\pi$ and $\eta_c\rho$ are
suppressed for the $1^+(1^{+-})$ tetraquarks with the $280_{cs}$
color-spin representation. Therefore, $Z_c(4200)^+$ is a good
candidate of the $1^+(1^{+-})$ tetraquarks with the $35_{cs}$
color-spin representation. If so, its P-wave decay mode $h_c\pi$ is
strongly suppressed in the heavy quark limit, while the P-wave decay
modes $\chi_{c0}\rho$ and $\chi_{c1}\rho$ are allowed.

%%%%%%%%%%%%%%%%%%%%%%%%%%%%%%%%%%%%%%%%%%%
\section{Summary}\label{sec7}
%%%%%%%%%%%%%%%%%%%%%%%%%%%%%%%%%%%%%%%%%%%

More and more charmonium-like XYZ states have been reported in the
recent years. Some of these exotic state are hard to be accommodated
into the charmonium spectrum in the conventional quark model. Some
of them are even charged. Various theoretical interpretations were
proposed such as the $c\bar c$ states distorted by couple-channel
effects, hybrid charmonium, di-meson molecules, tetraquarks, and
non-resonant kinematics artifacts caused by threshold effects. The
tetraquark scheme is attractive for those XYZ states which do not
sit on the di-meson threshold and are very broad with a width around
several hundred MeV.

The charm quark mass is much larger than $\Lambda_{QCD}$ and up and
down quark mass. The heavy quark symmetry is a very powerful tool to
study the charmonium-like states. In heavy quark symmetry limit, the
heavy quark spin is conserved during the decay process. The
color-spin rearrangement scheme is based on the conservations of
heavy spin, light spin, parity, C parity and G parity, which
provides us an effective framework to probe the inner structures of
these exotic states from their decay behaviors.

In this work we have performed an extensive investigations of the
decay patterns of the S-wave tetraquark and P-wave tetraquark states
with the color-spin rearrangement scheme. We considered two kinds of
P-wave tetraquarks where the P-wave excitation exists between the
diquark and anti-diquark pair or inside one of the diquark. We
notice the second type of P-wave tetraquark do not decay into
$J/\psi\pi^+\pi^-$ while the first type decays into this mode
easily. The $J/\psi\pi^+\pi^-$ mode is also strongly suppressed if
$Y(4260)$ is a molecular state. The decay patterns of the hybrid
charmonium and hidden-charm tetraquark states very similar. Both
these structures allow the open-charm decay modes, which have not
been observed experimentally. This feature is very puzzling. We also
discussed the decay patterns of $X(3872)$, $Y(4360)$, and several
charged states such as $Z_c(3900)$. The observation of in the
$h_c\pi$ mode disfavors the tetraquark assumption of $Z_c(4020)$.

The strong decay behaviors of the charmonium-like $XYZ$ states
encode important information on their inner structures. The
experiment measurement of their decay patterns will shed light on
their underlying dynamics. Hopefully the present study on the
tetraquark schemes will help us illuminate the puzzling situation of
the XYZ states.

\subsection*{Acknowledgments}

This project is supported by the National Natural Science Foundation
of China under Grant 11575008.

%%%%%%%%%%%%%%%
\section*{Appendix A: Color-spin wave functions of the S-wave tetraquarks}
%%%%%%%%%%%%%%%

%%%%%%%%%%%%%%%%%%
\fbox{$0^{++}$}
%%%%%%%%%%%%%%%%%%%

\begin{eqnarray}
|1_{cs},1_c,0\rangle_{(21,\bar{21})} &=& \frac{\sqrt{21}}{6}(1,1,0)_H^{-+}\otimes(1,1,0)_l^{-+}-\frac{\sqrt{7}}{14}(35,1,1)_H^{--}\otimes(35,1,1)_l^{--} \nonumber\\
&+&\frac{\sqrt{42}}{21}(35,8,0)_H^{-+}\otimes(35,8,0)_l^{-+}-\frac{\sqrt{14}}{7}(35,8,1)_H^{--}\otimes(35,8,1)_l^{--}, \label{eq27}\\
|405_{cs},1_c,0\rangle_{(21,\bar{21})} &=& -\frac{2\sqrt{42}}{21}(35,1,1)_H^{--}\otimes(35,1,1)_l^{--}+\frac{3\sqrt{7}}{14}(35,8,0)_H^{-+}\otimes(35,8,0)_l^{-+} \nonumber\\
&+&\frac{5\sqrt{21}}{42}(35,8,1)_H^{--}\otimes(35,8,1)_l^{--},\\
|1_{cs},1_c,0\rangle_{(15,\bar{15})} &=& \frac{\sqrt{15}}{6}(1,1,0)_H^{-+}\otimes(1,1,0)_l^{-+}-\frac{\sqrt{5}}{10}(35,1,1)_H^{--}\otimes(35,1,1)_l^{--} \nonumber\\
&-&\frac{\sqrt{30}}{15}(35,8,0)_H^{-+}\otimes(35,8,0)_l^{-+}-\frac{\sqrt{10}}{5}(35,8,1)_H^{--}\otimes(35,8,1)_l^{--},\\
|189_{cs},1_c,0\rangle_{(15,\bar{15})} &=& -\frac{2\sqrt{30}}{15}(35,1,1)_H^{--}\otimes(35,1,1)_l^{--}-\frac{3\sqrt{5}}{10}(35,8,0)_H^{-+}\otimes(35,8,0)_l^{-+} \nonumber\\
&-&\frac{\sqrt{15}}{30}(35,8,1)_H^{--}\otimes(35,8,1)_l^{--},
\end{eqnarray}

%%%%%%%%%%%%%%%
\fbox{$1^{++}$}
%%%%%%%%%%%%%%%%

\begin{eqnarray}
|35_{cs},1_c,1\rangle_{(21,\bar{21})} &=& -\frac{2\sqrt{2}}{3}(35,8,1)_H^{--}\otimes(35,8,1)_l^{--}-\frac{1}{3}(35,1,1)_H^{--}\otimes(35,1,1)_l^{--} \\
|280_{cs},1_c,1\rangle_{(21,\bar{15})} &=& -\frac{1}{3}(35,8,1)_H^{--}\otimes(35,8,1)_l^{--}+\frac{2\sqrt{2}}{3}(35,1,1)_H^{--}\otimes(35,1,1)_l^{--}\\
|35_{cs},1_c,1\rangle_{(15,\bar{15})} &=& -\frac{2\sqrt{2}}{3}(35,8,1)_H^{--}\otimes(35,8,1)_l^{--}-\frac{1}{3}(35,1,1)_H^{--}\otimes(35,1,1)_l^{--} \\
|280_{cs},1_c,1\rangle_{(15,\bar{21})} &=&
-\frac{1}{3}(35,8,1)_H^{--}\otimes(35,8,1)_l^{--}+\frac{2\sqrt{2}}{3}(35,1,1)_H^{--}\otimes(35,1,1)_l^{--},
\end{eqnarray}

%%%%%%%%%%%%%%%%%
\fbox{$1^{+-}$}
%%%%%%%%%%%%%%%%%

\begin{eqnarray}
|35_{cs},1_c,1\rangle_{(21,\bar{21})} &=& \frac{\sqrt{2}}{2}(35,1,1)_H^{--}\otimes(1,1,0)_l^{-+}-\frac{\sqrt{2}}{2}(1,1,0)_H^{-+}\otimes(35,1,1)_l^{--} \label{eq35}\\
|35_{cs},1_c,1\rangle_{(15,\bar{15})} &=& \frac{\sqrt{2}}{2}(35,1,1)_H^{--}\otimes(1,1,0)_l^{-+}-\frac{\sqrt{2}}{2}(1,1,0)_H^{-+}\otimes(35,1,1)_l^{--} \label{eq36}\\
|35_{cs},1_c,1\rangle_{(21,\bar{15})} &=& \frac{\sqrt{3}}{3}(35,1,1)_H^{--}\otimes(1,1,0)_l^{-+}+\frac{\sqrt{6}}{6}(35,8,1)_H^{--}\otimes(35,8,0)_l^{-+} \nonumber\\
&&+\frac{\sqrt{3}}{3}(1,1,0)_H^{-+}\otimes(35,1,1)_l^{--}+\frac{\sqrt{6}}{6}(35,8,0)_H^{-+}\otimes(35,8,1)_l^{--}\\
|35_{cs},1_c,1\rangle_{(15,\bar{21})} &=& \frac{\sqrt{6}}{6}(35,1,1)_H^{--}\otimes(1,1,0)_l^{-+}-\frac{\sqrt{3}}{3}(35,8,1)_H^{--}\otimes(35,8,0)_l^{-+} \nonumber\\
&&+\frac{\sqrt{6}}{6}(1,1,0)_H^{-+}\otimes(35,1,1)_l^{--}-\frac{\sqrt{3}}{3}(35,8,0)_H^{-+}\otimes(35,8,1)_l^{--}\\
|280_{cs},1_c,1\rangle_{(21,\bar{15})} &=& -\frac{\sqrt{2}}{2}(35,8,1)_H^{--}\otimes(35,8,0)_l^{-+}+\frac{\sqrt{2}}{2}(35,8,0)_H^{-+}\otimes(35,8,1)_l^{--} \\
|280_{cs},1_c,1\rangle_{(15,\bar{21})} &=&
-\frac{\sqrt{2}}{2}(35,8,1)_H^{--}\otimes(35,8,0)_l^{-+}+\frac{\sqrt{2}}{2}(35,8,0)_H^{-+}\otimes(35,8,1)_l^{--},
\end{eqnarray}

%%%%%%%%%%%%%%%
\section*{Appendix B: Color-spin wave functions of the P-wave tetraquarks}
%%%%%%%%%%%%%%%

%%%%%%%%%%%
\fbox{$1^{--}(^3P_1)$ of type II}
%%%%%%%%%%%
%If the P-wave excitation is inside the diquark or anti-diquark, then the tetraquark have the same C-parity with one without P-wave excitation.

\begin{eqnarray}
|35_{cs},1_c,1\rangle_{(21,\bar{21})} &=& \frac{\sqrt{2}}{2}(35,1,1)_H^{--}\otimes(1,1,1)_l^{++}-\frac{\sqrt{2}}{2}(1,1,0)_H^{-+}\otimes(35,1,1)_l^{+-} \\
|35_{cs},1_c,1\rangle_{(15,\bar{15})} &=& \frac{\sqrt{2}}{2}(35,1,1)_H^{--}\otimes(1,1,1)_l^{++}-\frac{\sqrt{2}}{2}(1,1,0)_H^{-+}\otimes(35,1,1)_l^{+-} \\
|35_{cs},1_c,1\rangle_{(21,\bar{15})} &=& \frac{\sqrt{3}}{3}(35,1,1)_H^{--}\otimes(1,1,1)_l^{++}+\frac{\sqrt{6}}{6}(35,8,1)_H^{--}\otimes(35,8,1)_l^{++} \nonumber\\
&&+\frac{\sqrt{3}}{3}(1,1,0)_H^{-+}\otimes(35,1,1)_l^{+-}+\frac{\sqrt{6}}{6}(35,8,0)_H^{-+}\otimes(35,8,1)_l^{+-}\\
|35_{cs},1_c,1\rangle_{(15,\bar{21})} &=& \frac{\sqrt{6}}{6}(35,1,1)_H^{--}\otimes(1,1,1)_l^{++}-\frac{\sqrt{3}}{3}(35,8,1)_H^{--}\otimes(35,8,1)_l^{++} \nonumber\\
&&+\frac{\sqrt{6}}{6}(1,1,0)_H^{-+}\otimes(35,1,1)_l^{+-}-\frac{\sqrt{3}}{3}(35,8,0)_H^{-+}\otimes(35,8,1)_l^{+-}\\
|280_{cs},1_c,1\rangle_{(21,\bar{15})} &=& -\frac{\sqrt{2}}{2}(35,8,1)_H^{--}\otimes(35,8,1)_l^{++}+\frac{\sqrt{2}}{2}(35,8,0)_H^{-+}\otimes(35,8,1)_l^{+-} \\
|280_{cs},1_c,1\rangle_{(15,\bar{21})} &=&
-\frac{\sqrt{2}}{2}(35,8,1)_H^{--}\otimes(35,8,1)_l^{++}+\frac{\sqrt{2}}{2}(35,8,0)_H^{-+}\otimes(35,8,1)_l^{+-},
\end{eqnarray}

%%%%%%%%%%%%%%%
\fbox{$1^{--}(^1P_1)$ of type I}
%%%%%%%%%%%%%%%%

%If the P-wave excitation is between the diquark and anti-diquark, then the tetraquark have the different C-parity with one without P-wave excitation.

\begin{eqnarray}
|1_{cs},1_c,0\rangle_{(21,\bar{21})} &=& \frac{\sqrt{21}}{6}(1,1,0)_H^{-+}\otimes(1,1,1)_l^{+-}-\frac{\sqrt{7}}{42}(35,1,1)_H^{--}\otimes(35,1,0)_l^{++} \nonumber\\
&&+\frac{\sqrt{21}}{42}(35,1,1)_H^{--}\otimes(35,1,1)_l^{++}-\frac{\sqrt{35}}{42}(35,1,1)_H^{--}\otimes(35,1,2)_l^{++} \nonumber\\
&&+\frac{\sqrt{42}}{21}(35,8,0)_H^{-+}\otimes(35,8,1)_l^{+-}-\frac{\sqrt{7}}{42}(35,8,1)_H^{--}\otimes(35,8,0)_l^{++}\nonumber\\
&&+\frac{\sqrt{21}}{42}(35,8,1)_H^{--}\otimes(35,8,1)_l^{++}-\frac{\sqrt{35}}{42}(35,8,1)_H^{--}\otimes(35,8,2)_l^{++}\\
|405_{cs},1_c,0\rangle_{(21,\bar{21})} &=& -\frac{2\sqrt{42}}{63}(35,1,1)_H^{--}\otimes(35,1,0)_l^{++}+\frac{2\sqrt{14}}{21}(35,1,1)_H^{--}\otimes(35,1,1)_l^{++}\nonumber\\
&&-\frac{2\sqrt{210}}{63}(35,1,1)_H^{--}\otimes(35,1,2)_l^{++}+\frac{3\sqrt{7}}{14}(35,8,0)_H^{-+}\otimes(35,8,1)_l^{+-}\nonumber\\
&&-\frac{5\sqrt{21}}{126}(35,8,1)_H^{--}\otimes(35,8,0)_l^{++}-\frac{5\sqrt{7}}{42}(35,8,1)_H^{--}\otimes(35,8,1)_l^{++}\nonumber\\
&&+\frac{5\sqrt{105}}{126}(35,8,1)_H^{--}\otimes(35,8,2)_l^{++}\\
|1_{cs},1_c,0\rangle_{(15,\bar{15})} &=& \frac{\sqrt{15}}{6}(1,1,0)_H^{-+}\otimes(1,1,1)_l^{+-}-\frac{\sqrt{5}}{30}(35,1,1)_H^{--}\otimes(35,1,0)_l^{++} \nonumber\\
&&-\frac{\sqrt{15}}{30}(35,1,1)_H^{--}\otimes(35,1,1)_l^{++}+\frac{1}{6}(35,1,1)_H^{--}\otimes(35,1,2)_l^{++} \nonumber\\
&&-\frac{\sqrt{30}}{15}(35,8,0)_H^{-+}\otimes(35,8,1)_l^{+-}+\frac{\sqrt{10}}{15}(35,8,1)_H^{--}\otimes(35,8,0)_l^{++}\nonumber\\
&&-\frac{\sqrt{30}}{15}(35,8,1)_H^{--}\otimes(35,8,1)_l^{++}+\frac{\sqrt{2}}{3}(35,8,1)_H^{--}\otimes(35,8,2)_l^{++}\\
|189_{cs},1_c,0\rangle_{(15,\bar{15})} &=& -\frac{2\sqrt{30}}{45}(35,1,1)_H^{--}\otimes(35,1,0)_l^{++}+\frac{2\sqrt{10}}{15}(35,1,1)_H^{--}\otimes(35,1,1)_l^{++}\nonumber\\
&&-\frac{2\sqrt{6}}{9}(35,1,1)_H^{--}\otimes(35,1,2)_l^{++}-\frac{3\sqrt{5}}{10}(35,8,0)_H^{-+}\otimes(35,8,1)_l^{+-}\nonumber\\
&&-\frac{\sqrt{15}}{90}(35,8,1)_H^{--}\otimes(35,8,0)_l^{++}+\frac{\sqrt{5}}{30}(35,8,1)_H^{--}\otimes(35,8,1)_l^{++}\nonumber\\
&&-\frac{\sqrt{3}}{18}(35,8,1)_H^{--}\otimes(35,8,2)_l^{++},
\end{eqnarray}

%%%%%%%%%%%%%%%
\fbox{$1^{--}(^3P_1)$ of type I}
%%%%%%%%%%%%%%%%

\begin{eqnarray}
|35_{cs},1_c,1\rangle_{(21,\bar{21})} &=& \frac{2\sqrt{6}}{9}(35,8,1)_H^{--}\otimes(35,8,0)_l^{++}-\frac{\sqrt{2}}{3}(35,8,1)_H^{--}\otimes(35,8,1)_l^{++}\nonumber\\
&&-\frac{\sqrt{30}}{9}(35,8,1)_H^{--}\otimes(35,8,2)_l^{++}\nonumber\\
&&+\frac{\sqrt{3}}{9}(35,1,1)_H^{--}\otimes(35,1,0)_l^{++}-\frac{1}{6}(35,1,1)_H^{--}\otimes(35,1,1)_l^{++} \nonumber\\
&&-\frac{\sqrt{15}}{18}(35,1,1)_H^{--}\otimes(35,1,2)_l^{++}\\
|35_{cs},1_c,1\rangle_{(15,\bar{15})} &=& \frac{2\sqrt{6}}{9}(35,8,1)_H^{--}\otimes(35,8,0)_l^{++}-\frac{\sqrt{2}}{3}(35,8,1)_H^{--}\otimes(35,8,1)_l^{++}\nonumber\\
&&-\frac{\sqrt{30}}{9}(35,8,1)_H^{--}\otimes(35,8,2)_l^{++}\nonumber\\
&&+\frac{\sqrt{3}}{9}(35,1,1)_H^{--}\otimes(35,1,0)_l^{++}-\frac{1}{6}(35,1,1)_H^{--}\otimes(35,1,1)_l^{++} \nonumber\\
&&-\frac{\sqrt{15}}{18}(35,1,1)_H^{--}\otimes(35,1,2)_l^{++}\\
|280_{cs},1_c,1\rangle_{(21,\bar{15})} &=& \frac{\sqrt{3}}{9}(35,8,1)_H^{--}\otimes(35,8,0)_l^{++}-\frac{1}{6}(35,8,1)_H^{--}\otimes(35,8,1)_l^{++}\nonumber\\
&&-\frac{\sqrt{15}}{18}(35,8,1)_H^{--}\otimes(35,8,2)_l^{++}\nonumber\\
&&-\frac{2\sqrt{6}}{9}(35,1,1)_H^{--}\otimes(35,1,0)_l^{++}+\frac{\sqrt{2}}{3}(35,1,1)_H^{--}\otimes(35,1,1)_l^{++} \nonumber\\
&&+\frac{\sqrt{30}}{9}(35,1,1)_H^{--}\otimes(35,1,2)_l^{++}\\
|280_{cs},1_c,1\rangle_{(15,\bar{21})} &=& \frac{\sqrt{3}}{9}(35,8,1)_H^{--}\otimes(35,8,0)_l^{++}-\frac{1}{6}(35,8,1)_H^{--}\otimes(35,8,1)_l^{++}\nonumber\\
&&-\frac{\sqrt{15}}{18}(35,8,1)_H^{--}\otimes(35,8,2)_l^{++}\nonumber\\
&&-\frac{2\sqrt{6}}{9}(35,1,1)_H^{--}\otimes(35,1,0)_l^{++}+\frac{\sqrt{2}}{3}(35,1,1)_H^{--}\otimes(35,1,1)_l^{++} \nonumber\\
&&+\frac{\sqrt{30}}{9}(35,1,1)_H^{--}\otimes(35,1,2)_l^{++},
\end{eqnarray}

%%%%%%%%%%%%%%%
\fbox{$1^{--}(^5P_1)$ of type I}
%%%%%%%%%%%%%%%%

\begin{eqnarray}
|405_{cs},1_c,1\rangle_{(21,\bar{21})} &=& \frac{\sqrt{15}}{9}(35,8,1)_H^{--}\otimes(35,8,0)_l^{++}+\frac{\sqrt{5}}{6}(35,8,1)_H^{--}\otimes(35,8,1)_l^{++}\nonumber\\
&&+\frac{\sqrt{3}}{18}(35,8,1)_H^{--}\otimes(35,8,2)_l^{++}\nonumber\\
&&+\frac{\sqrt{30}}{9}(35,1,1)_H^{--}\otimes(35,1,0)_l^{++}-\frac{\sqrt{10}}{6}(35,1,1)_H^{--}\otimes(35,1,1)_l^{++} \nonumber\\
&&+\frac{\sqrt{6}}{18}(35,1,1)_H^{--}\otimes(35,1,2)_l^{++}\\
|189_{cs},1_c,1\rangle_{(15,\bar{15})} &=& \frac{\sqrt{30}}{9}(35,8,1)_H^{--}\otimes(35,8,0)_l^{++}+\frac{\sqrt{10}}{6}(35,8,1)_H^{--}\otimes(35,8,1)_l^{++}\nonumber\\
&&+\frac{\sqrt{6}}{18}(35,8,1)_H^{--}\otimes(35,8,2)_l^{++}\nonumber\\
&&-\frac{\sqrt{15}}{9}(35,1,1)_H^{--}\otimes(35,1,0)_l^{++}-\frac{\sqrt{5}}{6}(35,1,1)_H^{--}\otimes(35,1,1)_l^{++} \nonumber\\
&&-\frac{\sqrt{3}}{18}(35,1,1)_H^{--}\otimes(35,1,2)_l^{++},
\end{eqnarray}

%%%%%%%%%%%%%%%
\fbox{$1^{-+}$ of type I}
%%%%%%%%%%%%%%%%

\begin{eqnarray}
|35_{cs},1_c,1\rangle_{(21,\bar{21})} &=& \frac{\sqrt{2}}{2}(35,1,1)_H^{--}\otimes(1,1,1)_l^{+-}-\frac{\sqrt{2}}{2}(1,1,0)_H^{-+}\otimes(35,1,1)_l^{++} \\
|35_{cs},1_c,1\rangle_{(15,\bar{15})} &=& \frac{\sqrt{2}}{2}(35,1,1)_H^{--}\otimes(1,1,1)_l^{+-}-\frac{\sqrt{2}}{2}(1,1,0)_H^{-+}\otimes(35,1,1)_l^{++} \\
|35_{cs},1_c,1\rangle_{(21,\bar{15})} &=& \frac{\sqrt{3}}{3}(35,1,1)_H^{--}\otimes(1,1,1)_l^{+-}+\frac{\sqrt{6}}{6}(35,8,1)_H^{--}\otimes(35,8,1)_l^{+-} \nonumber\\
&&+\frac{\sqrt{3}}{3}(1,1,0)_H^{-+}\otimes(35,1,1)_l^{++}+\frac{\sqrt{6}}{6}(35,8,0)_H^{-+}\otimes(35,8,1)_l^{++}\\
|35_{cs},1_c,1\rangle_{(15,\bar{21})} &=& \frac{\sqrt{6}}{6}(35,1,1)_H^{--}\otimes(1,1,1)_l^{+-}-\frac{\sqrt{3}}{3}(35,8,1)_H^{--}\otimes(35,8,1)_l^{+-} \nonumber\\
&&+\frac{\sqrt{6}}{6}(1,1,0)_H^{-+}\otimes(35,1,1)_l^{++}-\frac{\sqrt{3}}{3}(35,8,0)_H^{-+}\otimes(35,8,1)_l^{++}\\
|280_{cs},1_c,1\rangle_{(21,\bar{15})} &=& -\frac{\sqrt{2}}{2}(35,8,1)_H^{--}\otimes(35,8,1)_l^{+-}+\frac{\sqrt{2}}{2}(35,8,0)_H^{-+}\otimes(35,8,1)_l^{++} \\
|280_{cs},1_c,1\rangle_{(15,\bar{21})} &=&
-\frac{\sqrt{2}}{2}(35,8,1)_H^{--}\otimes(35,8,1)_l^{+-}+\frac{\sqrt{2}}{2}(35,8,0)_H^{-+}\otimes(35,8,1)_l^{++},
\end{eqnarray}

%%%%%%%%%%%%%%%
\fbox{$1^{-+}(^1P_1)$ of type II}
%%%%%%%%%%%%%%%%

%If the P-wave excitation is inside the diquark or anti-diquark. $1^{-+}(^1P_1)$

\begin{eqnarray}
|1_{cs},1_c,0\rangle_{(21,\bar{21})} &=& \frac{\sqrt{21}}{6}(1,1,0)_H^{-+}\otimes(1,1,1)_l^{++}-\frac{\sqrt{7}}{42}(35,1,1)_H^{--}\otimes(35,1,0)_l^{+-} \nonumber\\
&&+\frac{\sqrt{21}}{42}(35,1,1)_H^{--}\otimes(35,1,1)_l^{+-}-\frac{\sqrt{35}}{42}(35,1,1)_H^{--}\otimes(35,1,2)_l^{+-} \nonumber\\
&&+\frac{\sqrt{42}}{21}(35,8,0)_H^{-+}\otimes(35,8,1)_l^{++}-\frac{\sqrt{7}}{42}(35,8,1)_H^{--}\otimes(35,8,0)_l^{+-}\nonumber\\
&&+\frac{\sqrt{21}}{42}(35,8,1)_H^{--}\otimes(35,8,1)_l^{+-}-\frac{\sqrt{35}}{42}(35,8,1)_H^{--}\otimes(35,8,2)_l^{+-}\\
|405_{cs},1_c,0\rangle_{(21,\bar{21})} &=& -\frac{2\sqrt{42}}{63}(35,1,1)_H^{--}\otimes(35,1,0)_l^{+-}+\frac{2\sqrt{14}}{21}(35,1,1)_H^{--}\otimes(35,1,1)_l^{+-}\nonumber\\
&&-\frac{2\sqrt{210}}{63}(35,1,1)_H^{--}\otimes(35,1,2)_l^{+-}+\frac{3\sqrt{7}}{14}(35,8,0)_H^{-+}\otimes(35,8,1)_l^{++}\nonumber\\
&&-\frac{5\sqrt{21}}{126}(35,8,1)_H^{--}\otimes(35,8,0)_l^{+-}-\frac{5\sqrt{7}}{42}(35,8,1)_H^{--}\otimes(35,8,1)_l^{+-}\nonumber\\
&&+\frac{5\sqrt{105}}{126}(35,8,1)_H^{--}\otimes(35,8,2)_l^{+-}\\
|1_{cs},1_c,0\rangle_{(15,\bar{15})} &=& \frac{\sqrt{15}}{6}(1,1,0)_H^{-+}\otimes(1,1,1)_l^{++}-\frac{\sqrt{5}}{30}(35,1,1)_H^{--}\otimes(35,1,0)_l^{+-} \nonumber\\
&&-\frac{\sqrt{15}}{30}(35,1,1)_H^{--}\otimes(35,1,1)_l^{+-}+\frac{1}{6}(35,1,1)_H^{--}\otimes(35,1,2)_l^{+-} \nonumber\\
&&-\frac{\sqrt{30}}{15}(35,8,0)_H^{-+}\otimes(35,8,1)_l^{++}+\frac{\sqrt{10}}{15}(35,8,1)_H^{--}\otimes(35,8,0)_l^{+-}\nonumber\\
&&-\frac{\sqrt{30}}{15}(35,8,1)_H^{--}\otimes(35,8,1)_l^{+-}+\frac{\sqrt{2}}{3}(35,8,1)_H^{--}\otimes(35,8,2)_l^{+-}\\
|189_{cs},1_c,0\rangle_{(15,\bar{15})} &=& -\frac{2\sqrt{30}}{45}(35,1,1)_H^{--}\otimes(35,1,0)_l^{+-}+\frac{2\sqrt{10}}{15}(35,1,1)_H^{--}\otimes(35,1,1)_l^{+-}\nonumber\\
&&-\frac{2\sqrt{6}}{9}(35,1,1)_H^{--}\otimes(35,1,2)_l^{+-}-\frac{3\sqrt{5}}{10}(35,8,0)_H^{-+}\otimes(35,8,1)_l^{++}\nonumber\\
&&-\frac{\sqrt{15}}{90}(35,8,1)_H^{--}\otimes(35,8,0)_l^{+-}+\frac{\sqrt{5}}{30}(35,8,1)_H^{--}\otimes(35,8,1)_l^{+-}\nonumber\\
&&-\frac{\sqrt{3}}{18}(35,8,1)_H^{--}\otimes(35,8,2)_l^{+-},
\end{eqnarray}

%%%%%%%%%%%%%%%
\fbox{$1^{-+}(^3P_1)$ of type II}
%%%%%%%%%%%%%%%%

\begin{eqnarray}
|35_{cs},1_c,1\rangle_{(21,\bar{21})} &=& \frac{2\sqrt{6}}{9}(35,8,1)_H^{--}\otimes(35,8,0)_l^{+-}-\frac{\sqrt{2}}{3}(35,8,1)_H^{--}\otimes(35,8,1)_l^{+-}\nonumber\\
&&-\frac{\sqrt{30}}{9}(35,8,1)_H^{--}\otimes(35,8,2)_l^{+-}\nonumber\\
&&+\frac{\sqrt{3}}{9}(35,1,1)_H^{--}\otimes(35,1,0)_l^{+-}-\frac{1}{6}(35,1,1)_H^{--}\otimes(35,1,1)_l^{+-} \nonumber\\
&&-\frac{\sqrt{15}}{18}(35,1,1)_H^{--}\otimes(35,1,2)_l^{+-}\\
|35_{cs},1_c,1\rangle_{(15,\bar{15})} &=& \frac{2\sqrt{6}}{9}(35,8,1)_H^{--}\otimes(35,8,0)_l^{+-}-\frac{\sqrt{2}}{3}(35,8,1)_H^{--}\otimes(35,8,1)_l^{+-}\nonumber\\
&&-\frac{\sqrt{30}}{9}(35,8,1)_H^{--}\otimes(35,8,2)_l^{+-}\nonumber\\
&&+\frac{\sqrt{3}}{9}(35,1,1)_H^{--}\otimes(35,1,0)_l^{+-}-\frac{1}{6}(35,1,1)_H^{--}\otimes(35,1,1)_l^{+-}\nonumber \\
&&-\frac{\sqrt{15}}{18}(35,1,1)_H^{--}\otimes(35,1,2)_l^{+-}\\
|280_{cs},1_c,1\rangle_{(21,\bar{15})} &=& \frac{\sqrt{3}}{9}(35,8,1)_H^{--}\otimes(35,8,0)_l^{+-}-\frac{1}{6}(35,8,1)_H^{--}\otimes(35,8,1)_l^{+-}\nonumber\\
&&-\frac{\sqrt{15}}{18}(35,8,1)_H^{--}\otimes(35,8,2)_l^{+-}\nonumber\\
&&-\frac{2\sqrt{6}}{9}(35,1,1)_H^{--}\otimes(35,1,0)_l^{+-}+\frac{\sqrt{2}}{3}(35,1,1)_H^{--}\otimes(35,1,1)_l^{++} \nonumber\\
&&+\frac{\sqrt{30}}{9}(35,1,1)_H^{--}\otimes(35,1,2)_l^{+-}\\
|280_{cs},1_c,1\rangle_{(15,\bar{21})} &=& \frac{\sqrt{3}}{9}(35,8,1)_H^{--}\otimes(35,8,0)_l^{+-}-\frac{1}{6}(35,8,1)_H^{--}\otimes(35,8,1)_l^{+-}\nonumber\\
&&-\frac{\sqrt{15}}{18}(35,8,1)_H^{--}\otimes(35,8,2)_l^{+-}\nonumber\\
&&-\frac{2\sqrt{6}}{9}(35,1,1)_H^{--}\otimes(35,1,0)_l^{+-}+\frac{\sqrt{2}}{3}(35,1,1)_H^{--}\otimes(35,1,1)_l^{+-}\nonumber \\
&&+\frac{\sqrt{30}}{9}(35,1,1)_H^{--}\otimes(35,1,2)_l^{+-},
\end{eqnarray}

%%%%%%%%%%%%%%%
\fbox{$1^{-+}(^5P_1)$ of type II}
%%%%%%%%%%%%%%%%

\begin{eqnarray}
|405_{cs},1_c,1\rangle_{(21,\bar{21})} &=& \frac{\sqrt{15}}{9}(35,8,1)_H^{--}\otimes(35,8,0)_l^{+-}+\frac{\sqrt{5}}{6}(35,8,1)_H^{--}\otimes(35,8,1)_l^{+-}\nonumber\\
&&+\frac{\sqrt{3}}{18}(35,8,1)_H^{--}\otimes(35,8,2)_l^{+-}\nonumber\\
&&+\frac{\sqrt{30}}{9}(35,1,1)_H^{--}\otimes(35,1,0)_l^{+-}-\frac{\sqrt{10}}{6}(35,1,1)_H^{--}\otimes(35,1,1)_l^{+-} \nonumber\\
&&+\frac{\sqrt{6}}{18}(35,1,1)_H^{--}\otimes(35,1,2)_l^{+-}\\
|189_{cs},1_c,1\rangle_{(15,\bar{15})} &=& \frac{\sqrt{30}}{9}(35,8,1)_H^{--}\otimes(35,8,0)_l^{+-}+\frac{\sqrt{10}}{6}(35,8,1)_H^{--}\otimes(35,8,1)_l^{+-}\nonumber\\
&&+\frac{\sqrt{6}}{18}(35,8,1)_H^{--}\otimes(35,8,2)_l^{+-}\nonumber\\
&&-\frac{\sqrt{15}}{9}(35,1,1)_H^{--}\otimes(35,1,0)_l^{+-}-\frac{\sqrt{5}}{6}(35,1,1)_H^{--}\otimes(35,1,1)_l^{+-}\nonumber \\
&&-\frac{\sqrt{3}}{18}(35,1,1)_H^{--}\otimes(35,1,2)_l^{+-},
\end{eqnarray}

%%%%%%%%%%%%%%%
\fbox{$0^{-+}$ of type I}
%%%%%%%%%%%%%%%%

\begin{eqnarray}
|35_{cs},1_c,0\rangle_{(21,\bar{21})} &=& \frac{\sqrt{2}}{2}(35,1,1)_H^{--}\otimes(1,1,1)_l^{+-}-\frac{\sqrt{2}}{2}(1,1,0)_H^{-+}\otimes(35,1,0)_l^{++} \\
|35_{cs},1_c,0\rangle_{(15,\bar{15})} &=& \frac{\sqrt{2}}{2}(35,1,1)_H^{--}\otimes(1,1,1)_l^{+-}-\frac{\sqrt{2}}{2}(1,1,0)_H^{-+}\otimes(35,1,0)_l^{++} \\
|35_{cs},1_c,0\rangle_{(21,\bar{15})} &=& \frac{\sqrt{3}}{3}(35,1,1)_H^{--}\otimes(1,1,1)_l^{+-}+\frac{\sqrt{6}}{6}(35,8,1)_H^{--}\otimes(35,8,1)_l^{+-} \nonumber\\
&&+\frac{\sqrt{3}}{3}(1,1,0)_H^{-+}\otimes(35,1,0)_l^{++}+\frac{\sqrt{6}}{6}(35,8,0)_H^{-+}\otimes(35,8,0)_l^{++}\\
|35_{cs},1_c,0\rangle_{(15,\bar{21})} &=& \frac{\sqrt{6}}{6}(35,1,1)_H^{--}\otimes(1,1,1)_l^{+-}-\frac{\sqrt{3}}{3}(35,8,1)_H^{--}\otimes(35,8,1)_l^{+-} \nonumber\\
&&+\frac{\sqrt{6}}{6}(1,1,0)_H^{-+}\otimes(35,1,0)_l^{++}-\frac{\sqrt{3}}{3}(35,8,0)_H^{-+}\otimes(35,8,0)_l^{++}\\
|280_{cs},1_c,0\rangle_{(21,\bar{15})} &=& -\frac{\sqrt{2}}{2}(35,8,1)_H^{--}\otimes(35,8,1)_l^{+-}+\frac{\sqrt{2}}{2}(35,8,0)_H^{-+}\otimes(35,8,0)_l^{++} \\
|280_{cs},1_c,0\rangle_{(15,\bar{21})} &=&
-\frac{\sqrt{2}}{2}(35,8,1)_H^{--}\otimes(35,8,1)_l^{+-}+\frac{\sqrt{2}}{2}(35,8,0)_H^{-+}\otimes(35,8,0)_l^{++},
\end{eqnarray}

%%%%%%%%%%%%%%%
\fbox{$0^{-+}$ of type II}
%%%%%%%%%%%%%%%%

\begin{eqnarray}
|35_{cs},1_c,1\rangle_{(21,\bar{21})} &=& -\frac{2\sqrt{2}}{3}(35,8,1)_H^{--}\otimes(35,8,1)_l^{+-}-\frac{1}{3}(35,1,1)_H^{--}\otimes(35,1,1)_l^{+-} \\
|280_{cs},1_c,1\rangle_{(21,\bar{15})} &=& -\frac{1}{3}(35,8,1)_H^{--}\otimes(35,8,1)_l^{+-}+\frac{2\sqrt{2}}{3}(35,1,1)_H^{--}\otimes(35,1,1)_l^{+-}\\
|35_{cs},1_c,1\rangle_{(15,\bar{15})} &=& -\frac{2\sqrt{2}}{3}(35,8,1)_H^{--}\otimes(35,8,1)_l^{+-}-\frac{1}{3}(35,1,1)_H^{--}\otimes(35,1,1)_l^{+-} \\
|280_{cs},1_c,1\rangle_{(15,\bar{21})} &=&
-\frac{1}{3}(35,8,1)_H^{--}\otimes(35,8,1)_l^{+-}+\frac{2\sqrt{2}}{3}(35,1,1)_H^{--}\otimes(35,1,1)_l^{+-},
\end{eqnarray}

%%%%%%%%%%%%%%%
\fbox{$0^{--}$ of type I}
%%%%%%%%%%%%%%%%

\begin{eqnarray}
|35_{cs},1_c,1\rangle_{(21,\bar{21})} &=& -\frac{2\sqrt{2}}{3}(35,8,1)_H^{--}\otimes(35,8,1)_l^{++}-\frac{1}{3}(35,1,1)_H^{--}\otimes(35,1,1)_l^{++} \\
|280_{cs},1_c,1\rangle_{(21,\bar{15})} &=& -\frac{1}{3}(35,8,1)_H^{--}\otimes(35,8,1)_l^{++}+\frac{2\sqrt{2}}{3}(35,1,1)_H^{--}\otimes(35,1,1)_l^{++}\\
|35_{cs},1_c,1\rangle_{(15,\bar{15})} &=& -\frac{2\sqrt{2}}{3}(35,8,1)_H^{--}\otimes(35,8,1)_l^{++}-\frac{1}{3}(35,1,1)_H^{--}\otimes(35,1,1)_l^{++} \\
|280_{cs},1_c,1\rangle_{(15,\bar{21})} &=&
-\frac{1}{3}(35,8,1)_H^{--}\otimes(35,8,1)_l^{++}+\frac{2\sqrt{2}}{3}(35,1,1)_H^{--}\otimes(35,1,1)_l^{++},
\end{eqnarray}

%%%%%%%%%%%%%%%
\fbox{$0^{--}$ of type II}
%%%%%%%%%%%%%%%%

\begin{eqnarray}
|35_{cs},1_c,0\rangle_{(21,\bar{21})} &=& \frac{\sqrt{2}}{2}(35,1,1)_H^{--}\otimes(1,1,1)_l^{++}-\frac{\sqrt{2}}{2}(1,1,0)_H^{-+}\otimes(35,1,0)_l^{+-} \\
|35_{cs},1_c,0\rangle_{(15,\bar{15})} &=& \frac{\sqrt{2}}{2}(35,1,1)_H^{--}\otimes(1,1,1)_l^{++}-\frac{\sqrt{2}}{2}(1,1,0)_H^{-+}\otimes(35,1,0)_l^{+-} \\
|35_{cs},1_c,0\rangle_{(21,\bar{15})} &=& \frac{\sqrt{3}}{3}(35,1,1)_H^{--}\otimes(1,1,1)_l^{++}+\frac{\sqrt{6}}{6}(35,8,1)_H^{--}\otimes(35,8,1)_l^{++} \nonumber\\
&&+\frac{\sqrt{3}}{3}(1,1,0)_H^{-+}\otimes(35,1,0)_l^{+-}+\frac{\sqrt{6}}{6}(35,8,0)_H^{-+}\otimes(35,8,0)_l^{+-}\\
|35_{cs},1_c,0\rangle_{(15,\bar{21})} &=& \frac{\sqrt{6}}{6}(35,1,1)_H^{--}\otimes(1,1,1)_l^{++}-\frac{\sqrt{3}}{3}(35,8,1)_H^{--}\otimes(35,8,1)_l^{++} \nonumber\\
&&+\frac{\sqrt{6}}{6}(1,1,0)_H^{-+}\otimes(35,1,0)_l^{+-}-\frac{\sqrt{3}}{3}(35,8,0)_H^{-+}\otimes(35,8,0)_l^{+-}\\
|280_{cs},1_c,0\rangle_{(21,\bar{15})} &=& -\frac{\sqrt{2}}{2}(35,8,1)_H^{--}\otimes(35,8,1)_l^{++}+\frac{\sqrt{2}}{2}(35,8,0)_H^{-+}\otimes(35,8,0)_l^{+-} \\
|280_{cs},1_c,0\rangle_{(15,\bar{21})} &=&
-\frac{\sqrt{2}}{2}(35,8,1)_H^{--}\otimes(35,8,1)_l^{++}+\frac{\sqrt{2}}{2}(35,8,0)_H^{-+}\otimes(35,8,0)_l^{+-},
\end{eqnarray}

%%%%%%%%%%%%%%%
\section*{Appendix C: The parameters $A_{(1-24)}$ and $B_{(1-24)}$ in Table \ref{tab:3}.}
%%%%%%%%%%%%%%%

\begin{eqnarray*}
A_1 &=& -\frac{\sqrt{7}}{126}H_{\rho}^{10}-\frac{\sqrt{7}}{42}H_{\rho}^{11}-\frac{5\sqrt{7}}{126}H_{\rho}^{12}-\frac{\sqrt{7}}{126}\tilde{H}_{\rho}^{10}-\frac{\sqrt{7}}{42}\tilde{H}_{\rho}^{11}-\frac{5\sqrt{7}}{126}\tilde{H}_{\rho}^{12}, \\
A_2 &=& \frac{\sqrt{21}}{126}H_{\rho}^{10}+\frac{\sqrt{21}}{84}H_{\rho}^{11}-\frac{5\sqrt{21}}{252}H_{\rho}^{12}+\frac{\sqrt{21}}{126}\tilde{H}_{\rho}^{10}+\frac{\sqrt{21}}{84}\tilde{H}_{\rho}^{11}-\frac{5\sqrt{21}}{252}\tilde{H}_{\rho}^{12}, \\
A_3 &=&
-\frac{\sqrt{35}}{126}H_{\rho}^{10}+\frac{\sqrt{35}}{84}H_{\rho}^{11}-\frac{\sqrt{35}}{252}H_{\rho}^{12}-\frac{\sqrt{35}}{126}\tilde{H}_{\rho}^{10}+\frac{\sqrt{35}}{84}\tilde{H}_{\rho}^{11}-\frac{\sqrt{35}}{252}\tilde{H}_{\rho}^{12},
\end{eqnarray*}

\begin{eqnarray*}
A_4 &=& -\frac{2\sqrt{42}}{189}H_{\rho}^{10}-\frac{2\sqrt{42}}{63}H_{\rho}^{11}-\frac{10\sqrt{42}}{189}H_{\rho}^{12}-\frac{5\sqrt{21}}{378}\tilde{H}_{\rho}^{10}+\frac{5\sqrt{21}}{126}\tilde{H}_{\rho}^{11}+\frac{25\sqrt{21}}{378}\tilde{H}_{\rho}^{12}, \\
A_5 &=& \frac{2\sqrt{14}}{63}H_{\rho}^{10}+\frac{\sqrt{14}}{21}H_{\rho}^{11}-\frac{5\sqrt{14}}{63}H_{\rho}^{12}+\frac{5\sqrt{7}}{126}\tilde{H}_{\rho}^{10}-\frac{5\sqrt{7}}{84}\tilde{H}_{\rho}^{11}+\frac{25\sqrt{7}}{252}\tilde{H}_{\rho}^{12}, \\
A_6 &=&
-\frac{2\sqrt{210}}{189}H_{\rho}^{10}+\frac{\sqrt{210}}{63}H_{\rho}^{11}-\frac{\sqrt{210}}{189}H_{\rho}^{12}-\frac{5\sqrt{105}}{378}\tilde{H}_{\rho}^{10}-\frac{5\sqrt{105}}{252}\tilde{H}_{\rho}^{11}+\frac{5\sqrt{105}}{756}\tilde{H}_{\rho}^{12},
\end{eqnarray*}

\begin{eqnarray*}
A_7 &=& -\frac{\sqrt{5}}{90}H_{\rho}^{10}+\frac{\sqrt{5}}{30}H_{\rho}^{11}+\frac{\sqrt{5}}{18}H_{\rho}^{12}+\frac{\sqrt{10}}{45}\tilde{H}_{\rho}^{10}+\frac{\sqrt{10}}{15}\tilde{H}_{\rho}^{11}+\frac{\sqrt{10}}{9}\tilde{H}_{\rho}^{12}, \\
A_8 &=& \frac{\sqrt{15}}{90}H_{\rho}^{10}-\frac{\sqrt{15}}{60}H_{\rho}^{11}+\frac{\sqrt{15}}{36}H_{\rho}^{12}-\frac{\sqrt{30}}{45}\tilde{H}_{\rho}^{10}-\frac{\sqrt{30}}{30}\tilde{H}_{\rho}^{11}+\frac{\sqrt{30}}{18}\tilde{H}_{\rho}^{12}, \\
A_9 &=&
-\frac{1}{18}H_{\rho}^{10}-\frac{1}{12}H_{\rho}^{11}+\frac{1}{36}H_{\rho}^{12}+\frac{\sqrt{2}}{9}\tilde{H}_{\rho}^{10}-\frac{\sqrt{2}}{6}\tilde{H}_{\rho}^{11}+\frac{\sqrt{2}}{18}\tilde{H}_{\rho}^{12},
\end{eqnarray*}

\begin{eqnarray*}
A_{10} &=& -\frac{2\sqrt{30}}{135}H_{\rho}^{10}-\frac{2\sqrt{30}}{45}H_{\rho}^{11}-\frac{2\sqrt{30}}{27}H_{\rho}^{12}-\frac{\sqrt{15}}{270}\tilde{H}_{\rho}^{10}-\frac{\sqrt{15}}{90}\tilde{H}_{\rho}^{11}-\frac{\sqrt{15}}{54}\tilde{H}_{\rho}^{12},\\
A_{11} &=& \frac{2\sqrt{10}}{45}H_{\rho}^{10}+\frac{\sqrt{10}}{15}H_{\rho}^{11}-\frac{\sqrt{10}}{9}H_{\rho}^{12}+\frac{\sqrt{5}}{90}\tilde{H}_{\rho}^{10}+\frac{\sqrt{5}}{10}\tilde{H}_{\rho}^{11}+\frac{\sqrt{5}}{36}\tilde{H}_{\rho}^{12},\\
A_{12} &=&
-\frac{2\sqrt{6}}{27}H_{\rho}^{10}+\frac{\sqrt{6}}{9}H_{\rho}^{11}-\frac{\sqrt{6}}{27}H_{\rho}^{12}-\frac{\sqrt{3}}{54}\tilde{H}_{\rho}^{10}+\frac{\sqrt{3}}{36}\tilde{H}_{\rho}^{11}-\frac{\sqrt{3}}{108}\tilde{H}_{\rho}^{12},
\end{eqnarray*}

\begin{eqnarray*}
A_{13} &=& \frac{\sqrt{3}}{27}H_{\rho}^{10}+\frac{\sqrt{3}}{18}H_{\rho}^{11}-\frac{5\sqrt{3}}{54}H_{\rho}^{12}+\frac{2\sqrt{6}}{27}\tilde{H}_{\rho}^{10}+\frac{\sqrt{6}}{9}\tilde{H}_{\rho}^{11}-\frac{5\sqrt{6}}{27}\tilde{H}_{\rho}^{12},\\
A_{14} &=& -\frac{1}{9}H_{\rho}^{10}-\frac{1}{12}H_{\rho}^{11}-\frac{5}{36}H_{\rho}^{12}-\frac{2\sqrt{2}}{9}\tilde{H}_{\rho}^{10}-\frac{\sqrt{2}}{6}\tilde{H}_{\rho}^{11}-\frac{5\sqrt{2}}{18}\tilde{H}_{\rho}^{12},\\
A_{15} &=&
\frac{\sqrt{15}}{27}H_{\rho}^{10}-\frac{\sqrt{15}}{36}H_{\rho}^{11}-\frac{\sqrt{15}}{108}H_{\rho}^{12}+\frac{2\sqrt{30}}{27}\tilde{H}_{\rho}^{10}-\frac{\sqrt{30}}{18}\tilde{H}_{\rho}^{11}-\frac{\sqrt{30}}{54}\tilde{H}_{\rho}^{12},
\end{eqnarray*}

\begin{eqnarray*}
A_{16} &=& -\frac{2\sqrt{6}}{27}H_{\rho}^{10}-\frac{\sqrt{6}}{9}H_{\rho}^{11}+\frac{5\sqrt{6}}{27}H_{\rho}^{12}+\frac{\sqrt{3}}{27}\tilde{H}_{\rho}^{10}+\frac{\sqrt{3}}{18}\tilde{H}_{\rho}^{11}+\frac{5\sqrt{3}}{54}\tilde{H}_{\rho}^{12},\\
A_{17} &=& \frac{2\sqrt{2}}{9}H_{\rho}^{10}+\frac{\sqrt{2}}{6}H_{\rho}^{11}+\frac{5\sqrt{2}}{18}H_{\rho}^{12}-\frac{1}{9}\tilde{H}_{\rho}^{10}-\frac{1}{12}\tilde{H}_{\rho}^{11}-\frac{5}{36}\tilde{H}_{\rho}^{12},\\
A_{18} &=&
-\frac{2\sqrt{30}}{27}H_{\rho}^{10}+\frac{\sqrt{30}}{18}H_{\rho}^{11}+\frac{\sqrt{30}}{54}H_{\rho}^{12}+\frac{\sqrt{15}}{27}\tilde{H}_{\rho}^{10}-\frac{\sqrt{15}}{36}\tilde{H}_{\rho}^{11}-\frac{\sqrt{15}}{108}\tilde{H}_{\rho}^{12},
\end{eqnarray*}

\begin{eqnarray*}
A_{19} &=& \frac{\sqrt{30}}{27}H_{\rho}^{10}+\frac{\sqrt{30}}{18}H_{\rho}^{11}+\frac{\sqrt{30}}{54}H_{\rho}^{12}+\frac{\sqrt{15}}{27}\tilde{H}_{\rho}^{10}-\frac{\sqrt{15}}{18}\tilde{H}_{\rho}^{11}+\frac{\sqrt{15}}{54}\tilde{H}_{\rho}^{12},\\
A_{20} &=& -\frac{\sqrt{10}}{9}H_{\rho}^{10}-\frac{\sqrt{10}}{12}H_{\rho}^{11}+\frac{\sqrt{10}}{36}H_{\rho}^{12}-\frac{\sqrt{5}}{9}\tilde{H}_{\rho}^{10}+\frac{\sqrt{5}}{12}\tilde{H}_{\rho}^{11}+\frac{\sqrt{5}}{36}\tilde{H}_{\rho}^{12},\\
A_{21} &=&
\frac{5\sqrt{6}}{27}H_{\rho}^{10}+\frac{5\sqrt{6}}{36}H_{\rho}^{11}+\frac{\sqrt{6}}{108}H_{\rho}^{12}+\frac{5\sqrt{3}}{27}\tilde{H}_{\rho}^{10}+\frac{5\sqrt{3}}{36}\tilde{H}_{\rho}^{11}+\frac{\sqrt{3}}{108}\tilde{H}_{\rho}^{12},
\end{eqnarray*}

\begin{eqnarray*}
A_{22} &=& -\frac{\sqrt{15}}{27}H_{\rho}^{10}+\frac{\sqrt{15}}{18}H_{\rho}^{11}-\frac{\sqrt{15}}{54}H_{\rho}^{12}+\frac{\sqrt{30}}{27}\tilde{H}_{\rho}^{10}-\frac{\sqrt{30}}{18}\tilde{H}_{\rho}^{11}+\frac{\sqrt{30}}{54}\tilde{H}_{\rho}^{12},\\
A_{23} &=& \frac{\sqrt{5}}{9}H_{\rho}^{10}+\frac{\sqrt{5}}{12}H_{\rho}^{11}-\frac{\sqrt{5}}{36}H_{\rho}^{12}-\frac{\sqrt{10}}{9}\tilde{H}_{\rho}^{10}+\frac{\sqrt{10}}{12}\tilde{H}_{\rho}^{11}+\frac{\sqrt{10}}{36}\tilde{H}_{\rho}^{12},\\
A_{24} &=&
-\frac{5\sqrt{3}}{27}H_{\rho}^{10}-\frac{5\sqrt{3}}{36}H_{\rho}^{11}-\frac{\sqrt{3}}{108}H_{\rho}^{12}+\frac{5\sqrt{6}}{27}\tilde{H}_{\rho}^{10}+\frac{5\sqrt{6}}{36}\tilde{H}_{\rho}^{11}+\frac{\sqrt{6}}{108}\tilde{H}_{\rho}^{12},
\end{eqnarray*}

\begin{eqnarray*}
A_{25} &=& -\frac{\sqrt{15}}{27}H_{\rho}^{10}+\frac{\sqrt{15}}{18}H_{\rho}^{11}-\frac{\sqrt{15}}{54}H_{\rho}^{12}+\frac{\sqrt{30}}{27}\tilde{H}_{\rho}^{10}-\frac{\sqrt{30}}{18}\tilde{H}_{\rho}^{11}+\frac{\sqrt{30}}{54}\tilde{H}_{\rho}^{12},\\
A_{23} &=& \frac{\sqrt{5}}{9}H_{\rho}^{10}+\frac{\sqrt{5}}{12}H_{\rho}^{11}-\frac{\sqrt{5}}{36}H_{\rho}^{12}-\frac{\sqrt{10}}{9}\tilde{H}_{\rho}^{10}+\frac{\sqrt{10}}{12}\tilde{H}_{\rho}^{11}+\frac{\sqrt{10}}{36}\tilde{H}_{\rho}^{12},\\
A_{24} &=&
-\frac{5\sqrt{3}}{27}H_{\rho}^{10}-\frac{5\sqrt{3}}{36}H_{\rho}^{11}-\frac{\sqrt{3}}{108}H_{\rho}^{12}+\frac{5\sqrt{6}}{27}\tilde{H}_{\rho}^{10}+\frac{5\sqrt{6}}{36}\tilde{H}_{\rho}^{11}+\frac{\sqrt{6}}{108}\tilde{H}_{\rho}^{12},
\end{eqnarray*}

\begin{eqnarray*}
B_1 &=& -\frac{\sqrt{21}}{84}H_{\pi}^{21}-\frac{\sqrt{105}}{84}H_{\pi}^{21}-\frac{\sqrt{21}}{84}\tilde{H}_{\pi}^{21}-\frac{\sqrt{105}}{84}\tilde{H}_{\pi}^{21}, \\
B_2 &=&  \frac{\sqrt{63}}{84}H_{\pi}^{21}-\frac{\sqrt{35}}{84}H_{\pi}^{21}+\frac{\sqrt{63}}{84}\tilde{H}_{\pi}^{21}-\frac{\sqrt{35}}{84}\tilde{H}_{\pi}^{21}, \\
B_3 &=&  -\frac{\sqrt{14}}{21}H_{\pi}^{21}-\frac{\sqrt{70}}{21}H_{\pi}^{21}+\frac{5\sqrt{7}}{84}\tilde{H}_{\pi}^{21}+\frac{5\sqrt{35}}{84}\tilde{H}_{\pi}^{21}, \\
B_4 &=& \frac{\sqrt{42}}{21}H_{\pi}^{21}-\frac{\sqrt{210}}{63}H_{\pi}^{21}-\frac{5\sqrt{21}}{84}\tilde{H}_{\pi}^{21}+\frac{5\sqrt{105}}{252}\tilde{H}_{\pi}^{21},\\
B_5 &=& \frac{\sqrt{15}}{60}H_{\pi}^{21}+\frac{\sqrt{3}}{12}H_{\pi}^{21}+\frac{\sqrt{30}}{30}\tilde{H}_{\pi}^{21}+\frac{\sqrt{6}}{6}\tilde{H}_{\pi}^{21},\\
B_6 &=&
-\frac{\sqrt{15}}{20}H_{\pi}^{21}+\frac{1}{12}H_{\pi}^{21}-\frac{\sqrt{10}}{10}\tilde{H}_{\pi}^{21}+\frac{\sqrt{2}}{6}\tilde{H}_{\pi}^{21},
\end{eqnarray*}

\begin{eqnarray*}
B_7 &=& -\frac{\sqrt{10}}{15}H_{\pi}^{21}-\frac{\sqrt{2}}{3}H_{\pi}^{21}-\frac{\sqrt{5}}{60}\tilde{H}_{\pi}^{21}-\frac{1}{12}\tilde{H}_{\pi}^{21},\\
B_8 &=& \frac{\sqrt{30}}{15}H_{\pi}^{21}-\frac{\sqrt{6}}{9}H_{\pi}^{21}+\frac{\sqrt{15}}{60}\tilde{H}_{\pi}^{21}-\frac{\sqrt{3}}{36}\tilde{H}_{\pi}^{21},\\
B_9 &=& \frac{1}{12}H_{\pi}^{21}-\frac{\sqrt{5}}{12}H_{\pi}^{21}+\frac{\sqrt{2}}{6}\tilde{H}_{\pi}^{21}-\frac{\sqrt{10}}{6}\tilde{H}_{\pi}^{21},\\
B_{10} &=& -\frac{\sqrt{3}}{12}H_{\pi}^{21}-\frac{\sqrt{15}}{36}H_{\pi}^{21}-\frac{\sqrt{6}}{6}\tilde{H}_{\pi}^{21}-\frac{\sqrt{30}}{18}\tilde{H}_{\pi}^{21},\\
B_{11} &=& -\frac{\sqrt{2}}{6}H_{\pi}^{21}+\frac{\sqrt{10}}{6}H_{\pi}^{21}+\frac{1}{12}\tilde{H}_{\pi}^{21}-\frac{\sqrt{5}}{12}\tilde{H}_{\pi}^{21},\\
B_{12} &=&
\frac{\sqrt{6}}{6}H_{\pi}^{21}+\frac{\sqrt{30}}{18}H_{\pi}^{21}-\frac{\sqrt{3}}{12}\tilde{H}_{\pi}^{21}-\frac{\sqrt{15}}{36}\tilde{H}_{\pi}^{21},
\end{eqnarray*}

\begin{eqnarray*}
B_{13} &=& \frac{\sqrt{10}}{12}H_{\pi}^{21}+\frac{\sqrt{2}}{12}H_{\pi}^{21}-\frac{\sqrt{5}}{12}\tilde{H}_{\pi}^{21}+\frac{1}{12}\tilde{H}_{\pi}^{21},\\
B_{14} &=& -\frac{\sqrt{30}}{12}H_{\pi}^{21}+\frac{\sqrt{6}}{36}H_{\pi}^{21}+\frac{\sqrt{15}}{12}\tilde{H}_{\pi}^{21}+\frac{\sqrt{3}}{36}\tilde{H}_{\pi}^{21},\\
B_{15} &=& \frac{\sqrt{5}}{12}H_{\pi}^{21}-\frac{1}{12}H_{\pi}^{21}-\frac{\sqrt{10}}{12}\tilde{H}_{\pi}^{21}+\frac{\sqrt{2}}{12}\tilde{H}_{\pi}^{21},\\
B_{16} &=&
-\frac{\sqrt{15}}{12}H_{\pi}^{21}-\frac{\sqrt{3}}{36}H_{\pi}^{21}+\frac{\sqrt{30}}{12}\tilde{H}_{\pi}^{21}+\frac{\sqrt{6}}{36}\tilde{H}_{\pi}^{21},
\end{eqnarray*}

\begin{eqnarray*}
B_{17} &=& -\frac{\sqrt{2}}{4}H_{\pi}^{21},\\
B_{18} &=& \frac{\sqrt{6}}{4}H_{\pi}^{21},\\
B_{19} &=&-\frac{\sqrt{3}}{6}H_{\pi}^{21}-\frac{\sqrt{6}}{12}\tilde{H}_{\pi}^{21},\\
B_{20} &=&\frac{1}{2}H_{\pi}^{21}+\frac{\sqrt{2}}{4}\tilde{H}_{\pi}^{21},\\
B_{21} &=&-\frac{\sqrt{6}}{12}H_{\pi}^{21}+\frac{\sqrt{3}}{6}\tilde{H}_{\pi}^{21},\\
B_{22} &=&\frac{\sqrt{2}}{4}H_{\pi}^{21}-\frac{1}{2}\tilde{H}_{\pi}^{21},\\
B_{23} &=&-\frac{\sqrt{2}}{4}\tilde{H}_{\pi}^{21},\\
B_{24} &=&\frac{\sqrt{6}}{4}\tilde{H}_{\pi}^{21},
\end{eqnarray*}

%%%%%%%%%%%%%%%
\section*{Appendix D: Spin configurations of the open charm final states}
%%%%%%%%%%%%%%%

When the tetraquarks decay into the open charm modes, the decay
matrix elements contain many terms. In order to avoid the
complicated and lengthy expressions, we only collect the spin
configurations of various open charm final states below. One can
first compare the spin configurations of the S-wave and P-wave
hidden-charm tetraquark states in Appendix A and different open
charm final states. If the initial and final states have one or more
common spin configurations, such a strong mode is allowed under the
heavy quark symmetry. Otherwise, such a decay is suppressed.

%%%%%%%%%%%%%%%%%%
\fbox{S-wave $D^{(*)} \bar D^{(*)}$}
%%%%%%%%%%%%%%%%%%%

\begin{eqnarray*}
B\bar{B}\,\,(0^{++}) &=& \frac{1}{2}(0_H\otimes0_l)_0^{++}+\frac{\sqrt{3}}{2}(1_H\otimes1_l)_0^{++},\\
B^{\ast}\bar{B}^{\ast}\,\,(0^{++}) &=& \frac{\sqrt{3}}{2}(0_H\otimes0_l)_0^{++}-\frac{1}{2}(1_H\otimes1_l)_0^{++},\\
B\bar{B}^{\ast}\,\,(1^{+-}) &=& \frac{\sqrt{2}}{2}(0_H\otimes1_l)_1^{+-}-\frac{\sqrt{2}}{2}(1_H\otimes0_l)_1^{+-},\\
B^{\ast}\bar{B}^{\ast}\,\,(1^{+-}) &=& \frac{\sqrt{2}}{2}(0_H\otimes1_l)_1^{+-}+\frac{\sqrt{2}}{2}(1_H\otimes0_l)_1^{+-},\\
B\bar{B}^{\ast}\,\,(1^{++}) &=& (1_H\otimes1_l)_1^{++},\\
B^{\ast}\bar{B}^{\ast}\,\,(2^{++}) &=& (1_H\otimes1_l)_2^{++},
\end{eqnarray*}

%%%%%%%%%%%%%%%%%%
\fbox{P-wave $D^{(*)} \bar D^{(*)}$}
%%%%%%%%%%%%%%%%%%%

\begin{eqnarray*}
D\bar{D}\,\,\{^1P_1\} &=& \frac{1}{2}(0_H\otimes1_l)_1^{--}+\frac{\sqrt{3}}{6}(1_H\otimes0_l)_1^{--}-\frac{1}{2}(1_H\otimes1_l)_1^{--}+\frac{\sqrt{15}}{6}(1_H\otimes2_l)_1^{--},\\
D^{\ast}\bar{D}^{\ast}\,\,\{^1P_1\} &=& \frac{\sqrt{3}}{2}(0_H\otimes1_l)_1^{--}-\frac{1}{6}(1_H\otimes0_l)_1^{--}+\frac{\sqrt{3}}{6}(1_H\otimes1_l)_1^{--}-\frac{\sqrt{5}}{6}(1_H\otimes2_l)_1^{--},\\
D\bar{D}^{\ast}\,\,\{^3P_1\} &=& \frac{\sqrt{2}}{2}(0_H\otimes1_l)_1^{--}-\frac{\sqrt{2}}{2}(1_H\otimes1_l)_1^{--},\\
D^{\ast}\bar{D}^{\ast}\,\,\{^5P_1\} &=&
\frac{1}{3}(1_H\otimes0_l)_1^{--}-\frac{\sqrt{3}}{3}(1_H\otimes1_l)_1^{--}+\frac{\sqrt{5}}{3}(1_H\otimes2_l)_1^{--},
\end{eqnarray*}
where the label ${^1P_1}$ in the above denote that the spins of $D$
and $\bar{D}$ are coupled into the total spin $0$, then the total
spin $0$ is coupled with the P-wave orbital angular momentum into
the total angular momentum $1$. The labels ${^3P_1}$ and ${^5P_1}$
have the similar meanings with ${^1P_1}$.

%%%%%%%%%%%%%%%%%%
\fbox{S-wave $D_{0,1,2}^{(*)} \bar D^{(*)}$}
%%%%%%%%%%%%%%%%%%%

\begin{eqnarray*}
D_0\bar{D}^{\ast}+c.c. &=& \frac{1}{\sqrt{2}}[(0_H\otimes1_l)_1^{--}+(1_H\otimes1_l)_1^{--}],\\
D'_1\bar{D}^{\ast}+c.c. &=& (0_H\otimes1_l)_1^{--},\\
D_1\bar{D}^{\ast}+c.c. &=& \frac{1}{\sqrt{10}}[-(0_H\otimes1_l)_1^{--}+3(1_H\otimes1_l)_1^{--}],\\
D_2\bar{D}^{\ast}+c.c. &=& -\frac{1}{\sqrt{2}}[(0_H\otimes1_l)_1^{--}+(1_H\otimes1_l)_1^{--}],\\
D'_1\bar{D}+c.c. &=& \frac{1}{\sqrt{2}}[-(0_H\otimes1_l)_1^{--}+(1_H\otimes1_l)_1^{--}],\\
D_1\bar{D}+c.c. &=&
\frac{1}{\sqrt{2}}[-(0_H\otimes1_l)_1^{--}+(1_H\otimes1_l)_1^{--}].
\end{eqnarray*}

\end{document}